\documentclass[12pt]{article}

\RequirePackage[OT1]{fontenc}
\usepackage{natbib}
\RequirePackage[colorlinks,citecolor=blue,urlcolor=blue]{hyperref}
\usepackage[english]{babel}
\usepackage{amsthm}
\usepackage{amsmath}
\usepackage{amsfonts}
\usepackage{amssymb}
\usepackage{graphicx}
\usepackage{enumerate}
\usepackage{color}
\usepackage{array}
\usepackage{psfrag,epsf}
\usepackage{url} 
\usepackage[export]{adjustbox}
\usepackage{bm}
\usepackage{titling}
\usepackage{algorithm}
\usepackage{algorithmic}
\usepackage{mathtools}

\newtheorem{theorem}{Theorem}
\newtheorem{remark}{Remark}
\newtheorem{lemma}{Lemma}

\begin{document}
\thispagestyle{empty}
\baselineskip=28pt
\vskip 5mm
\begin{center} {\large{\bf A semiparametric spatiotemporal Bayesian model for the bulk and extremes of the Fosberg Fire Weather Index}}
\end{center}


\baselineskip=12pt
\vskip 5mm

\begin{center}
Arnab Hazra$^1$, Brian J. Reich$^2$, Benjamin A. Shaby$^3$, and Ana-Maria Staicu$^2$
\end{center}

\footnotetext[1]{
\baselineskip=10pt Computer, Electrical and Mathematical Sciences and Engineering (CEMSE) Division, King Abdullah University of Science and Technology (KAUST), Thuwal 23955-6900, Saudi Arabia. \\ E-mail: arnab.hazra@kaust.edu.sa}

\footnotetext[2]{
\baselineskip=10pt Department of Statistics, North Carolina State University, Raleigh, United States. \\ E-mails: bjreich@ncsu.edu; astaicu@ncsu.edu}
\footnotetext[3]{
\baselineskip=10pt Department of Statistics, Colorado State University, Fort Collins, United States. \\ E-mail: bshaby@colostate.edu}


\baselineskip=17pt
\vskip 4mm
\centerline{\today}
\vskip 6mm

\begin{center}
{\large{\bf Abstract}}
\end{center}

Large wildfires pose a major environmental concern, and precise maps of fire risk can improve disaster relief planning. Fosberg Fire Weather Index (FFWI) is often used to measure wildfire risk; FFWI exhibits non-Gaussian marginal distributions as well as strong spatiotemporal extremal dependence and thus, modeling FFWI using geostatistical models like Gaussian processes is questionable. Extreme value theory (EVT)-driven models like max-stable processes are theoretically appealing but are computationally demanding and applicable only for threshold exceedances or block maxima. Disaster management policies often consider moderate-to-extreme quantiles of climate parameters and hence, joint modeling of the bulk and the tail of the data is required. In this paper, we consider a Dirichlet process mixture of spatial skew-$t$ processes that can flexibly model the bulk as well as the tail. The proposed model has nonstationary mean and covariance structure, and also nonzero spatiotemporal extremal dependence. A simulation study demonstrates that the proposed model has better spatial prediction performance compared to some competing models. We develop spatial maps of FFWI medians and extremes, and discuss the wildfire risk throughout the Santa Ana region of California.

\baselineskip=16pt

\par\vfill\noindent

\pagenumbering{arabic}
\baselineskip=25pt

\newpage

\section{Introduction}
Southern California is susceptible to catastrophic wildfires which are often caused by Santa Ana winds. During the late fall and winter, the Santa Ana winds originate from the Great Basin and heat up as they cross the mountains, move towards the coast due to offshore surface pressure gradients and often lead to wildfires \citep{raphael2003santa}. This phenomenon is most common in December \citep{hughes2010local}. As the whole Santa Ana region is a small geographic domain, the fire risk is likely to be high throughout the region on a particular day of extreme weather. Therefore, models for analyzing fire risk should be capable of exhibiting spatial dependence between the station-wise extremes.




Fosberg Fire Weather Index (FFWI) is a well-established measure that quantifies the potential influence of important weather parameters on fire risk \citep{fosberg1978ffwi}. It is a nonlinear function of air temperature, wind speed and relative humidity. The National Oceanic and Atmospheric Administration (NOAA) considers $\textrm{FFWI}$ larger than 50 to be significant. The Storm Prediction Center (SPC) fire weather verification scheme (\url{http://www.spc.noaa.gov}) uses FFWI for fire danger rating ranging between high to extremes.  Disaster management policies often consider moderate-to-extreme quantiles of weather parameters and hence require modeling of the bulk as well as the tail \citep{dey2016extreme} of the FFWI observations using a proper spatiotemporal model.



Gaussian processes (GPs) are by far the most common models in spatial statistics due to their good theoretical properties, tractability in high-dimensions and their computational ease \citep{gelfand2016spatial}. However, GPs are often criticized for modeling spatial extremes because the induced extremal dependence (a function of the probability that two sites are simultaneously extreme) between any two spatial locations is zero \citep{davison2013geostatistics}. Even after normalization of the data, if GPs are used to model the transformed data, the induced extremal dependence between two spatial locations remains zero \citep{schmidt2017spatiotemporal}. \cite{gelfand2005bayesian} propose a flexible nonparametric and nonstationary model based on Dirichlet process mixture (DPM) of GPs which relaxes the Gaussian assumption; however the spatial extremal dependence remains zero. As a result, when the data shows evidences of spatial extremal dependence, even if the model of \cite{gelfand2005bayesian} is appropriate for modeling the bulk of the spatial data, its suitability is questionable for modeling the extremes. 


\vspace{-1mm}

Literature on spatial modeling of extremes covers a number of approaches like Bayesian hierarchical models \citep{sang2009hierarchical, sang2010continuous, turkman2010asymptotic}, copula-based approaches \citep{ribatet2013extreme, fuentes2013nonparametric, genest2012copulas} and max-stable processes \citep{reich2012hierarchical, mathieu2013spatial, davison2015statistics}. \cite{davison2012statistical} compare many of these approaches and recommend max-stable processes for modeling of the spatial extremes. In spite of their appealing theoretical properties, max-stable processes are not preferred for drawing inference, because they involve dealing with the joint density of a multivariate generalized extreme value (GEV) distribution which is difficult to calculate unless the dimension is small \citep{padoan2010likelihood}. \cite{thibaud2016bayesian} propose a hierarchical Bayesian model approach for approximating the full joint distribution, though their approach is still computationally intensive. A nonparametric copula-based model is proposed by \cite{fuentes2013nonparametric} where the spatial dependence between the extremes is modeled by DPM of GPs with marginal distributions are modeled by GEV. \cite{krupskii2016factor} propose factor copula models that can model tail dependence and tail asymmetry though the model is parametric. A class of sub-asymptotic models is developed by \cite{huser2017bridging} using random scaling of GPs for modeling tail dependence, and further extended by \cite{huser2019modeling}, while \cite{engelke2019extremal} discuss different asymptotic regimes obtained by random scale mixtures. \cite{morris2017space} propose a spatiotemporal skew-$t$ process (STP) model, a location-scale mixture of GPs, for threshold exceedances.





\vspace{-1mm}

Mixture models have been used extensively in univariate extreme value analysis.  Models that splice a generalized Pareto tail to a different ``bulk distribution'' were proposed by \citet{behrens-2004a} and extended by \citet{carreau-2009a}, \citet{macdonald-2011a}, \citet{hu-2018a}, and others. \cite{naveau2016modeling} propose a family of distributions for modeling low, moderate as well as heavy rainfall intensities jointly without any thresholding; two tails follow Pareto-type decay but the bulk also share the same parameters. In the time series context, \citet{shaby-2016a} use a dependent mixture of a normal bulk and generalized Pareto tail to model heat waves. Nevertheless, the choice of ``high thresholds" is arbitrary - for example, \citet{shaby-2016a} define the threshold by the $0.98^{th}$ data quantile. Besides, their ideas of Markov switching are not readily applicable in spatial extremes, and computational burden is an issue.



In this paper, we propose a DPM of STPs to model spatiotemporal FFWI data obtained across the Santa Ana region of southern California, with particular emphasis on modeling the joint probability that multiple sites are at high risk. This approach differs from the current prevailing practice in extreme event analysis in that we model the entire spatial process above a very low threshold; the current state of the art either throws away or censors all the observations that are not considered extreme. Our mixture approach seeks the advantages of ``letting the tail speak for itself'' without sacrificing the ability to consider non-extreme events. It simultaneously probabilistically clusters events and estimates cluster-dependent parameters, so that events that land in the extreme cluster with high probability influence the fit in the tail, while events that land in other clusters with high probability do not. Taken together, the mixture components can both flexibly model the bulk and the tail in an uncontaminated way.  Furthermore, each mixture component is a STP, which is itself considerably more flexible than standard tools like GPs, and allows spatial extremal dependence. Inference is drawn based on Markov chain Monte Carlo sampling. We perform a simulation study and a leave-one-out cross-validation study to compare the proposed model with some competing models. Finally, we use our model to make spatial maps of FFWI medians and extremes.







\section{Fosberg Fire Weather Index data}
\label{data}
\begin{figure}
        \adjincludegraphics[height = 0.36\linewidth, width = 0.54\linewidth, trim = {{.0\width} {.0\width} 0 {.0\width}}, clip]{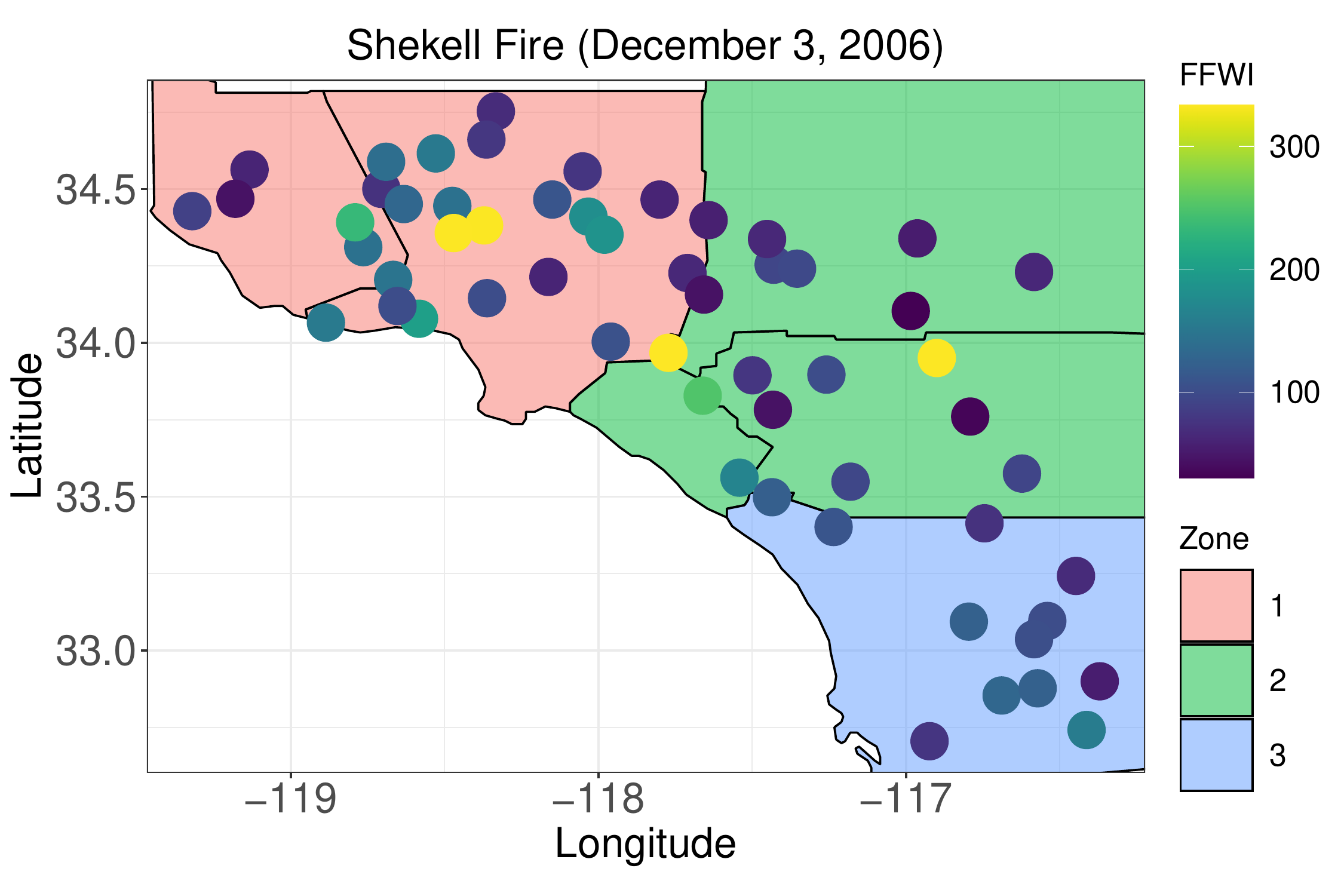}
        \adjincludegraphics[height = 0.36\linewidth, width = 0.45\linewidth, trim = {{.0\width} {0.0\width} 0 {0.0\width}}, clip]{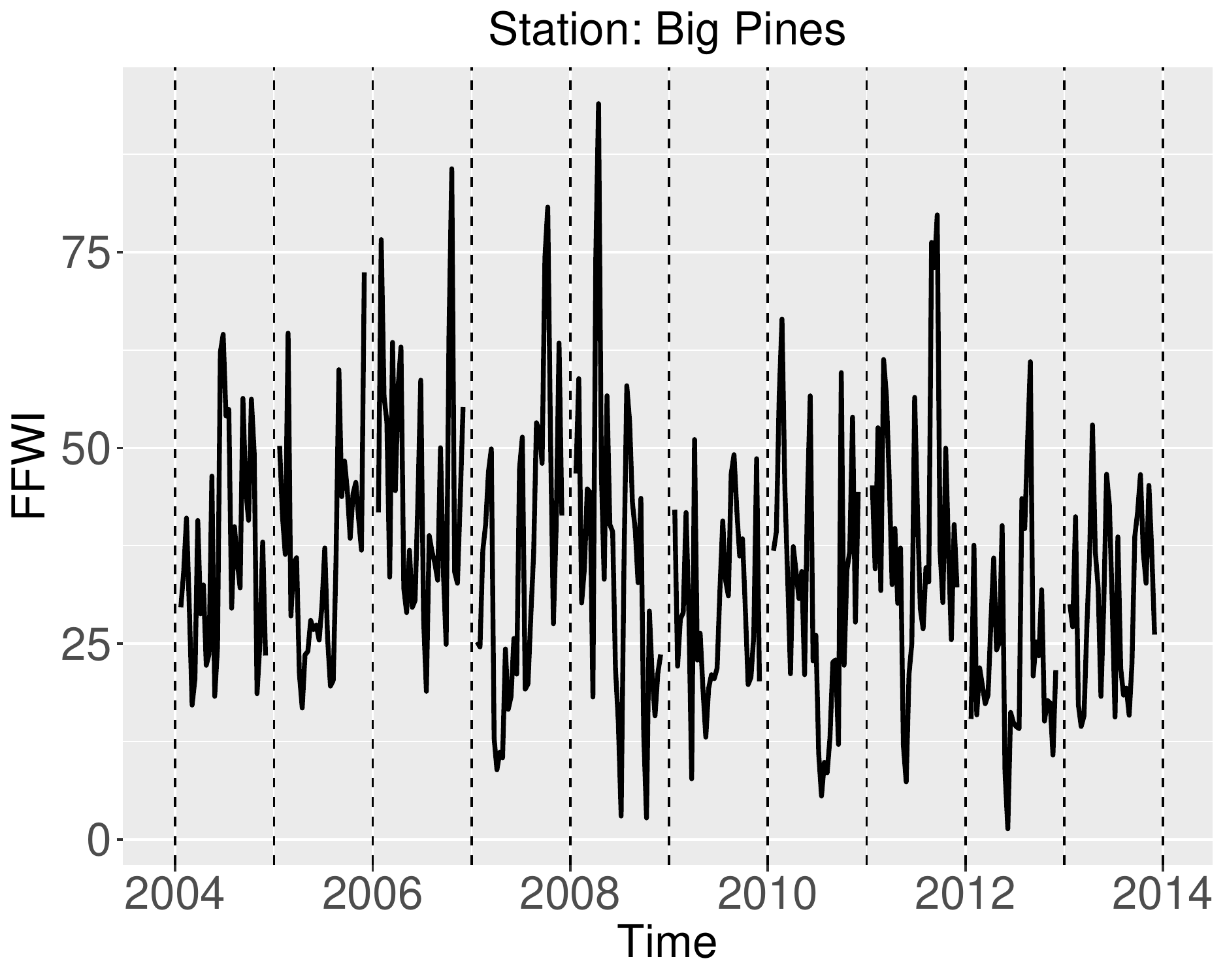}
        \vspace{-1cm}
        \caption{The FFWI observations at 61 RAWS on December 3, 2006, the day of Shekell Fire (left). The time series (only December) of the daily maximum FFWI at Big Pines, California (right).}
		\label{fig1}
		\vspace{-3mm}
	\end{figure}

Proposed by \cite{fosberg1978ffwi}, Fosberg Fire Weather Index (FFWI) is a nonlinear function of air temperature, wind speed and relative humidity (the functional form is provided in the Supplementary Materials (SM, henceforth) Appendix A).
We consider hourly FFWI data from 61 Remotely Allocated Monitoring Stations (RAWS) across the Santa Ana region of southern California, over the period December 31, 2003 through December 31, 2013. More details about RAWS are available at \url{https://www.wrh.noaa.gov/sto/obsmap.php}. Based on the definition of fire zones of the Santa Ana region by \cite{rolinski2016santa}, we divide the whole region into three zones (presented in Figure \ref{fig1})-- Zone 1 includes some parts of Ventura and Los Angeles Counties, Zone 2 includes some parts of Orange, San Bernardino and Riverside Counties, while Zone 3 includes a large region within the San Diego County. While FFWI is not truncated at 100 in general \citep{roads1991medium, sapsis2016fire}, some authors consider truncation at 100, considering it as a threshold for the extreme fire situation \citep{kambezidis2016fire}. Here we use the raw index without truncation. As a first step for data preprocessing, we consider daily maxima of the hourly observations. This step follows the protocols of National Fire Danger Rating System (NFDRS), where daily maxima are considered to be the representative of the combined typical active burning period conditions. Second, we consider only the observations in December as the probability of large katabatic forcing due to strong temperature gradient, the source of the Santa Ana winds, is highest in December \citep{hughes2010local}. Third, as mentioned by  \cite{sapsis2016fire}, even if colder conditions support fire growth in some areas of the United States, it is not true in California and hence, following \cite{sapsis2016fire}, we discard the observations where the recorded air temperature is less than 50$^\circ$F, for model fitting and prediction.




The observed FFWI values at 61 RAWS corresponding to the day of Shekell Fire (December 3, 2006) in the Ventura county, are provided in the left panel of Figure \ref{fig1}. Large values of FFWI across the stations indicate a strong correlation between high FFWI and fire events. In the right panel of Figure \ref{fig1}, the time series of FFWI (only December) at Big Pines, is provided. The time series appears to be fairly stationary across the years. Similar stationary pattern is observed for most of the stations. 

To motivate the need for a model with spatial and temporal extremal dependence, we compute empirical estimates of the extremal dependence in space and time. The extremal dependence between two random variables $Y_1$ and $Y_2$ is often quantified using the $\chi$-measure \citep{sibuya1960bivariate} defined as $\chi = \lim_{u \rightarrow 1} \chi_u$, where
\begin{eqnarray} \label{chi_def}
\chi_u = \textrm{Pr}\left[Y_1 > F_1^{-1}(u) | Y_2 > F_2^{-1}(u)\right],
\end{eqnarray}
where $F_1$ and $F_2$ are the marginal distribution functions of $Y_1$ and $Y_2$, respectively. A positive value of $\chi$ indicates asymptotic dependence, while $\chi=0$ defines asymptotic independence. 
For a spatiotemporal process, the extremal dependence between two spatial locations $\bm{s}_1$ and $\bm{s}_2$ is defined by $\chi(\bm{s}_1, \bm{s}_2)$. When the spatial extremal dependence is isotropic, it is defined by $\chi(h)$ for $h = \lVert \bm{s}_1 - \bm{s}_2 \rVert$, where $\lVert \cdot \rVert$ is some notion of distance (Mahalanobis distance, for example). Both the forms assume the spatial extremal dependence to be time-invariant. Similarly, for a spatial location $\bm{s}$, the temporal extremal dependence between two time points $t_1$ and $t_2$, is defined by $\chi_{\bm{s}}(t_1, t_2)$ and further, assuming the $\chi$-measure being stationary in time and spatially-invariant, we denote it by $\chi^{(\delta)}$,  where $\delta = \lvert t_1 - t_2 \rvert$. While $\chi$ is defined as a limit, for exploratory purposes, we need to fix $u$ to some value close to one and estimate $\chi$ empirically; we choose $u=0.95$.

\begin{figure}
		\adjincludegraphics[height = 0.33\linewidth, width = 0.5\linewidth, trim = {{.0\width} {.0\width} 0 {.0\width}}, clip]{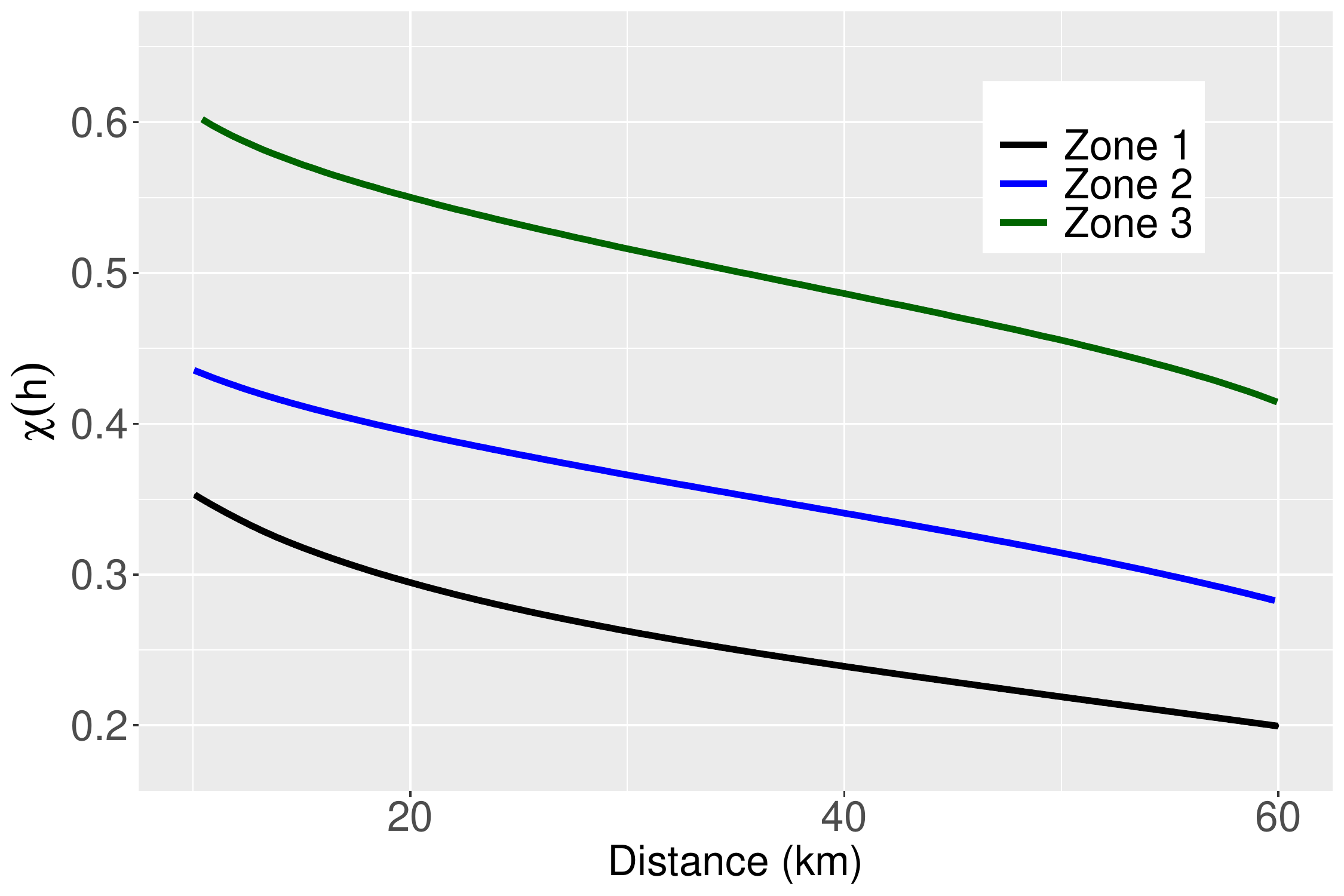}
		\adjincludegraphics[height = 0.33\linewidth, width = 0.5\linewidth, trim = {{.0\width} {.0\width} 0 {.0\width}}, clip]{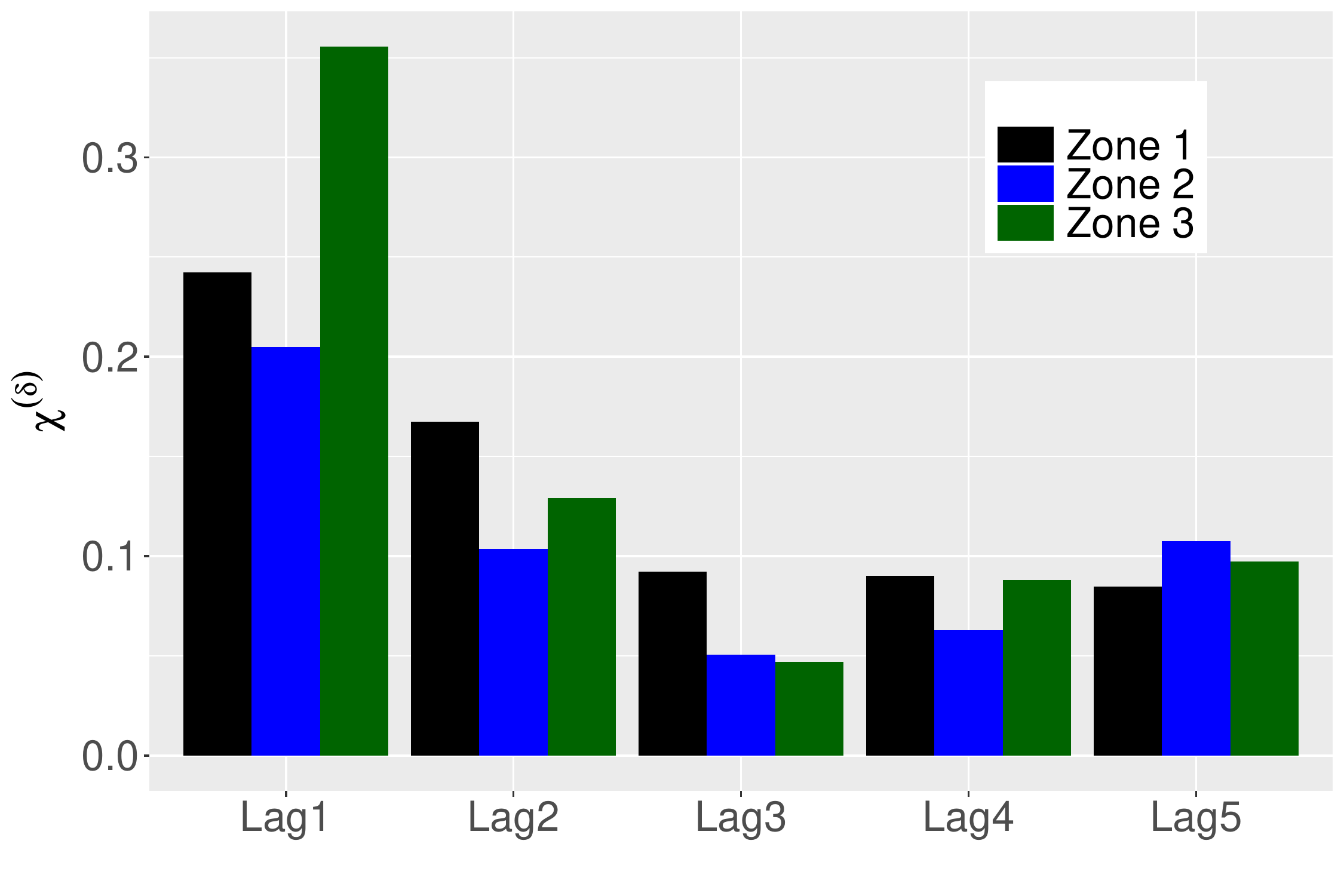}       
		\caption{The spatial extremal dependence (left panel) and temporal extremal dependence (right panel) considering different fire zones of Santa Ana region defined by \cite{rolinski2016santa}.}
		\label{fig2}
\end{figure}

The empirical spatial extremal dependence and (spatially averaged) temporal extremal dependence are presented in Figure \ref{fig2} considering the different fire zones defined by \cite{rolinski2016santa}. From Figure \ref{fig2}, it appears that $\chi(h)$ varies between 0.2 and 0.6, while $\chi^{(\delta)}$ varies between 0.05 and 0.36, and $\chi(h)$ varies more than $\chi^{(\delta)}$ across the zones. 
Due to the nonzero spatial and temporal extremal dependence, inferences based on models having no asymptotic dependence like GPs are questionable.

SM Appendix B contains additional exploratory analyses that show different spatial trends of the sample quantiles at moderate and high quantile levels, and also includes results based on fitting mixture normal distributions at each site, where multiple components are selected for 60 out of 61 sites. These motivate us to build a semiparametric model over a parametric model like a GP, with separate parameters in the bulk and tail, that allows spatial as well as temporal extremal dependence.

\section{Methodology}
\label{methodology}

In Section \ref{skewt} and Section \ref{dpm}, we develop a semiparametric hierarchical Bayesian spatial model using a Dirichlet process mixture of STPs assuming temporal independence. In Section \ref{spacetime}, we propose an extension to accommodate the temporal extremal dependence.


\subsection{Spatial skew-$t$ process}
\label{skewt}
Let $Y^*_t(\bm{s})$ be the (potentially transformed, see Section \ref{dpm}) observation at the monitoring site located at $\bm{s}$ and time $t$. 
We propose to model $Y_t^*(\cdot)$ using STP, which can accommodate heavy tails and asymmetry in the marginal distribution, and non-trivial asymptotic spatial dependence for extremes. Thus STP represent a more appealing alternative to GP, when dealing with data that exhibit such properties \citep{padoan2011multivariate}. We describe the STP using the formulation proposed by  \cite{morris2017space}, as a location-scale mixture of a GP that is inspired by additive processes \citep{azzalini2003distributions, azzalini2014skew}. Let $Y_t^*(\cdot)$ be a STP defined over the spatial domain of interest $\mathcal{D} \subset \mathbb{R}^2$, for given time $t$. At a spatial location $\bm{s}$, we write 
\begin{eqnarray} \label{stp_hierarchical}
Y^*_t(\bm{s}) = \mu(\bm{s}) + \lambda(\bm{s}) \sigma_t |z_t| + \sigma_t \epsilon_t(\bm{s}),
\end{eqnarray} 
where $\mu(\cdot)$ is the spatially-varying mean process, $\lambda(\cdot)$ is the spatially-varying skewness parameter, $z_t \overset{\textrm{iid}}{\sim} \textrm{Normal}(0, 1)$ and $\sigma_t^2 \overset{\textrm{iid}}{\sim} \textrm{Inverse-Gamma}(a/2, a b / 2)$. For a given $\bm{s}$, if $\epsilon_t(\bm{s})$ follows a standard normal distribution, then marginally over the random $z_t$ and $\sigma_t^2$, the distribution of $Y^*_t(\bm{s})$ is skew-$t$ with location $\mu(\bm{s})$, scale $b$, skewness $\lambda(\bm{s})$ and degrees of freedom $a$; see SM Appendix C. To account for the spatial dependence, let $\epsilon_t(\cdot)$ follow a standard GP with correlation $\textrm{cor}[\epsilon_t(\bm{s}_1), \epsilon_t(\bm{s}_2)] = r(\bm{s}_1, \bm{s}_2)$ and $r(\cdot, \cdot)$ is parametrized by a set of parameters $\Omega$.
    
For a set of locations $\mathcal{S} = \lbrace \bm{s}_1, \ldots, \bm{s}_n \rbrace$, let $\bm{Y}_t^* = [Y_t^*(\bm{s}_1), \ldots, Y_t^*(\bm{s}_n)]'$ denote the vector of observations, $\bm{\mu} = [\mu(\bm{s}_1), \ldots, \mu(\bm{s}_n)]'$ and $\tilde{\bm{\lambda}} = [\lambda(\bm{s}_1), \ldots, \lambda(\bm{s}_n)]'$. Then, marginally over $z_t$ and $\sigma_t^2$, the joint distribution of $\mathbf{Y}_t^*$ is 
\begin{eqnarray} \label{stp}
\mathbf{Y}_t^* \sim \textrm{ST}_n (\bm{\mu}, b (\bm{\Sigma} + \tilde{\bm{\lambda}} \tilde{\bm{\lambda}}') , \bm{\Sigma}^{-1} \tilde{\bm{\lambda}}, a),
\end{eqnarray}
where $\bm{\Sigma}$ is the $n\times n$-dimensional matrix obtained by evaluating $r(\cdot, \cdot)$ at $\mathcal{S} \times \mathcal{S}$; here we match the notations of \cite{azzalini2014skew}. The functional form of the joint density is provided in SM Appendix C.



In this paper, we consider a mixture of STPs (described later in Section \ref{dpm}), where we choose specific forms of $\lambda(\cdot)$ and $r(\cdot, \cdot)$ for each STP component. Considering fully spatially-varying $\lambda(\cdot)$ involves stability issues and also increased computational burden. Hence, for simplicity, we consider separate skewness parameters for the three fire zones-- $\lambda(\bm{s}) = \lambda_j$, if the location $\bm{s}$ is within the fire zone $j \in \{1,2,3\}$, which is in line with varying spatial extremal dependence across different fire zones (left panel of Figure \ref{fig2} and Section \ref{model_props}). Let $\bm{\lambda} = (\lambda_1, \lambda_2, \lambda_3)'$. Then, $\tilde{\bm{\lambda}} = \bm{A} \bm{\lambda}$, where $\bm{A}$ is a $(n \times 3)$-dimensional design matrix, with its $(i,j)$-th element-- $a_{i,j} = 1$ if $\bm{s}_i$ is within Zone $j$, and 0 otherwise. Henceforth, denoting the $i$-th row of $\bm{A}$ by $\bm{a}_i$, we will denote $\lambda(\bm{s}_i)$ by $\lambda(\bm{s}_i) = \bm{a}_i'\bm{\lambda}$. For a general location $\bm{s} \in \mathcal{D}$, we will denote $\bm{a}(\bm{s}) = [a_1(\bm{s}), a_2(\bm{s}), a_3(\bm{s})]'$, and $\lambda(\bm{s}) = \bm{a}(\bm{s})'\bm{\lambda}$, where $a_j(\bm{s}) = 1$ if $\bm{s}$ is within Zone $j$, and 0 otherwise. 



For $r(\cdot, \cdot)$, an isotropic correlation function choice would be questionable considering the specific direction of wind through the mountain passes, variation of altitude, distance from the Pacific ocean, and possibly several other factors. Considering the computational burden as well as the stability issues of allowing a nonstationary correlation function, we consider an anisotropic Mat\'ern correlation function \citep{haskard2007anisotropic}--
\begin{eqnarray} 
\label{cov_structure}
&& r(\bm{s}_1, \bm{s}_2) = \frac{\gamma}{\Gamma(\nu) 2^{\nu - 1}} \left( \frac{h}{\rho} \right)^{\nu} K_{\nu} \left( \frac{h}{\rho}  \right) + (1 - \gamma) \mathbb{I}(\bm{s}_1 = \bm{s}_2),
\end{eqnarray}
where $h = \sqrt{\tilde{\rho}^{2} [h_1 \cos(\psi) + h_2 \sin(\psi)]^2 + [h_1 \sin(\psi) - h_2 \cos(\psi)]^2}$ is the Mahalanobis distance between $\bm{s}_1$ and $\bm{s}_2$, with $(h_1, h_2) = \bm{s}_1 - \bm{s}_2$, the range ratio $\tilde{\rho} > 0$ and the angle between two principal axes $\psi \in [0, \pi/2]$. The other parameters-- $\rho > 0$, $\nu > 0$ and $\gamma \in [0, 1]$ are the range, smoothness and the ratio of spatial to total variation respectively. In (\ref{cov_structure}), $K_{\nu}$ is the modified Bessel function of degree $\nu$, and $\mathbb{I}(\bm{s}_1 = \bm{s}_2) = 1$ if $\bm{s}_1 = \bm{s}_2$, and 0 otherwise. Thus, we have $\Omega = \lbrace \rho, \nu, \gamma, \tilde{\rho}, \psi \rbrace$. Overall, we denote the set of parameters by $\Theta = \lbrace \bm{\mu} , \bm{\lambda}, a, b, \Omega \rbrace$, and denote the skew-$t$ density in (\ref{stp}) as $f_{\textrm{ST}}(\bm{y}^*| \Theta)$.


\subsection{Dirichlet process mixture (DPM) model}
\label{dpm}
In order to capture the characteristics of the bulk of the data as well as of its tail, we will consider mixture of STPs. Intuitively, we can imagine that each characteristic present in the data is described by a mixture STP component with component-specific parameters, and as a result, estimation of the parameters that characterize the distribution of the bulk of the data is expected to be minimally influenced by the estimation of the ones that characterize the tails.  For now, we assume that for all $t$, $\mathbf{Y}_t^*$ are \textit{iid} $n$-dimensional realizations from a DPM of STPs (henceforth, STP-DPM) with $K$ mixture components, for some positive integer $K$; the corresponding density is represented as
	\begin{equation} \label{dpdensity}
     f_{\textrm{DPM}}(\mathbf{y}^*) = \sum_{k=1}^{K} \pi_k f_{\textrm{ST}}(\mathbf{y}^* | \Theta_k),
	\end{equation}
where $\pi_k > 0$ are the mixture probabilities with $\sum_{k=1}^{K} \pi_k = 1$, $\Theta_k$ denotes the set of parameters of the $k$-th component, and $f_{\textrm{ST}}(\cdot)$ denotes the density function of a $n$-dimensional realization from the skew-$t$ process as described in Section \ref{skewt}. The density function $f_{\textrm{DPM}}(\cdot)$ denotes the density function of a DPM of multivariate skew-$t$ densities. For a fully nonparametric model, $K = \infty$.
 
 
An equivalent representation of (\ref{dpdensity}) is the clustering model described below. For given $t$, let $g_t \in \lbrace 1, \ldots,K \rbrace$ denote the cluster membership associated with $\mathbf{Y}_t^*$, and let $\textrm{Pr}(g_t = k) = \pi_k$; we have $\mathbf{Y}_t^* | g_t = k \sim f_{\textrm{ST}}(\cdot, \Theta_{k})$. Thus, the STP-DPM model assumes that the replications (e.g. days) that are similar, can be clustered together and their common distribution can be described by the same STP, but with cluster-specific parameters. In our Bayesian model, the cluster labels are treated as unknown parameters, and thus we account for uncertainty in cluster allocation.
	
	
Given the cluster label $g_t = k$, the conditional distribution of $Y_t^*(\bm{s})$ is    
\begin{eqnarray} \label{model}
	\nonumber && Y_t^*(\bm{s}) = \mu_{k}(\bm{s}) + \bm{a}(\bm{s})' \bm{\lambda}_{k} \sigma_t \lvert z_t \rvert + \sigma_t \epsilon_t(\bm{s}), \\
	\nonumber && \epsilon_t(\cdot) \sim \textrm{Gaussian process}\left(0, r_{k}(\cdot, \cdot)  \right), \\ 
    \nonumber && \sigma^2_t  \overset{\textrm{indep}}{\sim}  \textrm{Inverse-Gamma}\left(\frac{a_{k}}{2}, \frac{a_{k} b_{k}}{2}\right), \\ 
    && z_t \overset{\textrm{iid}}{\sim} \textrm{Normal}(0, 1),
	\end{eqnarray}
	where $r_k(\cdot,\cdot)$ is the spatial correlation defined in (\ref{cov_structure}) with the set of parameters $\Omega_{k}$. The overall set of parameters corresponding to time $t$ is $\Theta_{k} = \lbrace \bm{\mu}_{k}, \bm{\lambda}_{k}, a_{k}, b_{k}, \Omega_{k} \rbrace$, with $\bm{\mu}_{k} = [\mu_{k}(\bm{s}_1), \ldots, \mu_{k}(\bm{s}_n)]'$ and $\bm{\lambda}_{k} = (\lambda_{k,1}, \lambda_{k,2}, \lambda_{k,3})'$, for $k=1, \ldots, K$. 


The mixture probabilities, $\pi_k$'s, are sequentially constructed following the stick-breaking representation proposed by \cite{sethuraman1994constructive} so that they sum to one and hence the $\pi_k$'s ``break the stick" of unit length. The first mixture probability $\pi_1$ is modeled as $\pi_1 = V_1$, where $V_1 \sim \textrm{Beta}(1, \delta)$. Subsequently, the $k$-th mixture probability is constructed as $\pi_k = (1-\sum_{i=1}^{k-1}\pi_i)V_k$ where $1-\sum_{i=1}^{k-1}\pi_i$ is the probability not considered by the first $k-1$ components and $V_k \overset{\textrm{iid}}{\sim} \textrm{Beta}(1, \delta)$. In case of finite $K$, we set $V_K = 1$ so that $\pi_K = 1-\sum_{i=1}^{K-1}\pi_i$ which ensures that $\sum_{k=1}^{K} \pi_k = 1$ almost surely. The case of $K = \infty$ corresponds to the Dirichlet process prior \citep{ferguson1973bayesian,ferguson1974prior}. 
We put hyperpriors on the cluster-specific parameters $\Theta_{k}$'s, which are the atoms of the stick-breaking process. For $K = \infty$, the hyperpriors relate to the base measure of the corresponding Dirichlet process. We assume $\Theta_{k} \overset{\textrm{iid}}{\sim} G_{\Theta}$ and the components of $\Theta_{k}$'s are independent. 
Choices of the hyperpriors are discussed in Section \ref{computation}. 


While the model (\ref{model}) clusters the extreme observations probabilistically, all the mixture components are still STPs and hence, they are supported over the whole real line. Under suitable regularity conditions, the only possible non-degenerate limiting distribution of the renormalized block maxima is the GEV distribution \citep{de2007extreme} which has location, scale and shape parameters-- $\mu_y \in \mathbb{R}, \sigma_y>0$ and $ \xi_y \in \mathbb{R}$, respectively. The support of the $\textrm{GEV}(\mu_y, \sigma_y, \xi_y)$ distribution is $(\mu_y - \sigma_y / \xi_y, \infty)$ if $\xi_y > 0$, $\mathbb{R}$ if $\xi_y = 0$ and $(-\infty, \mu_y - \sigma_y / \xi_y)$ if $\xi_y < 0$. To accommodate bounded support like a GEV distribution, we include a transformation step. Let $Y_t(\bm{s})$ be the observed data at site $\bm{s}$ and time $t$. We consider the support (only, not the distribution) of $Y_t(\bm{s})$ to be the same as the support of the $\textrm{GEV}(\mu_y, \sigma_y, \xi_y)$ distribution, for all $\bm{s},t$. It is easier to implement Bayesian nonparametrics (BNP) when we assume the support is the whole real line, and hence we relate the observed and transformed data using a monotonically-increasing GEV-log transformation-- $Y_t^*(\bm{s}) = \xi_y^{-1} \log\left \lbrace 1 + \xi_y \sigma_y^{-1} (Y_t(\bm{s}) - \mu_y)\right \rbrace$. If $\xi_y =0$, the transformation is $Y_t^*(\bm{s}) = \sigma_y^{-1}(Y_t(\bm{s}) - \mu_y)$, defined in a limiting sense ($\xi \rightarrow 0$). The transformed outcome $Y_t^*(\bm{s})$ is then modeled flexibly using the described semiparametric approach in (\ref{model}). We emphasize that the parameters-- $\mu_y, \sigma_y$, and $ \xi_y $ are treated as unknown in our fully Bayesian analysis. More details are provided in SM Appendix D. 


\subsection{Model properties}
\label{model_props}
From the infinite mixture model representation ($K = \infty$) in (\ref{dpdensity}), it is evident that for any $n$ spatial locations ($n < \infty$), the class of $n$-dimensional joint densities is a superset of the class of priors of \cite{gelfand2005bayesian}; specifically, by setting $a_k = \infty$ and $\lambda_k = 0$ for each $k$ we obtain the priors of \cite{gelfand2005bayesian}. Thus, the prior-- DPM of multivariate skew-$t$ distributions spans the entire set of joint densities for any set of spatial locations \citep{gelfand2005bayesian, reich2015spatial}. \cite{wu2010l1} proved the posterior consistency of the DPM with multivariate Gaussian kernels, under suitable regularity conditions. The posterior consistency of the DPM with multivariate skew-$t$ kernels essentially relies on the fact that the skew-$t$ distribution can be obtained from the normal distribution by marginalizing the random location and scale; using arguments similar to Lemma B.11 of \cite{ghosal2017fundamentals}, it follows that the Kullback-Leibler divergence between the true density and the estimate based on our model is smaller than its value for the normal distribution. 

\begin{lemma} \label{rm1}
\normalfont{For the model (\ref{model}), given the cluster parameters $\Theta_k$, the conditional mean and covariances of $Y_t^*(\bm{s})$ (assuming $a_k > 2$ for each $k$ with $\pi_k > 0$) are
\begin{eqnarray}
\nonumber \textrm{E}\left[Y_t^*(\bm{s}) \right] &=& \sum_{k=1}^{K} \pi_k \left[ \mu_{k}(\bm{s}) + \bm{a}(\bm{s})' \bm{\lambda}_{k} \sqrt{\frac{a_k b_k}{\pi}} C(a_k) \right], \\
\nonumber \textrm{Cov}\left[Y_t^*(\bm{s}_1), Y_t^*(\bm{s}_2) \right] &=& \sum_{k=1}^{K} \pi_k \left[ \mu_{k}(\bm{s}_1) \mu_{k}(\bm{s}_2) +  \sqrt{\frac{a_k b_k}{\pi}} C(a_k) (\mu_{k}(\bm{s}_1) \bm{a}(\bm{s}_1)  + \mu_{k}(\bm{s}_2) \bm{a}(\bm{s}_2))'\bm{\lambda}_{k} \right. \\
\nonumber && \left. + \frac{a_k b_k}{a_k - 2} ((\bm{a}(\bm{s}_1)' \bm{\lambda}_{k}) (\bm{a}(\bm{s}_2)' \bm{\lambda}_{k}) + r_k( \bm{s}_1, \bm{s}_2 ) ) \right] - \textrm{E}\left[Y_t^*(\bm{s}_1) \right] \textrm{E}\left[Y_t^*(\bm{s}_2) \right],
\end{eqnarray}
where $C(a_k) = \Gamma\left(\frac{a_k - 1}{2} \right) / \Gamma\left(\frac{a_k}{2} \right)$.}
\end{lemma}
\begin{remark} \label{rm_zerocenter}
\normalfont{
The mean and covariance are both dependent on $\bm{s}_1$ and $\bm{s}_2$ and cannot be reduced to a function of $ \bm{s}_1 -  \bm{s}_2 $, irrespective of $r_k(\cdot, \cdot)$ being stationary or not, $\bm{s}_1$ and $\bm{s}_2$ being in the same fire zone or not, and hence the model has both nonstationary mean and covariance structure.


%

}
\end{remark}

\begin{remark} \label{rm0}
	\normalfont{By setting $\bm{\lambda}_k = 0$, $a_k = \infty$ and $b_k = 0$ for each $k$ with $K=\infty$, the model (\ref{model}) is a spatial Dirichlet process, where $Y_t^*(\bm{\cdot})$ has discrete support $\mu_k(\cdot)$'s with $\textrm{E}\left[Y_t^*(\bm{s}) \right] = \sum_{k=1}^{\infty} \pi_k \mu_{k}(\bm{s}) = \bar{\mu}(\bm{s})$ and $\textrm{Cov}\left[Y_t^*(\bm{s}_1), Y_t^*(\bm{s}_2) \right] = \sum_{k=1}^{\infty} \pi_k [\mu_{k}(\bm{s}_1) - \bar{\mu}(\bm{s})] [\mu_{k}(\bm{s}_2) -  \bar{\mu}(\bm{s}_2)]$, given the cluster-specific sets of parameters $\Theta_k$'s and the mixture probabilities $\pi_k$'s. The mean and covariance functions span those of any square-integrable stochastic process with continuous mean and covariance functions. The proof is provided in SM Appendix E.}
\end{remark}


Conditioning on the cluster-specific parameters $\Theta_k$ and the mixture probabilities $\pi_k$, the spatial extremal dependence (defined in Section \ref{data}), $\chi(\bm{s}_1, \bm{s}_2)$, of the model (\ref{model}) is given by the following theorem, when $\bm{s}_1$ and $\bm{s}_2$ are in the same fire zone. The proof is provided in SM Appendix F. A closed form of $\chi(\bm{s}_1, \bm{s}_2)$ does not exist when $\bm{s}_1$ and $\bm{s}_2$ are within two different fire zones. While some bounds can be obtained under certain assumptions, we do not discuss this case for the sake of simplicity. Suppose, both $\bm{s}_1$ and $\bm{s}_2$ are within Zone $j$; for convenience, we denote the corresponding skewness parameter (equal for both) of the $k$-th DPM component by $\lambda_{k,j} \equiv \lambda_{k}$, for $k=1, \ldots, K$.


\begin{theorem} \label{thm1}
The spatial extremal dependence of the model (\ref{model}) is given by
\begin{eqnarray} \label{extremal}
~~~~~~~~ \chi(\bm{s}_1, \bm{s}_2) = 2\frac{F_T\left( \lambda_m  \sqrt{\frac{2 a^{''}}{1 + r_m(\bm{s}_1, \bm{s}_2)}}; a^{''} \right) }{F_T(\lambda_m \sqrt{a'}; a')} \bar{F}_T\left( \sqrt{  \frac{a' (1 - r_m(\bm{s}_1, \bm{s}_2))}{1 + r_m(\bm{s}_1, \bm{s}_2) + 2 \lambda_m^2}}; a' \right)
\end{eqnarray}
with $m = \arg \min_k \left \lbrace a_k\right \rbrace$,  $\bar{F}_T(\cdot~; a) = 1 - F_T(\cdot~; a)$ is the survival function for a Student's $t$ distribution with $a$ degrees of freedom, $a' = a_m + 1$ and $a^{''} = a_m + 2$.
\end{theorem}

\begin{remark} \label{rm2}
\normalfont{The $\chi$-measure is dependent on $\bm{s}_1$ and $\bm{s}_2$ only through $r_m(\bm{s}_1, \bm{s}_2)$; thus, even if the model specification is very flexible with nonstationary mean and covariance structures (Lemma \ref{rm1}), the extremal dependence is stationary if $r_k(\cdot, \cdot)$'s are stationary. This characteristic may be appealing in many applications because data in the tail are sparse, and thus simple models are needed to provide stability. If the empirical estimates of $\chi$ (or, based on environmental or geographical nature, as mentioned in Section \ref{skewt}) appears to be (an)isotropic, using (an)isotropic Mat\'ern correlation function is reasonable. 
}
\end{remark}
\begin{remark} \label{rm3}
\normalfont{The $\chi$-measure depends only on the cluster with the smallest degrees of freedom, which is the component with the heaviest tail. Thus, the extreme observations are likely to be clustered into one component with the heaviest tail and as we allow different parameters for each cluster, the data appearing from other clusters with lighter tails influence the parameters of the \textit{extremal} cluster only minimally. Thus, STP-DPM model allows a probabilistic partitioning of the tail from the bulk and prevents the bulk from influencing on the inference about the extremes.

}
\end{remark}

\begin{remark} \label{rm4}
\normalfont{Similar to the upper-tail extremal dependence in (\ref{chi_def}), the lower-tail case is $\chi_L(\bm{s}_1, \bm{s}_2) = \lim_{u \rightarrow 1} \textrm{Pr}[Y(\bm{s}_1) < F_1^{-1}(u) | Y(\bm{s}_2) < F_2^{-1}(u)]$. For the STP-DPM model, $\chi_L(\bm{s}_1, \bm{s}_2)$ has a similar form as $\chi(\bm{s}_1, \bm{s}_2)$ in (\ref{extremal}) except $\lambda_k$ replaced by $-\lambda_k$. Thus, in both the tails, the $\chi$-measure depends only on the cluster with the smallest degrees of freedom and therefore, the observations in the lower tail can influence the estimates of the parameters in the upper tail. 
To bypass this issue, we censor the observations in the left tail (below the 0.1$^{th}$ data quantile, for the FFWI data).

}
\end{remark}

Some corollaries of Theorem \ref{thm1} for the extremal dependence in case of sub-models of (\ref{model}), e.g., GP ($a_k = \infty, \lambda_k = 0, K = 1$), Student's $t$ process (TP, $\lambda_k = 0, K = 1$), STP ($K = 1$), DPM of GPs (\cite{gelfand2005bayesian}, GP-DPM, $a_k = \infty, \lambda_k = 0$), and DPM of TPs (TP-DPM, $\lambda_k = 0$), are as follows. Since we consider only stationary choices of $r_k(\cdot, \cdot)$'s, $\chi(\bm{s}_1, \bm{s}_2)$ is stationary for each sub-model, and hence, we denote it simply by $\chi(\bm{h})$, where $\bm{h} = \bm{s}_1 - \bm{s}_2$.

\begin{figure}[t]
\centering
\includegraphics[height = 0.45\linewidth, width = 0.45\linewidth]{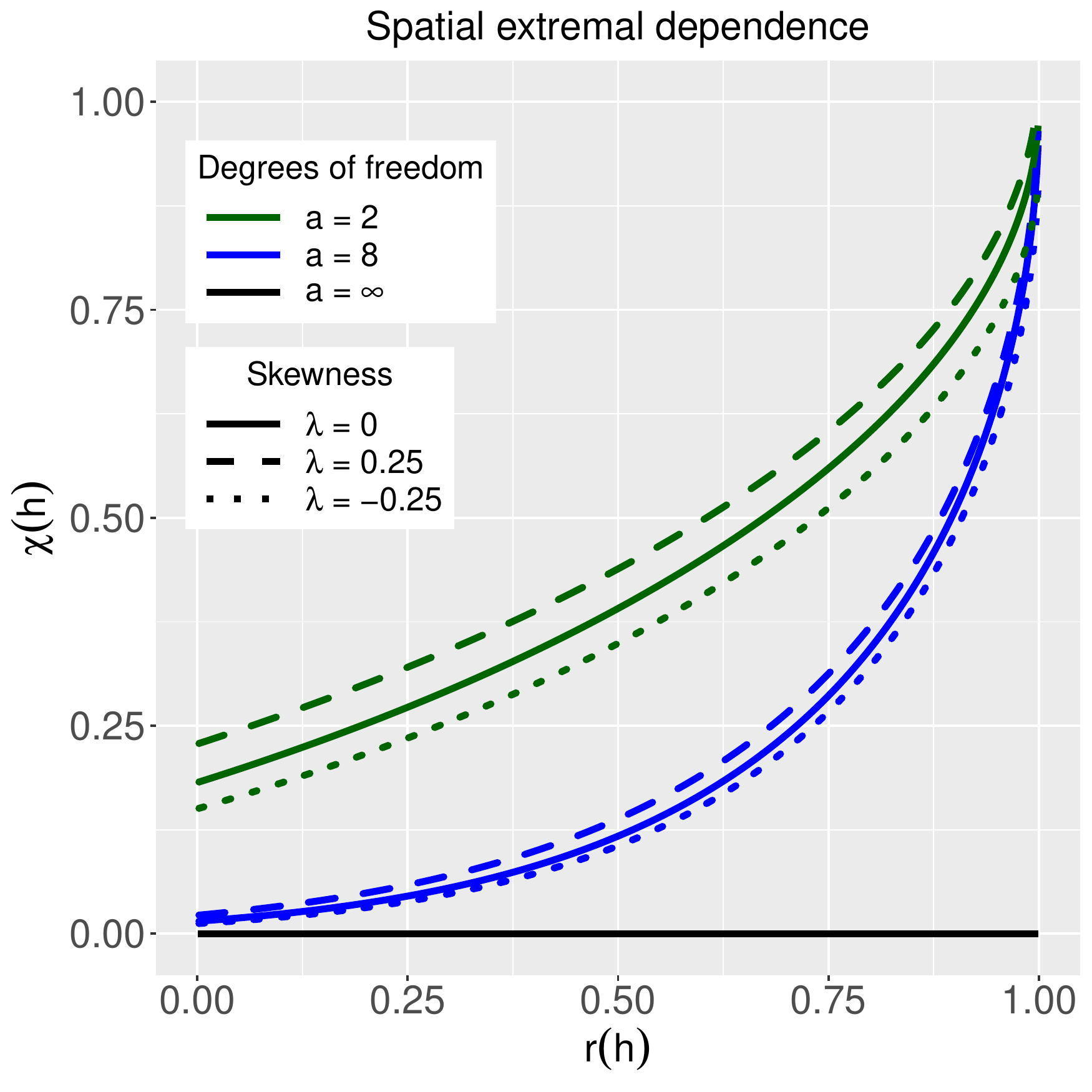}
\includegraphics[height = 0.45\linewidth, width = 0.45\linewidth]{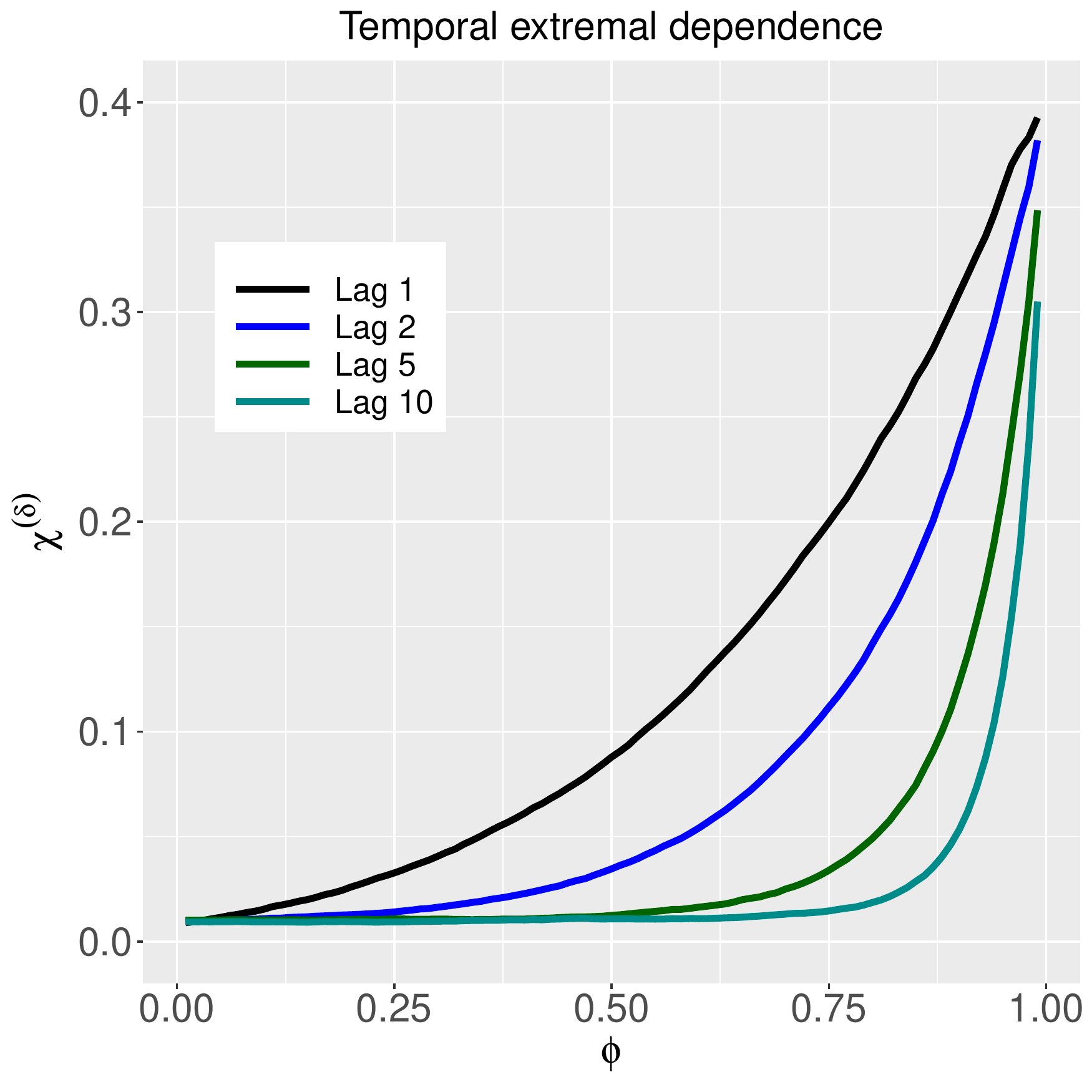}
\vspace{-2mm}
\caption{The values of $\chi(\bm{h})$ for STP-DPM at different values of $r(\bm{h})$ and some specific choices of the parameter values of the cluster with the smallest degrees of freedom (left). The cases with finite $a$ and $\lambda = 0$ correspond to symmetric-$t$ processes while $a=\infty$ corresponds to Gaussian processes ($\lambda = 0$) or skew-normal processes ($\lambda \neq 0$).  In the right panel, simulated lag-$\delta$ temporal $\chi$-measure ($\chi^{(\delta)}$) are provided, for varying levels of $\phi$ (we fix $u=0.99$ in \ref{chi_def}).}
\label{fig6}
\vspace{-2mm}
\end{figure}

\begin{itemize}
\item GP, GP-DPM: $\chi(\bm{h}) = 0$.
\item TP:  $\chi(\bm{h}) = 2 \bar{F}_T \left( \sqrt{a'} \sqrt{\frac{1 - r(\bm{h})}{1 + r(\bm{h})}}; a' \right)$.
\item STP:  $\chi(\bm{h}) = 2 F_T\left( \lambda \sqrt{a^{''}} \sqrt{\frac{2}{1 + r(\bm{h})}}; a^{''} \right) \bar{F}_T\left( \sqrt{ a' \frac{1 - r(\bm{h})}{1 + r(\bm{h}) + 2 \lambda^2}}; a' \right) \Big{/} F_T(\lambda \sqrt{a'}; a')$.
\item TP-DPM:  $\chi(\bm{h}) = 2 \bar{F}_T \left( \sqrt{a'} \sqrt{\frac{1 - r_m(\bm{h})}{1 + r_m(\bm{h})}}; a' \right)$ with $a' = a_m + 1$.
\end{itemize}
In Figure \ref{fig6} (left), we plot $r_m(\bm{h})$ versus $\chi(\bm{h})$ for several parameter choices (for convenience, we drop the subscript $m$ from $a_m$, $\lambda_m$ and $r_m(\bm{h})$). 
The extremal dependence decreases with increasing $a$ (for $a=\infty$, the extremal dependence remains zero for any $\lambda \in \mathbb{R}$, the case of a skew-normal process). For any finite $a$, $\chi(\bm{h})$ is positive, even if $r(\bm{h})$ is zero. The lines corresponding to $\lambda = 0.25$ and -0.25 show that a positive (negative) skewness increases (decreases) $\chi(\bm{h})$. 
The sub-asymptotic (without considering the limit) exceedance probabilities in (\ref{chi_def}), for a skew-$t$ process with different parameter choices, are discussed in SM Appendix G. 

We perform a simulation study to assess the performance of the model (\ref{model}) in spatial prediction of low through high marginal quantiles, and in estimation of extremal dependence. We compare the performances of GP, TP, STP, GP-DPM and TP-DPM with the proposed STP-DPM model. The simulation designs and the results are provided in SM Appendix H. Overall, STP-DPM performs equally well or better than other models, both in terms of prediction bias and prediction RMSE, as well as in the estimation of spatial extremal dependence, irrespective of the data generating model.

\subsection{Extension to the spatiotemporal Dirichlet process}
\label{spacetime}
Here we consider the scenario when the data exhibits temporal extremal dependence additional to the spatial extremal dependence. The literature on dependent DPM approaches is dominated by the autoregressive DPM approach \citep{beal2002infinite, fox2011sticky, storlie2014modeling} which considers a Markov model of the cluster indexes $g_t$'s in (\ref{model}). Although this approach adds temporal autocorrelation, the resulting process still exhibits temporal extremal independence. The proof is provided in SM Appendix I.


\begin{theorem}
	Consider the STP-DPM model in (\ref{model}). The temporal extremal dependence at any spatial location is zero if the temporal dependence in the spatiotemporal process $\lbrace Y_t(\cdot);t \geq 1 \rbrace$ is constructed only through the temporal dependence of the cluster indexes $g_t$, i.e., given $g_t$ and $g_{t+\delta}$, $Y_{t}(\bm{s})$ and $Y_{t+\delta}(\bm{s})$ are independent. The only exception is the trivial case of $b_k=0$ which leads to exact dependence.
\end{theorem}
    
Following \cite{morris2017space}, we consider an AR(1) structure for $z_t$ and $\sigma_t^2$. To ensure the spatial process is DPM of STPs, the inverse-gamma distribution of $\sigma_t^2$ with parameters $a_{g_t}$ and $b_{g_t}$, and the standard half-normal distribution of $\lvert z_t \rvert$ need to be preserved; it is done as follows. 
Suppose $\textrm{F}_{\textrm{HN}}$ and $\textrm{F}_{\textrm{IG}}$ denote the CDFs of standard half-normal and $\textrm{inverse-gamma}(a_{g_t}, b_{g_t})$ distributions, respectively. Thus, for each $t$, $\textrm{F}_{\textrm{HN}}(\lvert z_t \rvert) \sim \textrm{Uniform}(0, 1)$ and $\textrm{F}_{\textrm{IG}}(\sigma_t^2) \sim \textrm{Uniform}(0, 1)$, and so $z_t^* = \Phi^{-1}[\textrm{F}_{\textrm{HN}}(z_t)] \sim \textrm{Normal}(0, 1)$ and $\sigma_t^{2*} = \Phi^{-1}[\textrm{F}_{\textrm{IG}}(\sigma_t^2)] \sim \textrm{Normal}(0, 1)$, where $\Phi$ denotes the standard normal CDF. We specify an AR(1) structure as follows.
\begin{eqnarray} \label{timeseries}
\nonumber && z_1^* \sim \textrm{Normal}(0, 1);~~ z_{t+1}^* \sim \textrm{Normal}(\phi_z z_t^*, 1 - \phi_z^2), \\
 && \sigma_{1}^{2*} \sim \textrm{Normal}(0, 1);~~ \sigma_{t+1}^{2*} \sim \textrm{Normal}(\phi_\sigma \sigma_t^{2*}, 1 - \phi_\sigma^2).
\end{eqnarray}
This specification ensures the process is stationary across time. 


It is challenging to derive an analytical expression for the temporal extremal dependence. Based on simulated data (with $10^6$ replications), we calculate $\chi_u$ empirically, for $u = 0.99$ in (\ref{chi_def}); choosing $u$ to be larger than 0.99 involves numerical instability and hence not considered. 
We consider a model with $K=2$ mixture components and mixing probabilities-- $\pi_1 = \pi_2 = 0.5$. The vector of skew-$t$ parameters $(\mu, \lambda, a, b)$ for the two components are chosen to be $(0, 1.5, 4, 0.5)$ and $(0, 2, 6, 0.5)$, respectively ($\mu$ and $\lambda$ denote the location and skewness parameters, respectively, as in (\ref{stp_hierarchical}), ignoring the spatial notation). We generate lag-$\delta$ observations for $\delta = 1, 2, 5, 10$, from our model setting $\phi_\sigma = \phi_z = \phi$, for $\phi = 0, 0.01, 0.02, \ldots, 0.99$. 
The right panel of Figure \ref{fig6} suggests that $\chi^{(\delta)}$ increases as $\phi$ increases, and  decreases as the temporal lag $\delta$ increases.

\section{Computation}
\label{computation}
We use Markov chain Monte Carlo (MCMC) methods for model fitting and prediction. 
We consider conjugate priors for the parameters whenever possible which helps in updating the parameters using Gibbs sampling. Otherwise, we update individual parameters using random walk Metropolis-Hastings (M-H) algorithm. For the purpose of computation, we fix the number of components in the stick-breaking model at $K = 10$ by setting $V_K = 1$. The prior choices are as follows.
\begin{itemize}
\item The GEV parameters-- $\mu_y \sim \textrm{Normal}(0, 20^2)$, $\sigma_y^* = \log(\sigma_y) \sim \textrm{Normal}(-1, 2^2)$, and $\xi_y \sim \textrm{Normal}(0, 0.25^2)$ truncated below and above at -1 and 1, respectively. These choices cover a large class of supports of $Y_t(\bm{s})$.

\vspace{-2mm}

\item The spatially dependent means of the DPM components-- $\mu_k(\cdot), k=1, \ldots, K$, are assumed to be IID, and they follow a common Gaussian process with mean $\mathbf{B}'(\cdot) \bm{\beta}$, common marginal variance $\sigma_{\mu}^2$, and the spatial correlation is of the form (\ref{cov_structure}), with the set of anisotropic Mat\'ern correlation parameters-- $\Omega_{\mu} = \lbrace \rho_{\mu}, \nu_{\mu}, \gamma_{\mu}, \tilde{\rho}_{\mu}, \psi_{\mu} \rbrace$. Here $\bm{B}(\bm{s})$ is a vector of covariates corresponding to the spatial location $\bm{s}$. For simulation studies, we consider $\bm{B}(\bm{s}) = (1, s_1, s_2)'$ for $\bm{s} = (s_1, s_2)$, while for the FFWI data, we consider $\bm{B}(\bm{s}) = [1, \textrm{elevation}(\bm{s}), \textrm{slope}(\bm{s}), \textrm{aspect}(\bm{s})]'$. 

\vspace{-2mm}

\item The vectors of zone-specific skewness parameters-- $\bm{\lambda}_k, k=1, \ldots, K \overset{\textrm{iid}}{\sim} \textrm{Normal}_q(\bm{0}, \bm{I}_q)$, for $q$ homogeneous sub-regions of $\mathcal{D}$; in our application, $q=3$ for the three fire zones defined by \cite{rolinski2016santa}.

\vspace{-2mm}

\item The degrees of freedom parameters-- $a_k, k=1, \ldots, K \overset{\textrm{iid}}{\sim} \textrm{Uniform}(0.5, 40)$. While $a_k = 0.5$ corresponds to a very heavy-tailed process, the marginal and joint tails for $a_k = 40$ are practically equivalent with those for $a_k = \infty$, the Gaussian case. Here, no closed form expression is available for the posterior distributions of $a_k$'s. One way to bypass the requirement of M-H steps is by considering a discrete uniform prior-- $a_k \overset{\textrm{iid}}{\sim} \textrm{Discrete-Uniform}\lbrace 0.5, 0.6, \ldots, 39.9, 40.0 \rbrace$, for example, similar to \cite{gelfand2005bayesian}; although, it involves higher computational burden.

\vspace{-2mm}

\item The scale parameters-- $b_k, k=1, \ldots, K \overset{\textrm{iid}}{\sim} \textrm{Gamma}(1, 1)$. In case of the purely spatial model (\ref{model}), this choice is a conjugate prior, and hence $b_k$'s can be updated within MCMC using Gibbs sampling. Although, considering (\ref{timeseries}), no closed form expression of the posterior of $b_k$ is available and $b_k$'s are updated using M-H steps.


\vspace{-2mm}
 
\item For the component-specific anisotropic Mat\'ern parameters, we consider the priors-- $\rho_k \overset{\textrm{iid}}{\sim} \textrm{Uniform}(0, 2.5\Delta)$, where $\Delta$ is the largest Euclidean distance between two data locations, $\nu_k^* = \log(\nu_k) \overset{\textrm{iid}}{\sim} \textrm{Normal}(-1.2, 1^2)$ with $\nu_k$ is truncated above at 40, $\gamma_k \overset{\textrm{iid}}{\sim} \textrm{Uniform}(0, 1)$, $\tilde{\rho}_k \overset{\textrm{iid}}{\sim} \textrm{Gamma}(1, 1)$, and $\psi_k \overset{\textrm{iid}}{\sim} \textrm{Uniform}(0, \pi/2)$. The specific prior choice of $\nu_k^*$ ensures that $\nu_k$ is approximately centered around 0.5, the exponential correlation case, and $\nu_k > 40$ is practically equivalent with $\nu_k = \infty$, the squared exponential correlation scenario. Other choices are noninformative.

\vspace{-2mm}

\item For the model (\ref{timeseries}), we consider noninformative priors-- $\phi_z, \phi_{\sigma} \overset{\textrm{iid}}{\sim} \textrm{Uniform}(0, 1)$.

\vspace{-2mm}


\item For the hyperparameters of $\mu_k(\cdot)$'s, we consider the conjugate noninformative priors $\bm{\beta} \sim \textrm{Normal}_p(0, 2.5^2 \bm{I}_p)$, where $p$ is the length of $\bm{B}(\bm{s})$, $\sigma_{\mu}^2 \sim \textrm{Inverse-Gamma}(1, 1)$, and for the anisotropic Mat\'ern parameters $\Omega_{\mu}$, we consider $\rho_{\mu} \overset{\textrm{d}}{=} \rho_k, \nu_{\mu}^* = \log(\nu_{\mu}) \overset{\textrm{d}}{=} \nu_k^*, \gamma_{\mu} \overset{\textrm{d}}{=} \gamma_k, \tilde{\rho}_{\mu} \overset{\textrm{d}}{=} \tilde{\rho}_k,~\textrm{and}~ \psi_{\mu} \overset{\textrm{d}}{=} \psi_k$.


\vspace{-2mm}

\item For the Dirichlet process concentration parameter, we consider $\delta \sim \textrm{Gamma}(0.1, 0.1)$.
\end{itemize}



We censor the observations below the 0.1$^{th}$ data quantile at each station, as mentioned in Remark \ref{rm4}, and impute the censored as well as the missing data using Gibbs sampling. We tune the hyperparameters adaptively to allow better mixing within MCMC. The MCMC steps of model fitting and prediction are provided in SM Appendix J. For fitting the STP-DPM model to the FFWI data, we generate 70,000 posterior samples and discard the first 20,000 iterations as burn-in period. Subsequently, we thin the Markov chains by keeping one out of five consecutive samples and thus, we finally obtain 10,000 samples for drawing posterior inference. The computation time is 454 minutes.



\section{Data application}
\label{application}

\subsection{Spatial covariates}

For FFWI data, we consider $\bm{B}(\bm{s}) = [1, \textrm{elevation}(\bm{s}), \textrm{slope}(\bm{s}), \textrm{aspect}(\bm{s})]'$, following \cite{sapsis2016fire}, to incorporate topographic information. We obtain high-resolution elevation data across the study region, from the LANDFIRE website (\url{https://landfire.gov}). Further, we obtain slope and aspect from elevation using the function \texttt{terrain} from the package \texttt{raster} in \texttt{R}. We keep the covariate information for the monitoring sites and 3797 grid cells (with spatial resolution $2' \times 2'$) across the Santa Ana region. One grid cell represents an area of 1000 hectares, approximately.

\subsection{Model comparisons}


We compare the performances of the models discussed in Section \ref{model_props}, along with considering temporal dependence in some cases, focusing on the high quantiles. The STP-DPM model is assumed to have temporal dependence as described in Section \ref{spacetime}, where the AR(1) structure is considered for both $\lvert z_t \rvert$ and $\sigma_t^2$. A competing TP-DPM model involve random $\sigma_t^2$, but no random $\lvert z_t \rvert$ and thus, for allowing time dependence, we consider a similar AR(1) structure for $\sigma_t^2$. The GP-DPM model has no time-dependence as $\sigma_t^2$'s are non-random in that case. Considering data-specific features--different skewness parameters for fire zones and anisotropic Mat\'ern correlation of the mixture components, we only allow anisotropic Mat\'ern $r_k(\cdot, \cdot)$'s for the TP-DPM and GP-DPM models, as they do not have skewness parameters. Considering parametric alternatives, the AR(1) structure of STP and TP are constructed similar to STP-DPM and TP-DPM models, respectively. For GP, we consider the spatial error process to have AR(1) correlation structure in time. Along with temporally-dependent GP, TP and STP, we also fit censored GP, TP and STP methods treating the data below the 90$^{th}$ percentile as censored. 
GP is taken as reference and all the alternatives are assessed in terms of relative performance in high level quantile estimation using a leave-one-site-out cross-validation. The spatially-invariant parameters are estimated only once based on the full data for each model. For a model $M$, the prediction RMSE for a quantile $q$ is calculated as $\textrm{RMSE}_{M} = \sqrt{n^{-1} \sum_{i=1}^{n}[\tilde{F}_i^{-1}(q) - {F}_i^{-1}(q)]^2}$, where $\tilde{F}_i$ and ${F}_i$ denote the CDF of the the posterior predictive distribution and the empirical CDF at site $i$ respectively. The RMSE skill score for model $M$ is defined as
\begin{eqnarray}
 \Delta_{M} &=& \frac{\textrm{RMSE}_{GP} - \textrm{RMSE}_{M}}{\textrm{RMSE}_{GP}} \times 100 \%,
\end{eqnarray}
where a model with higher value of $\Delta_{M}$ is preferred.


The $\Delta_{M}$ values for the competing models are reported in Table \ref{table4}. Except the censored models, $\Delta_{M}$ are positive and generally increase with the quantile level, indicating that the competing models generally outperform GP, particularly in the tails which is natural because GPs have thin tail and tend to underestimate the high quantiles. Among the parametric models without censoring, STP has higher $\Delta_{M}$ values compared to TP. When we ignore the temporal dependence, $\Delta_M$ for the STP-DPM model are always smaller than a parametric STP model with temporal dependence indicating the importance to consider temporal extremal dependence. The censored parametric models fail to capture the spatial dependence structure leading to negative $\Delta_{M}$ for most of the cases. Below the 0.96$^{th}$ quantile, GP-DPM performs better than TP-DPM but above that quantile, TP-DPM gradually outperforms GP-DPM with the highest difference being at the 0.98$^{th}$ quantile. The STP-DPM model outperforms other competing models throughout the range of quantiles. Further, we discuss the bulk-inference as well as the tail-inference based on the proposed STP-DPM model.



\begin{table}
	\label{table4}
 \caption{Prediction RMSE skill score $\Delta_{M}$ at a few high quantiles for the models-- Student's $t$ process (TP), skew-$t$ process (STP), Dirichlet process mixture (DPM) of Gaussian processes (GP-DPM), DPM of TPs (TP-DPM), temporally independent DPM of STPs (STP-DPM-indep) and the final DPM of STPs (STP-DPM), considering Gaussian process (GP) as the base model. The subscripts $c(0.5)$ and $c(0.9)$ denote censoring below the 0.5$^{th}$ and 0.9$^{th}$ data quantiles, respectively. A model with larger $\Delta_{M}$ is preferred.}
 \vspace{1mm}
\centering 
\begin{tabular}{lllllllll}
  \hline
Model & $q(0.92)$ & $q(0.93)$ & $q(0.94)$ & $q(0.95)$ & $q(0.96)$ & $q(0.97)$ & $q(0.98)$ \\
  \hline
  GP$_{c(0.9)}$   & 1.00 & 0.25 & 0.76 & 0.22 & -1.21 & -0.26 & 0.97 \\ 
TP  & 3.57 & 3.99 & 4.75 & 5.71 & 6.61 & 6.55 & 6.54 \\ 
TP$_{c(0.9)}$   & -1.38 & -1.43 & -1.50 & -1.59 & -1.76 & -1.96 & -2.22 \\ 
  STP  & 5.16 & 5.83 & 5.92 & 7.63 & 7.94 & 8.85 & 8.43 \\ 
  STP$_{c(0.9)}$ & -1.00 & -0.98 & -0.95 & -0.92 & -0.91 & -0.93 & -2.79 \\ 
GP-DPM  & 5.93 & 6.56 & 7.64 & 8.36 & 8.37 & 9.27 & 7.84 \\ 
  TP-DPM  & 4.76 & 4.83 & 7.31 &  8.08 & 9.39 & 10.51 & 11.06 \\ 
  STP-DPM-indep  & 3.30 & 4.29 & 5.55 & 6.18 & 6.84 & 7.33 & 7.81 \\
  STP-DPM & \textbf{9.43} & \textbf{10.22} & \textbf{11.48} & \textbf{12.03} & \textbf{12.51} & \textbf{13.25} & \textbf{11.80} \\ 
   \hline
\end{tabular}
\end{table}

\subsection{Model diagnostics}
We provide detailed goodness-of-fit diagnostics for the STP-DPM model in SM Appendix K. The proposed STP-DPM model allows finite first and second order moments only when the parameters $a_k \geq 2$ for all $k=1, \ldots, K$, but considering the priors we choose, theoretically, the posterior probability of such parameters being less than two is nonzero. Hence, we discuss results based on the median, along with low--through--high quantiles instead of the mean, the inter-quartile range (IQR) instead of the standard deviation, and the Spearman's rank correlation instead of the Pearson's correlation. While these assess the performance of the model in fitting the bulk of the data, we also discuss the estimated pairwise tail-dependence measure $\chi_u$ for $u=0.95$. To compare the performances, we also provide the results based on the sub-model GP as a reference.

To summarize, the medians as well as the moderate quantiles, and the IQR profiles are very well estimated overall, as demonstrated by the kernel densities of the differences between their corresponding empirical and fitted model-based estimates. For a few very high quantiles, the differences are large, although they are smaller than those for GP. While comparing the correlations, the differences are generally closer to zero for the STP-DPM model, but GP underestimates the correlation heavily. The uncertainty involved in the empirical $\chi_u$ are generally high due to the sparsity of extreme events; while the variability is high, the kernel densities of the differences are approximately centered around zero for the STP-DPM model, while GP underestimates $\chi_u$ for most of the pairs.




\subsection{Bulk-inference}

\begin{figure}[h]
\centering
\adjincludegraphics[width = 0.49\linewidth, trim = {{.0\width} {.0\width} 0 {.0\width}}, clip]{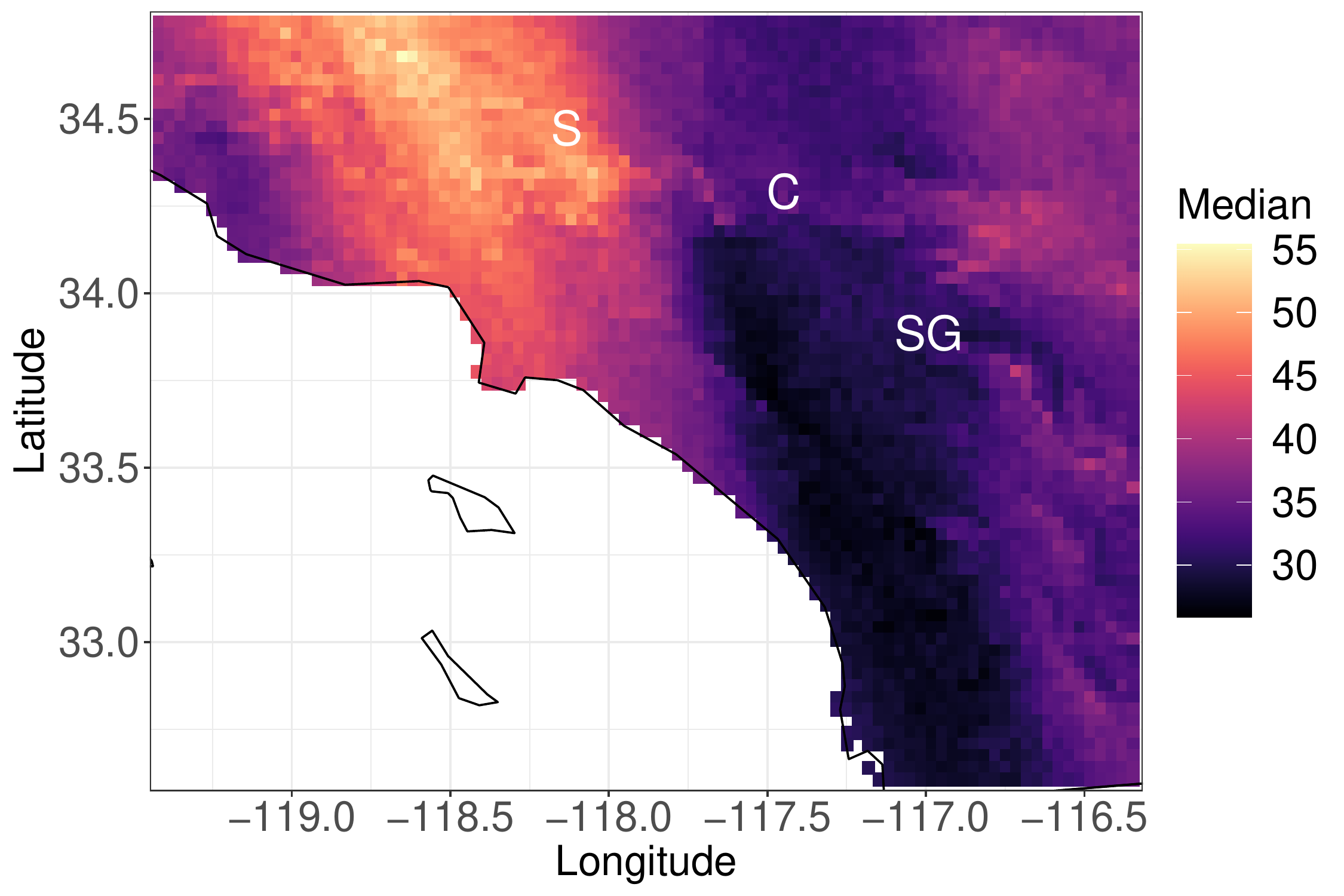}
		\adjincludegraphics[width = 0.49\linewidth, trim = {{.0\width} {.0\width} 0 {.0\width}}, clip]{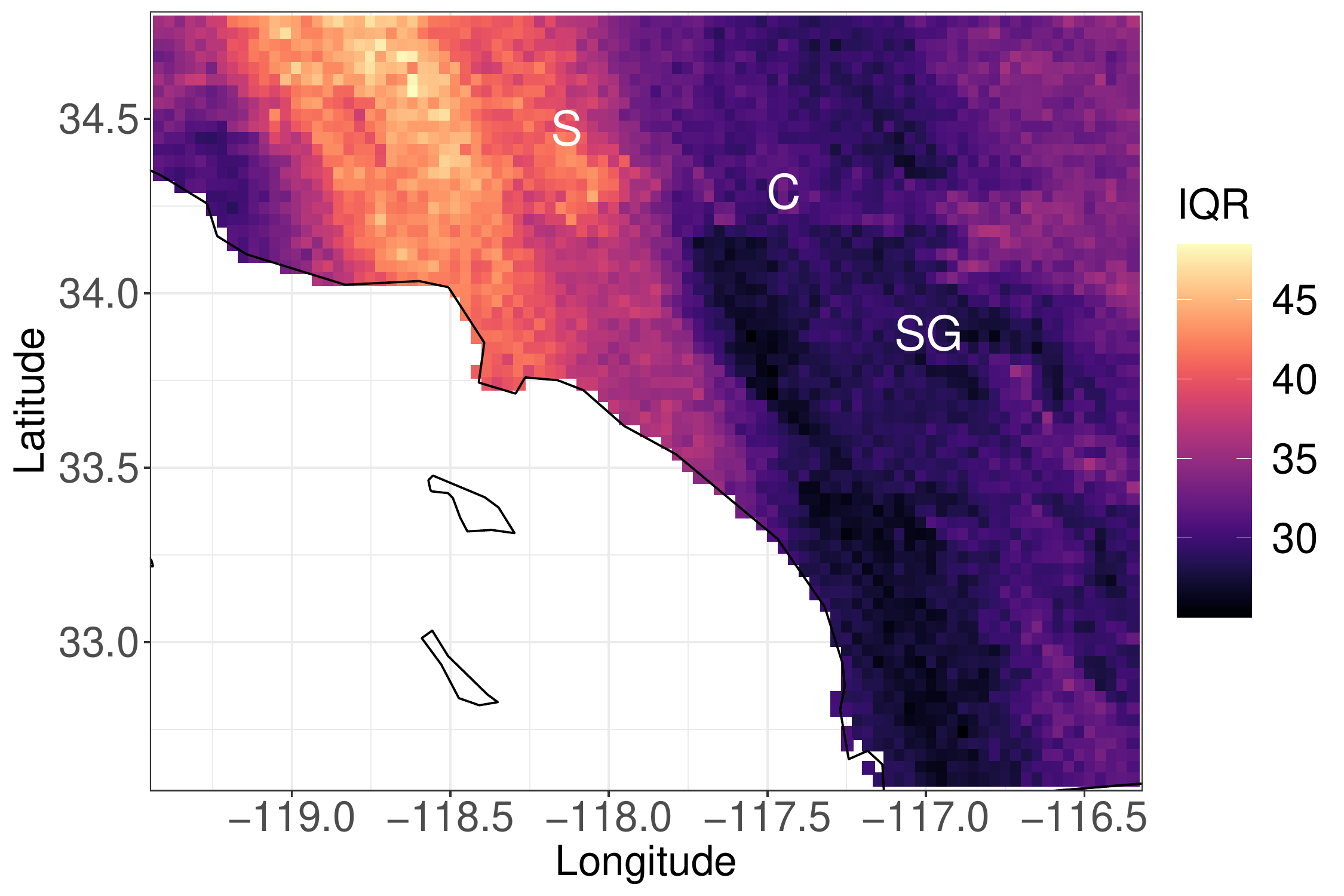}
\vspace{-3mm}		
\caption{Spatial maps of the FFWI medians (left) and IQRs (right). Here `S', `C', and `SG' denote the locations of the Soldead Pass, Cajon Pass, and San Gorgonio Pass, respectively.}
\label{fig_bulk_inference}
\vspace{-3mm}
\end{figure}

The spatial maps of the FFWI medians, and IQRs are provided in Figure \ref{fig_bulk_inference}. Both the spatial profiles are highly nonstationary; the medians and IQRs are lower near the Cleveland National Forest, while they are higher near the Los Padres National Forest. The locations of the three mountain passes that funnel offshore Santa Ana winds, as mentioned by \cite{moritz2010spatial}, namely the Soledad Pass, the Cajon Pass, and the San Gorgonio Pass, are also presented in the figure. We observe that the krigged medians and IQRs  clearly indicate the funneling from the west towards the Pacific Ocean through the mountain passes, and hence, the proposed STP-DPM model realistically predict FFWI medians and IQRs at the unobserved locations. The highest values are observed in the northwestern region, near the Soledad pass; this matches with the fire centroids presented by \cite{moritz2010spatial} based on the fire history of the Santa Ana region over the years 1950--2007. It also matches with the location of the Thomas Fire in 2017.


\vspace{-2mm}

\subsection{Tail-inference}

\begin{figure}[h]
\centering
\adjincludegraphics[width = 0.49\linewidth, trim = {{.0\width} {.0\width} 0 {.0\width}}, clip]{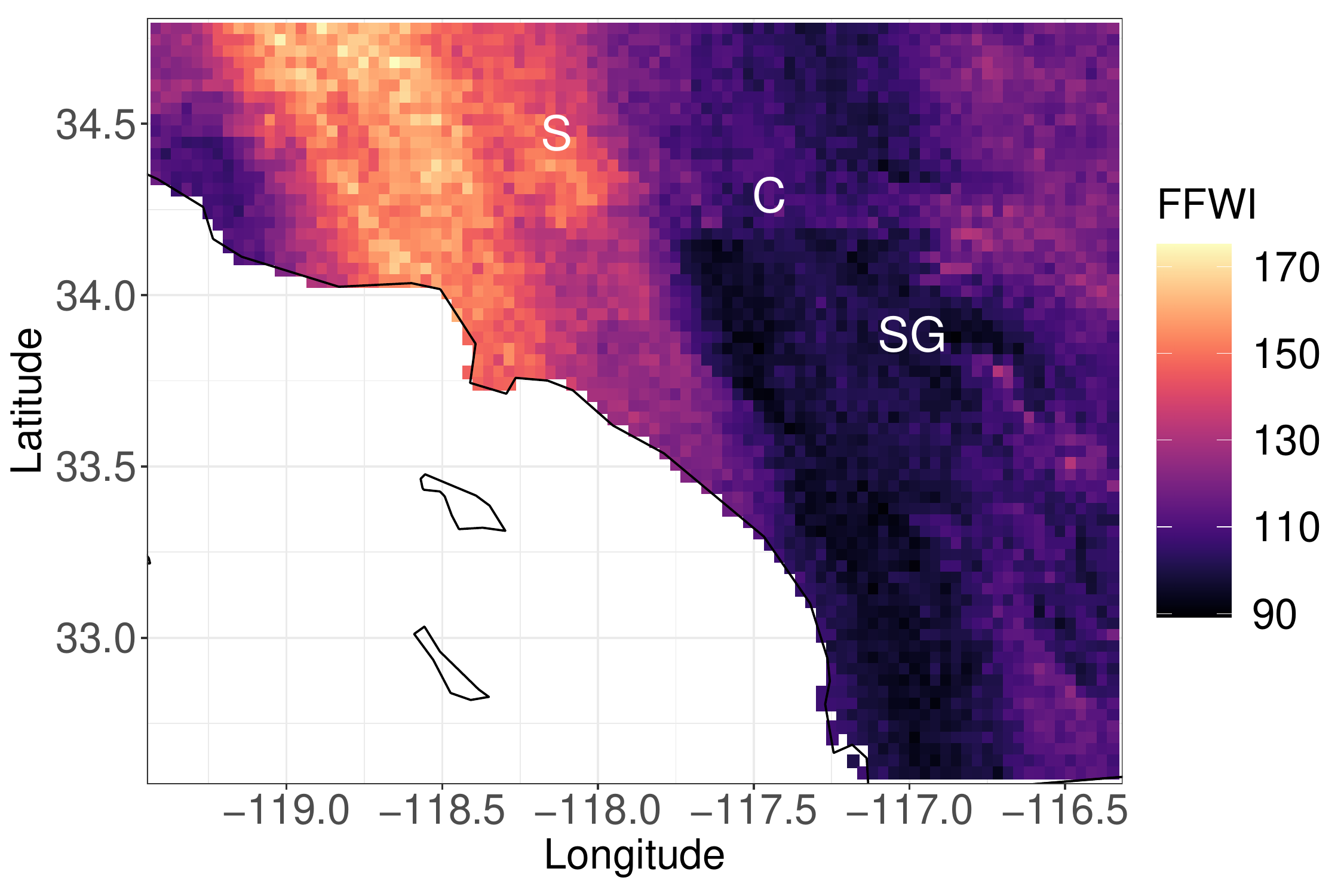}
		\adjincludegraphics[width = 0.49\linewidth, trim = {{.0\width} {.0\width} 0 {.0\width}}, clip]{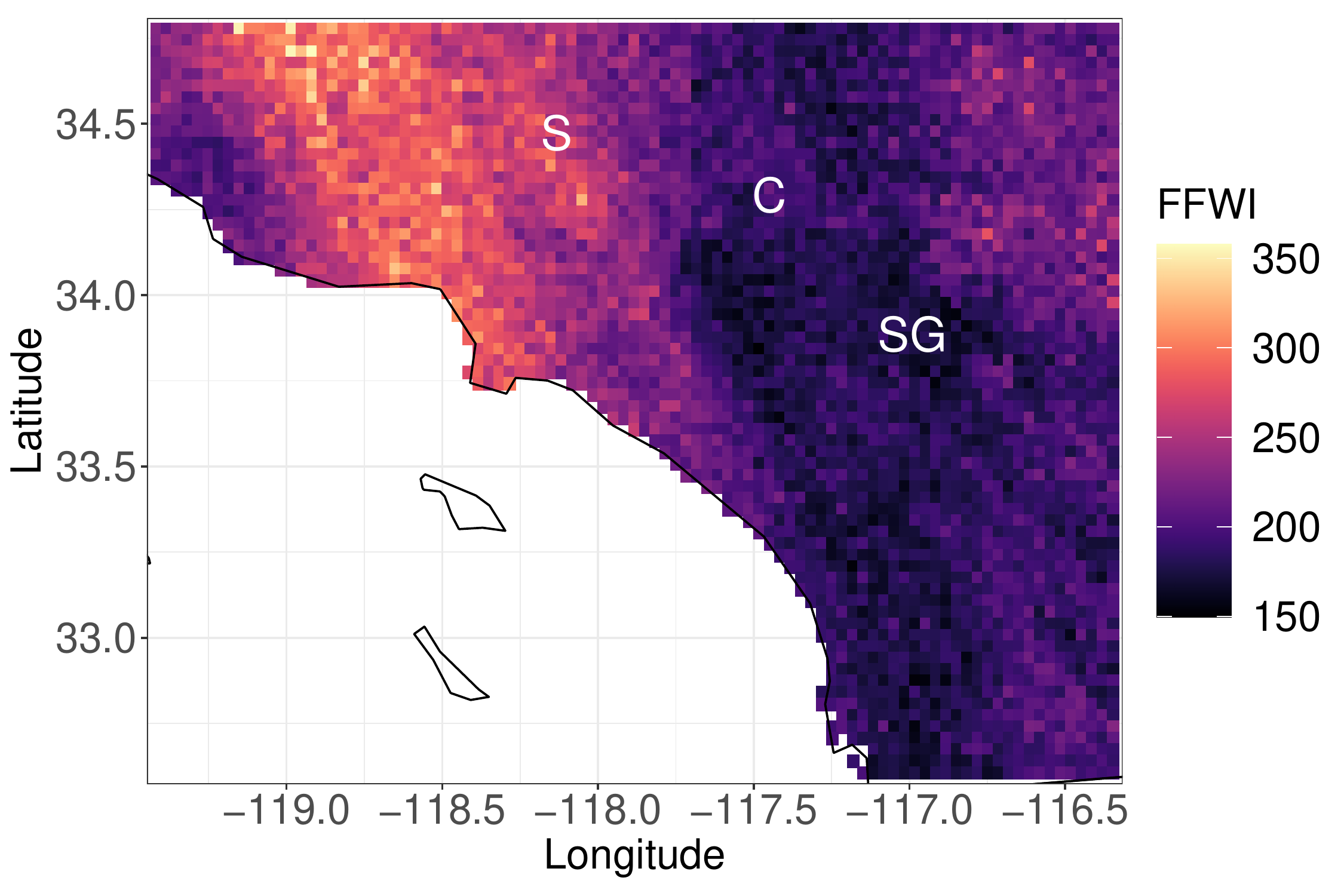}
\vspace{-3mm}		
\caption{Spatial maps of the 1-year (left) and 10-year (right) return levels of FFWI. The notations `S', `C', and `SG' are same as in Figure \ref{fig_bulk_inference}.}
\label{fig_tail_inference}
\vspace{-3mm}
\end{figure}

For tail-inference, a $r$-year return level is calculated as $[1 - 1/(31r)]^{th}$ quantile of the posterior predictive distribution, considering December only. The spatial maps of the 1-year and 10-year return levels of FFWI are provided in Figure  \ref{fig_tail_inference}. The spatial patterns in the return level maps are similar to the maps of medians and IQRs, although both the return-level profiles appear to be more jittery-- this is realistic as the spatial correlation gradually decreases near the tails. The return level maps indicate that the FFWI values are larger than 90 for at least one day per December and larger than 150 for at least one December day per decade (probabilistically), throughout the region, while it can be as high as 350 near the Los Padres National Forest.


\begin{figure}
\centering
\adjincludegraphics[width = 0.49\linewidth, trim = {{.04\width} {.0\width} {.04\width} {.0\width}}, clip]{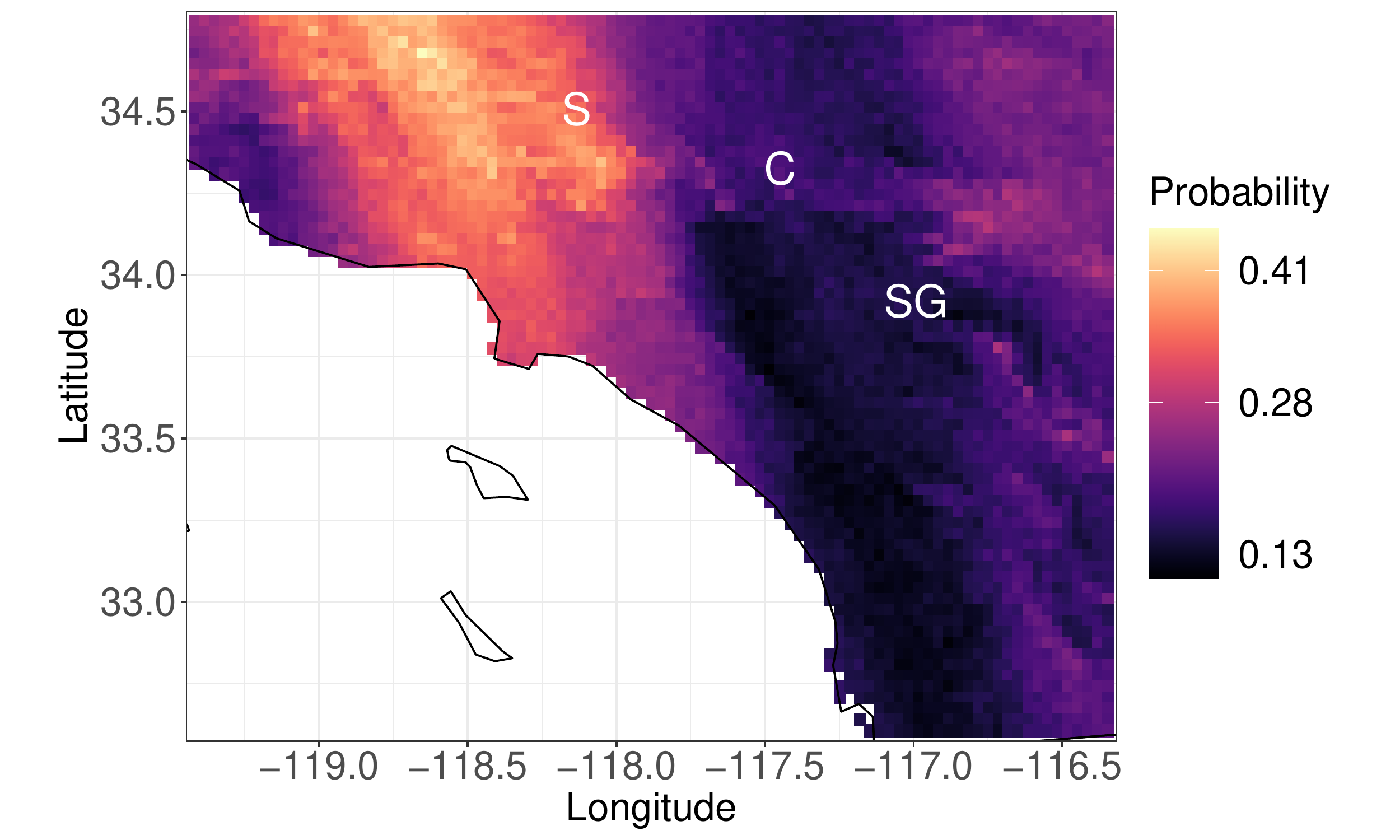}
		\adjincludegraphics[width = 0.49\linewidth, trim = {{.04\width} {.0\width} {.04\width} {.0\width}}, clip]{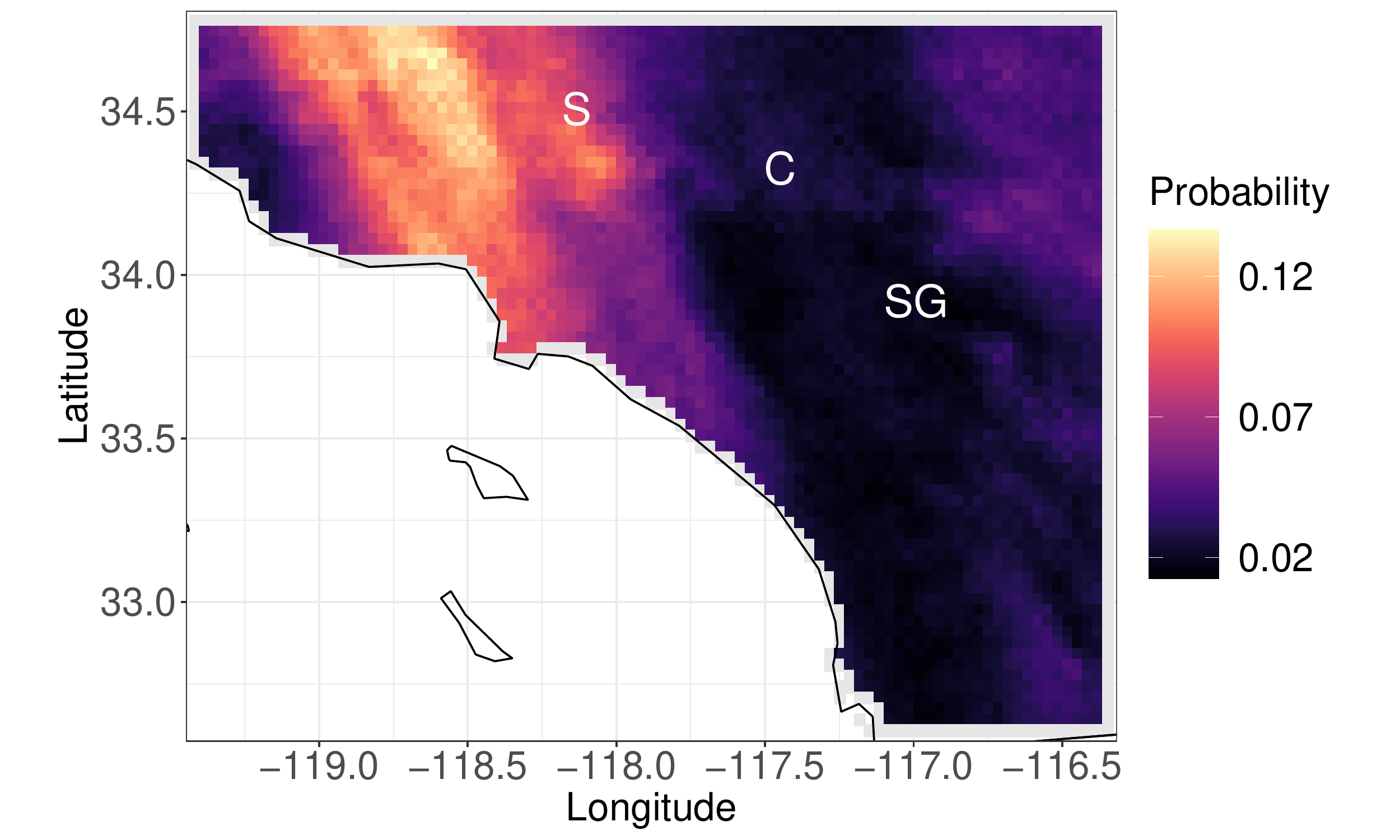}
\vspace{-3mm}		
\caption{Spatial maps of the marginal probabilities of FFWI exceeding 60 (left) and the joint probabilities of FFWI exceeding 60 together at all the four neighboring grid cells along with the one at the center (right). The notations `S', `C', and `SG' are as in Figure \ref{fig_bulk_inference}.}
\label{fig_exceed_inference}
\vspace{-3mm}
\end{figure}

\cite{sapsis2016fire} consider an event of FFWI being larger than 60 to be an extreme fire weather event for the southern California. Suppose $Y_{0}(\bm{s}^*_i)$ denotes a generic copy of FFWI at a grid cell $\bm{s}^*_i$.  We provide the spatial map of the estimated marginal exceedance probabilities, $p_i^0 = \textrm{Pr}(Y_{0}(\bm{s}^*_i) > 60)$, in the left panel of Figure \ref{fig_exceed_inference}. While the marginal probabilities indicate the severity of the local fire weather, the chance of a wildfire burning a large region is high when there is strong spatial association and FFWI are large jointly at the neighboring locations. Suppose $\mathcal{N}_i$ denotes the set of the four first order neighbors (ignoring the edges, where some neighbors are missing) of $\bm{s}^*_i$, along with $\bm{s}^*_i$. To quantify spatial association, we estimate the joint exceedance probabilities $p_i^1 = \textrm{Pr}( \cap_{\bm{s}_j \in \mathcal{N}_i} \{ Y_{0}(\bm{s}^*_j) > 60 \})$, for all the grid cells $\bm{s}^*_i$ except for those at the edges. We present the estimated $p_i^1$ for each $\bm{s}^*_i$, in the right panel of Figure \ref{fig_exceed_inference}. The estimated $p_i^0$'s are high near the Los Padres National Forest similar to the previous cases, while the estimates vary between 0.1060 and 0.4504. Considering the estimated $p_i^1$'s, the spatial pattern is approximately similar while the estimates vary between 0.0126 and 0.1366-- this indicates a strong spatial association and hence indicates the high chance of a large area facing the fire weather condition at the same time. 

\section{Discussions and Conclusions}
\label{discussions}

In this paper, we propose a flexible semiparametric Bayesian spatiotemporal model for analyzing the bulk and the tail jointly. The model automatically clusters the temporal replications (possibly dependent across time) and sets the extreme observations into one component with the heaviest tail. Allowing separate parameters for the mixture components, the data from the bulk minimally influence the parameters of the component of extremes. Considering infinite components, the proposed model spans the entire set of joint densities for any set of spatial locations \citep{gelfand2005bayesian, reich2015spatial}, the mean and covariance of the model are nonstationary, and spans all possible continuous mean and covariance over the spatial domain of interest. Allowing stationary correlation functions within each component, the extremal dependence measure $\chi$ is also stationary. We recommend this choice as the data is sparse for the component of extremes and an over-parametrized correlation function may lack stability. Using a simulation study, we demonstrate that the proposed model outperforms several parametric and semiparametric alternatives. Additional to spatial extremal dependence, in case of daily measurements, the data may exhibit temporal extremal dependence. Ignoring the temporal dependence can affect the model performance and hence, we extend our model for independent temporal replications to accommodate temporal extremal dependence.


The proposed model can be further generalized. A shortcoming of our model is that the spatial extremal dependence is nonzero throughout the entire spatial domain of interest. Thus, in case of a large spatial domain like the entire United States, a random partitioning of the spatial domain is required, as considered by \cite{morris2017space}. Although our exploratory analysis strongly suggest the FFWI data exhibit asymptotic dependence, \cite{huser2017bridging} consider a more general class of scale mixtures that bridges asymptotic dependence and independence, although, the marginal and joint distributions may not have closed form expressions. Thus, replacing the inverse-gamma distribution with the specific scaling distribution proposed in \cite{huser2017bridging}, our proposed model would allow a larger class of sub-asymptotic dependence. The temporal extremal dependence structure we consider does not allow Gibbs sampling for a number of model parameters and hence, any existence of stationary gamma and half-normal processes that allow Gibbs sampling, can lead to computational advantages. Because our proposed method involves a sampling-based inference technique (similar to other nonparametric Bayesian models like \cite{gelfand2005bayesian}), to obtain return level maps for higher return periods and to allow less sampling variability, we need to draw sufficient post-burn-in samples. 

We provide very high-resolution ($2' \times 2'$) spatial maps of the estimated FFWI medians, IQRs, return levels, as well as marginal and joint probabilities of the extreme event (according to \cite{sapsis2016fire}) of FFWI being higher than 60. The spatial patterns we obtain are realistic and similarities with the analysis of \cite{moritz2010spatial} are observed. The return-level maps and the exceedance probabilities indicate frequent and severe extreme fire weather conditions. The Los Padres National Forest region requires serious environmental concern and disaster management planning. 


\section*{Acknowledgement}

The authors thank Tim Brown at the Desert Research Institute for providing the data, David Sapsis at CalFIRE for suggesting the analysis of FFWI and Rapha{\"e}l Huser at KAUST for some valuable suggestions regarding the methodology. This work was partially supported by NSF grants DMS-2001433 and DMS-1454942, NOAA grant Z17-20337, DOE grant DE-AC02-05CH11231, DOI grant 14-1-04-9 and NIH grants R01ES027892 and 5P01 CA142538-09.

\newpage



\begin{center} {\large{\bf \textit{Supplementary Materials for} \\ A semiparametric spatiotemporal Bayesian model for the bulk and extremes of the Fosberg Fire Weather Index}}
\end{center}

\section*{Appendix A: Fosberg Fire Weather Index}
\label{fosberg}
Proposed by \cite{fosberg1978ffwi}, Fosberg Fire Weather Index (FFWI) is a nonlinear function of three important weather parameters-- air temperature, wind speed and relative humidity. The functional form is given by the following equation
$$\textrm{FFWI}= \eta \sqrt{1 + U^2}/0.3002,$$
where $U$ is the wind speed in miles per hour (mph). The moisture
damping coefficient, $\eta$, is given by
$$\eta = 1-2 (m/30) + 1.5 (m/30)^2 - 0.5 ( m/30)^3.$$
The equilibrium moisture content ($m$) is given as a function of temperature in degrees Fahrenheit ($T$) and relative humidity in percent ($h$):
\[
  \text{$m$} =
  \left\{
  \begin{aligned}
     & \text{$0.03229+0.281073h-0.000578hT$}  & \text{for $h < 10\%$} \\
    & \text{$2.22749+0.160107h-0.01478T$} & \text{for $10\% < h \leq 50\%$} \\
    & \text{$21.0606+ 0.005565h^2 -0.00035hT- 0.483199h$} & \text{for $h > 50\%$}.
  \end{aligned}
  \right.
\]
A value of $U = 30$ mph and $m = 0$ returns FFWI equal to 100 that is considered to be extreme weather condition for wildfire, and a larger value of $U$ leads to a higher value of FFWI. Hence, we do not consider any truncation at 100, following \cite{sapsis2016fire}.

\section*{Appendix B: Exploratory analysis}
\label{exploratory}
\begin{figure}[ht]
\centering
\adjincludegraphics[width = 0.49\linewidth,  trim = {{.0\width} {.0\width} 0 {.0\width}}, clip]{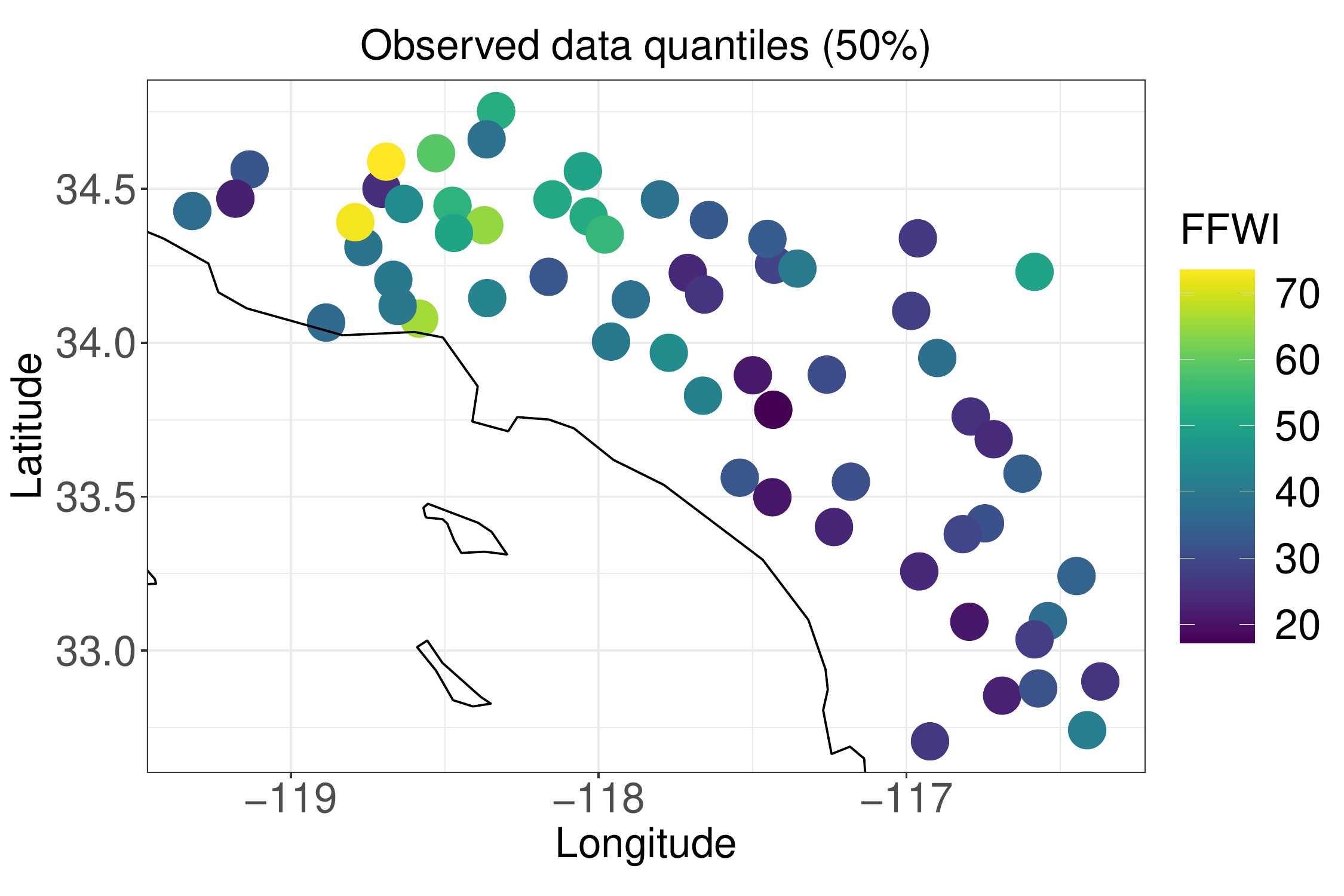}
		\adjincludegraphics[width = 0.49\linewidth, trim = {{.0\width} {.0\width} 0 {.0\width}}, clip]{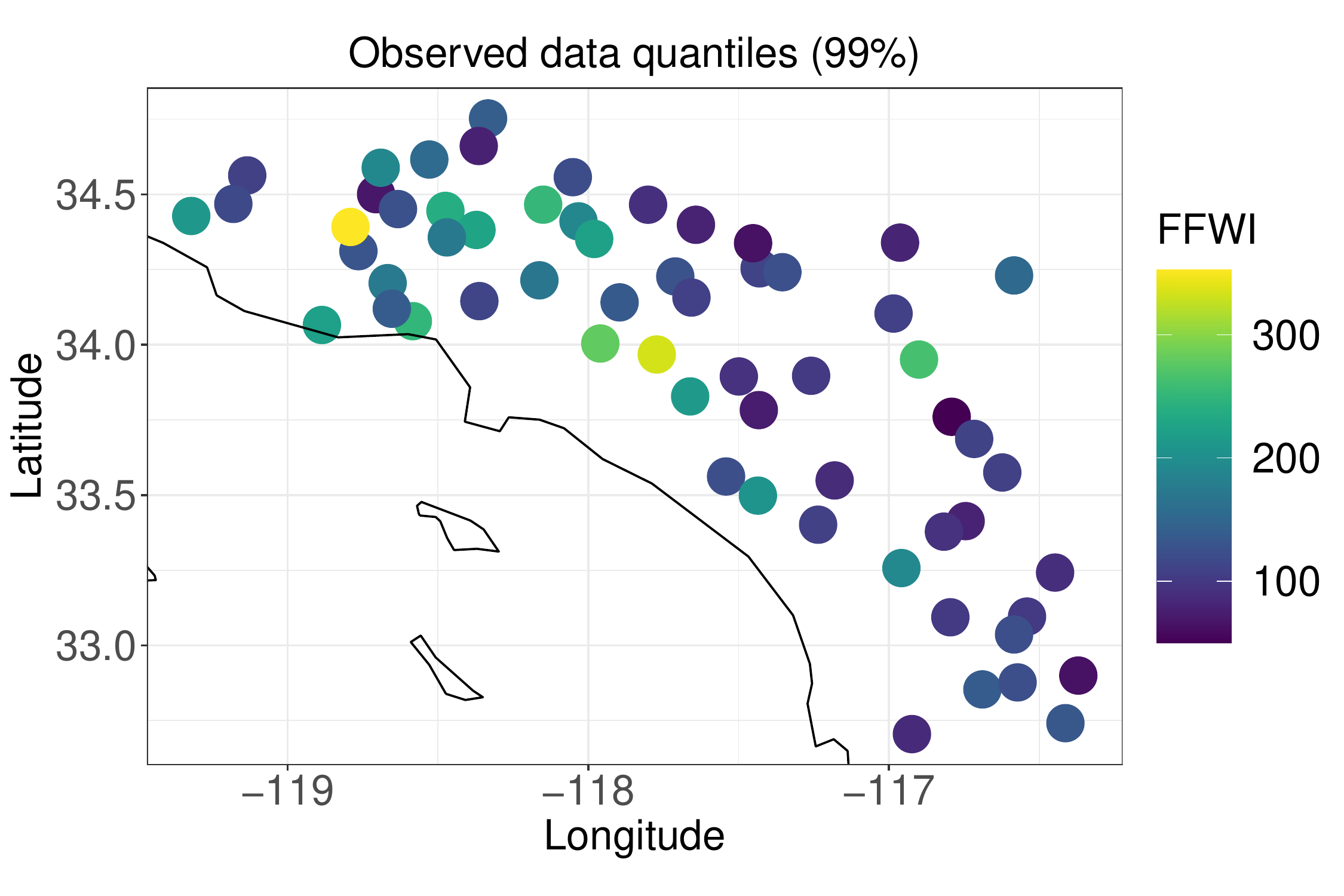}
\vspace{-3mm}		
\caption{Station-wise medians (left) and 0.99$^{th}$ data quantiles (right) for December.}
\label{fig1_supp}
\end{figure}

The plots of the medians and the 0.99$^{th}$ quantiles at each RAWS are provided in Figure \ref{fig1_supp}. The higher values of the quantiles for both the levels are more frequent in Zone 1. A careful inspection identifies that while the higher values of the medians are mainly observed within the Los Angeles county, the higher values of the 0.99$^{th}$ quantiles are observed in some parts of the coastal region additional to the Los Angeles county (consider the staion Talega, for example, with longitude 117.49$^\circ$ W and latitude 33.48$^\circ$ N). Thus, fitting a model that leads to parallel spatial surfaces across different quantile levels (any parametric model like GP, TP or STP which has spatially-invariant scale) is not apt for modeling the FFWI dataset.

\begin{figure}[ht]
\centering
\adjincludegraphics[height = 0.325\linewidth, width = 0.45\linewidth, trim = {{.0\width} {0.0\width} 0 {0.0\width}}, clip]{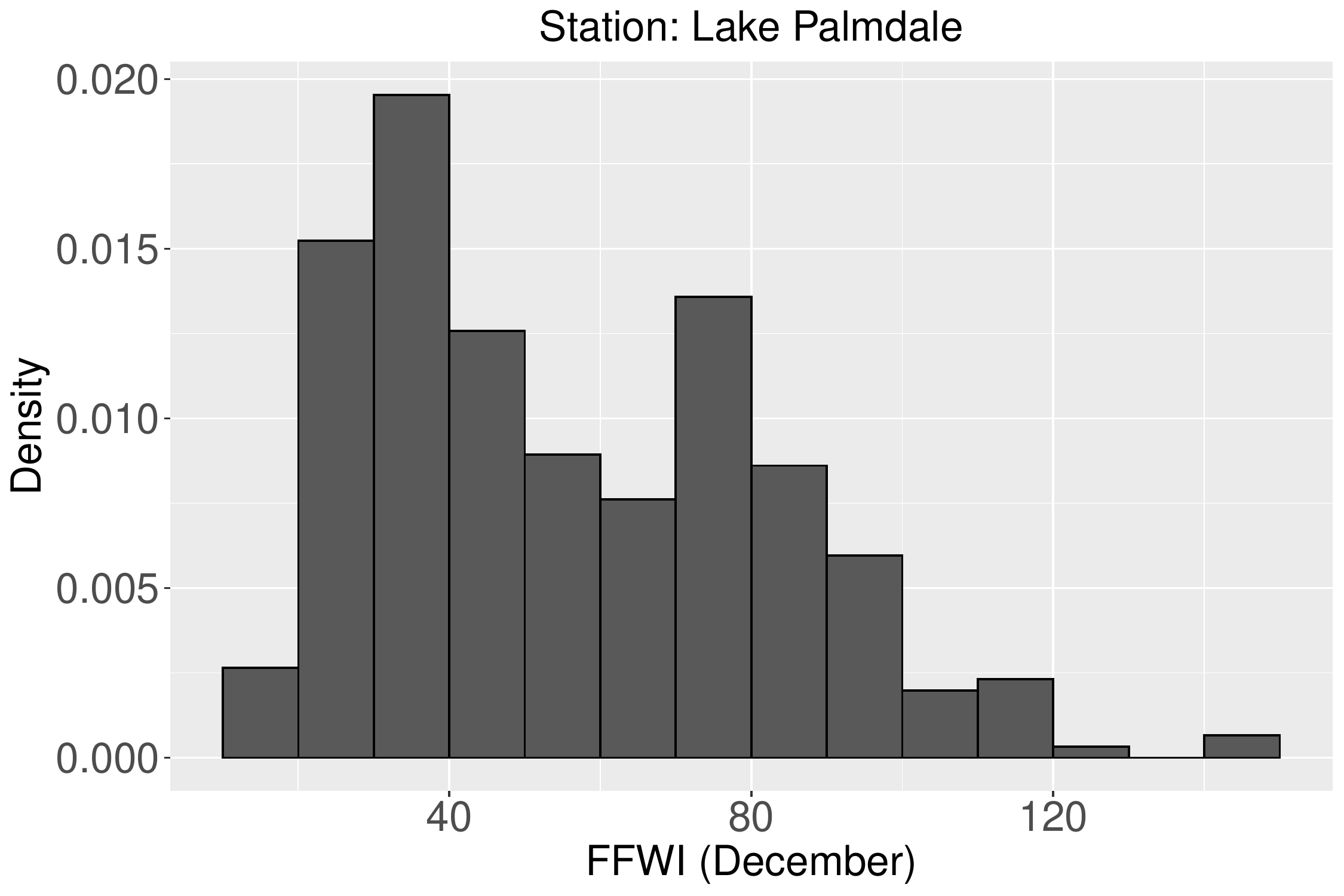}
\adjincludegraphics[width = 0.54\linewidth, trim = {{.0\width} {.02\width} 0 {.02\width}}, clip]{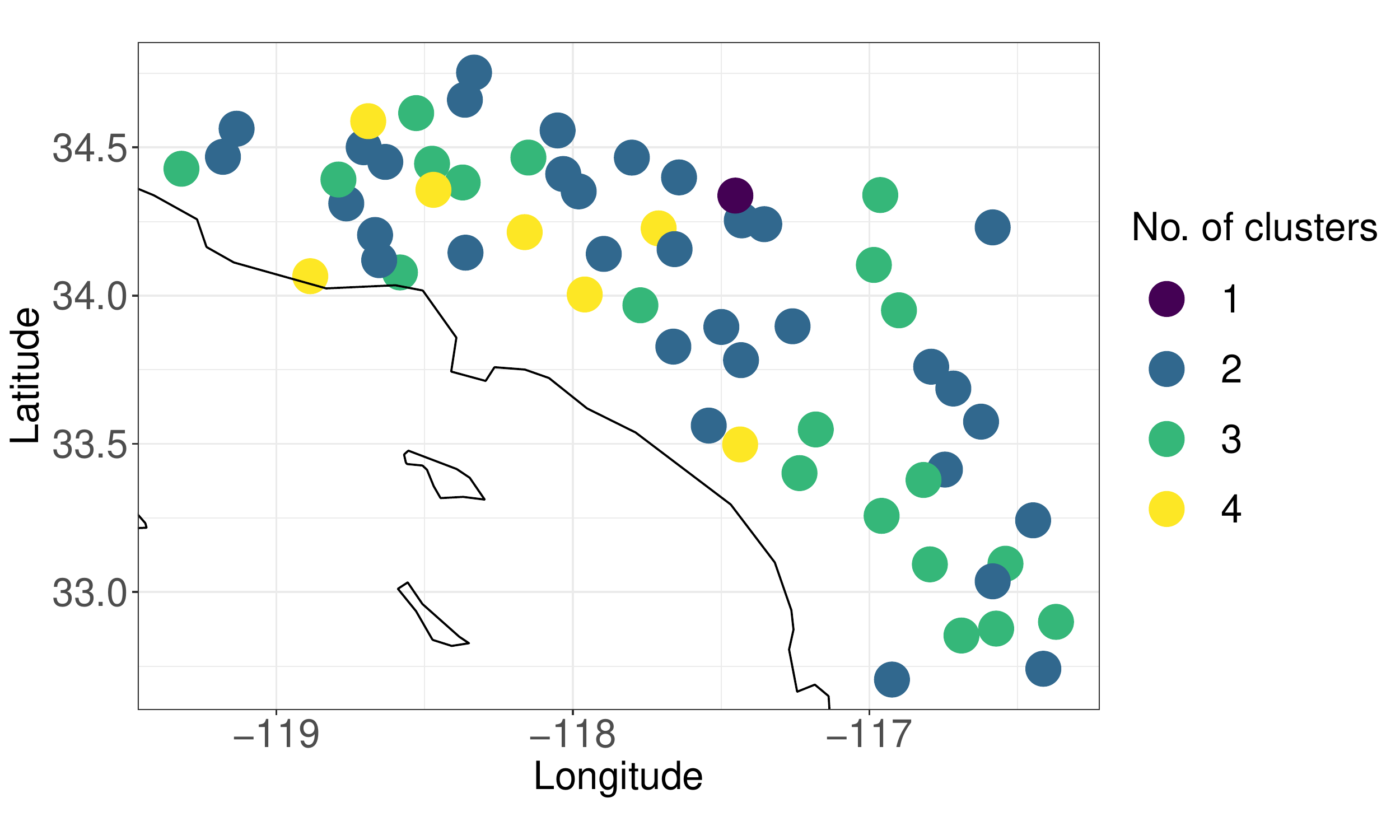}
 \vspace{-6mm}       
\caption{Histogram of FFWI observations in December at Lake Palmdale (left) and the number of components selected using Bayesian Information Criterion by fitting Gaussian mixture model at each station separately (right).}
\label{fig2_supp}
\end{figure}

As a representative station, the histogram at Lake Palmdale is provided in the left panel of Figure \ref{fig2_supp}. The histogram appears to be multi-modal. For each station separately, we fit a univariate Gaussian mixture model and identify the number of mixture components based on Bayesian Information Criterion (BIC) using the R package \texttt{mclust} \citep{fraley2006mclust}. The results are shown in the right panel of Figure \ref{fig2_supp}. Except for one station (Mormon Rocks, with longitude 117.48$^\circ$ W and latitude 34.37$^\circ$ N), multiple components are selected for the rest of the stations. These indicate the need for a semiparametric, or a nonparametric model.


\section*{Appendix C: Marginal and joint distributions of a Skew-t process} 

\noindent \underline{Univariate skew-$t$ distribution}\\
We call $Y$ to follow a univariate skew-$t$ distribution with parameters $(\mu, \lambda, a, b)$ if $Y| z, \sigma^2 \sim \textrm{Normal}(\mu + \lambda \lvert z \rvert, \sigma^2)$ with $z| \sigma^2 \sim \textrm{Normal}(0, \sigma^2)$ and $\sigma^2 \sim \textrm{Inverse-Gamma}(a/2, ab / 2)$. The density function of $Y$ is 
$$f_Y(y) = 2 \frac{1}{\sqrt{b (1 + \lambda^2)}} f_T\left(\frac{y - \mu}{\sqrt{b (1 + \lambda^2)}}; a \right) F_T\left( \lambda \frac{y - \mu}{\sqrt{b (1 + \lambda^2)}} \sqrt{\frac{a + 1}{a + \frac{(y - \mu)^2}{b (1 + \lambda^2)}}}; a + 1 \right), $$
where $f_T(\cdot; a)$ and $F_T(\cdot;a)$ are density and distribution functions of the univariate Student's $t$ distribution (location = 0 and scale = 1) with $a$ degrees of freedom, respectively.\\
	
\noindent \underline{Multivariate skew-$t$ distribution} \\
We call $\bm{Y}$ to follow a $n$-variate skew-$t$ $(\bm{\mu}, \bm{\Sigma}, \bm{\lambda}, a, b)$ distribution, if $\bm{Y}| z, \sigma^2 \sim  \textrm{Normal}_n(\bm{\mu} + \bm{\lambda} \lvert z \rvert , \sigma^2 \bm{\Sigma})$ with $z| \sigma^2 \sim  \textrm{Normal}(0, \sigma^2)$ and $\sigma^2 \sim \textrm{Inverse-Gamma}(a/2, ab / 2)$. The density function of $\bm{Y}$ is
$$f_{\bm{Y}}(\bm{y}) = \frac{2}{\lvert \bm{W} \rvert} f_{T_n}\left(\bm{z}; \bm{\Sigma_Y}, a \right) F_T\left( \bm{\lambda}' \bm{\Sigma}^{-1} \bm{z} \sqrt{\frac{a + n}{a + \bm{z}'\bm{\Sigma_Y}^{-1} \bm{z}}}; a + n \right), $$
where $\bm{z} = \bm{W}(\bm{y} - \bm{\mu})$ with $\bm{W} = \left( b (1 + \bm{\lambda}' \bm{\Sigma}^{-1} \bm{\lambda}) \right)^{-1/2} \bm{I}_n$, $\lvert \bm{W} \rvert$ is the determinant of $\bm{W}$,  $f_{T_n}(\cdot; \bm{\Sigma_Y}, a)$ is the density function of $n$-variate Student's $t$ distribution with location $\bm{0}_n$, shape matrix $\bm{\Sigma_Y}$ and $a$ degrees of freedom and $F_T(\cdot;a)$ is the CDF of the univariate Student's $t$ distribution (location = 0 and scale = 1) with $a$ degrees of freedom. The matrix $\bm{\Sigma_Y}$ is given by $\bm{\Sigma_Y} = (1 + \bm{\lambda}' \bm{\Sigma}^{-1} \bm{\lambda})^{-1} \left( \bm{\Sigma} + \bm{\lambda} \bm{\lambda}' \right)$.
	
\begin{figure}[h]
\centering
\adjincludegraphics[height = 0.3\linewidth, width = 0.24\linewidth , trim = {{.0\width} {.0\width} 0 {.0\width}}, clip]{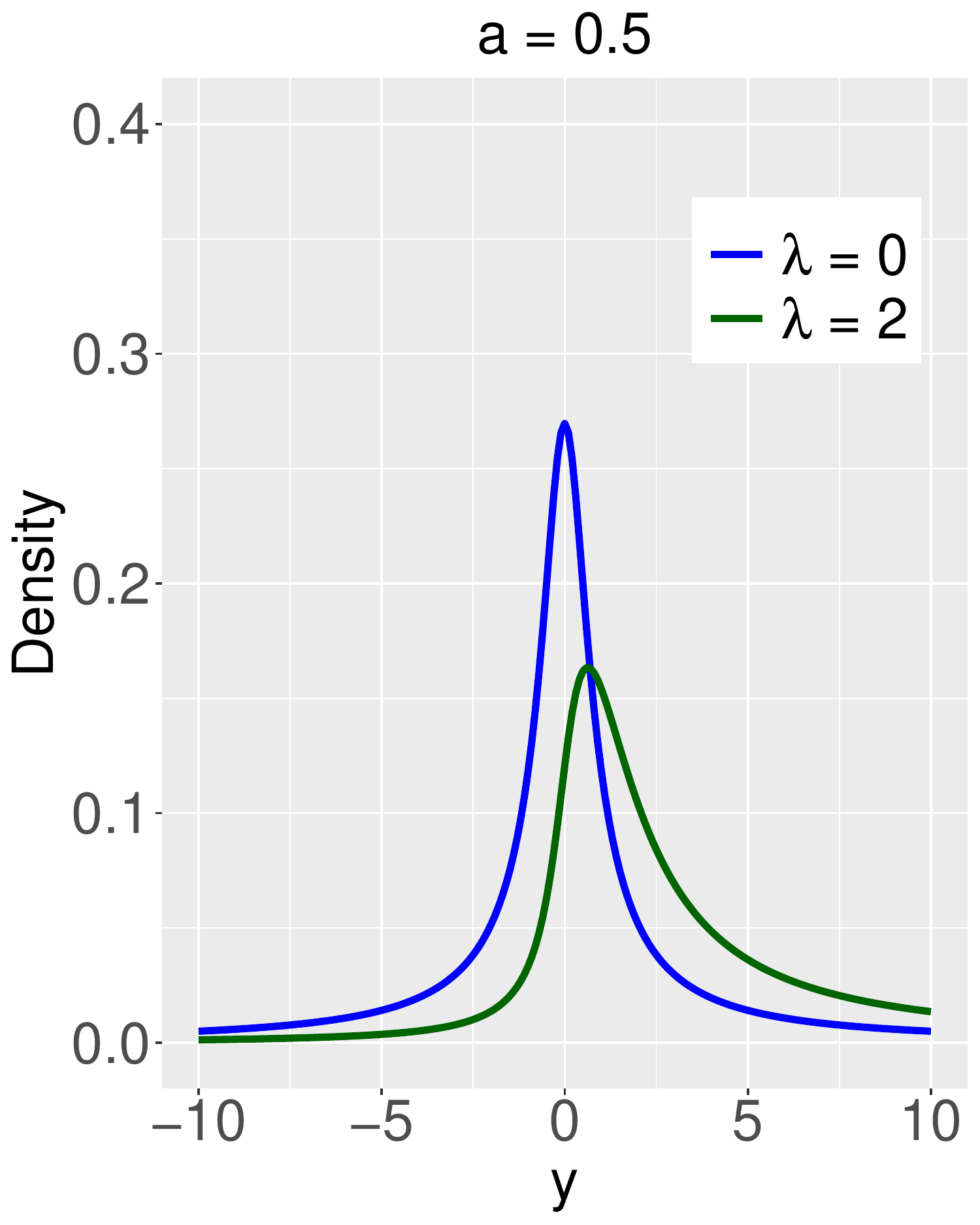}
\adjincludegraphics[height = 0.3\linewidth, width = 0.24\linewidth, trim = {{.0\width} {.0\width} 0 {.0\width}}, clip]{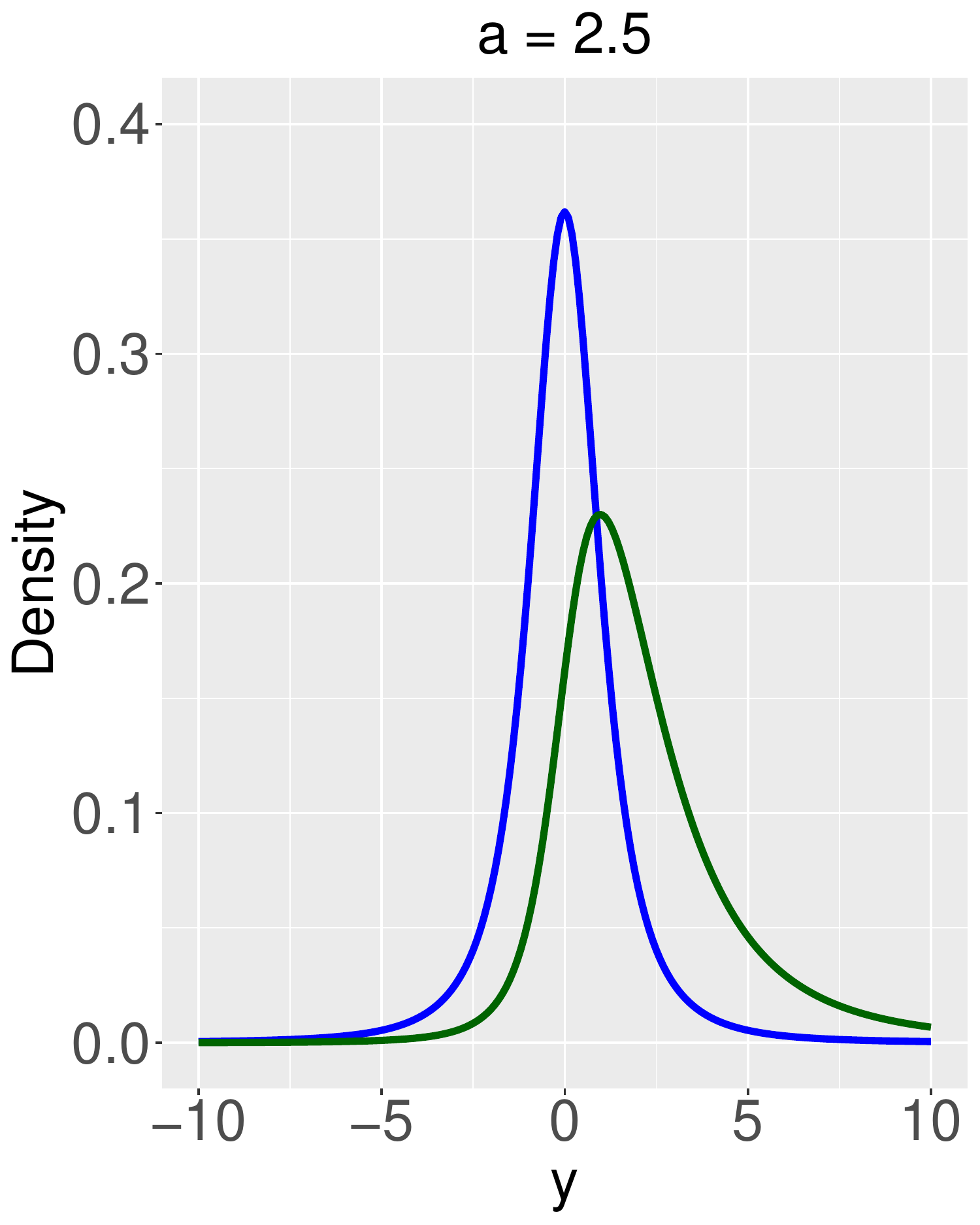}
\adjincludegraphics[height = 0.3\linewidth, width = 0.24\linewidth, trim = {{.0\width} {.0\width} 0 {.0\width}}, clip]{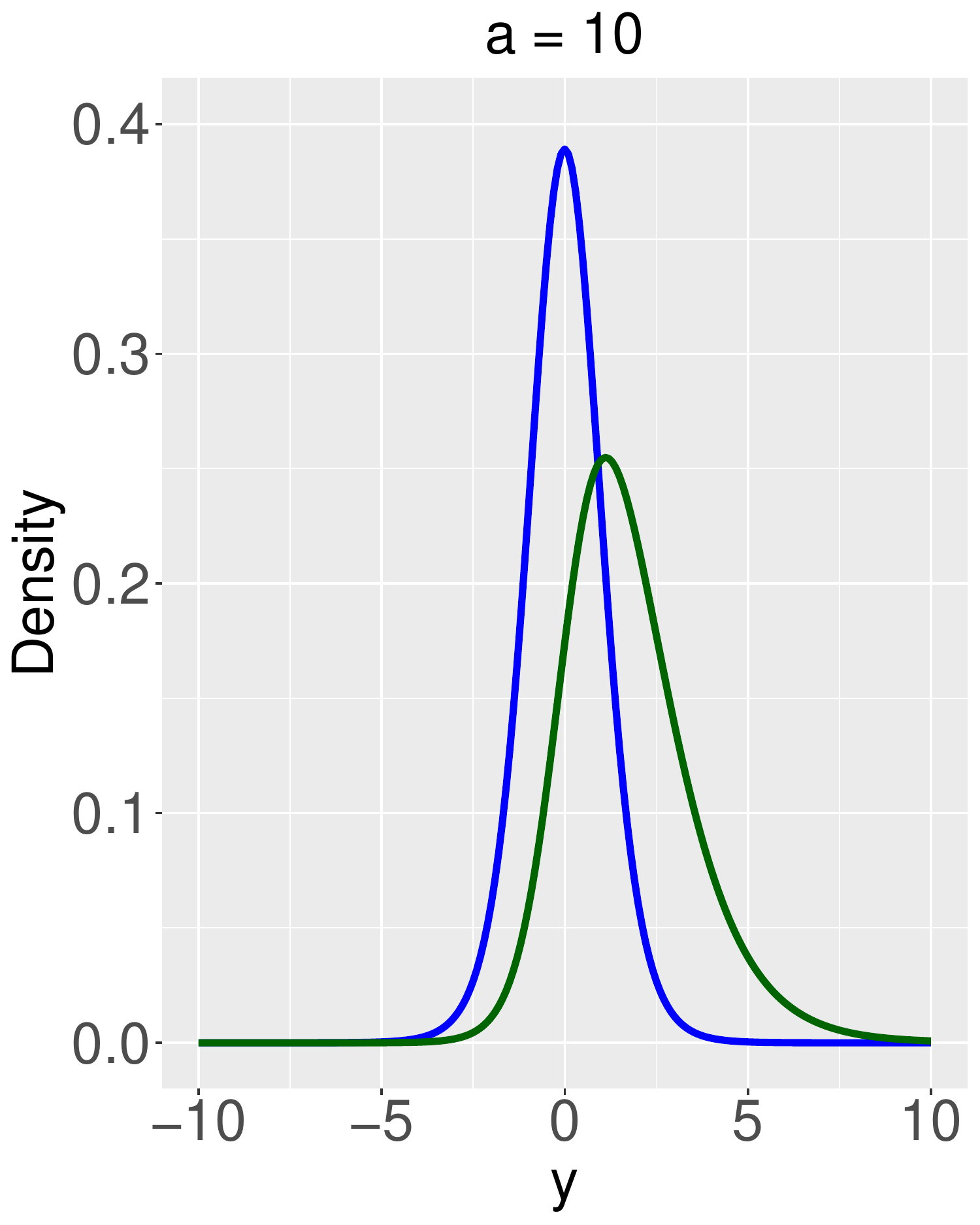}
\adjincludegraphics[height = 0.3\linewidth, width = 0.24\linewidth, trim = {{.0\width} {.0\width} 0 {.0\width}}, clip]{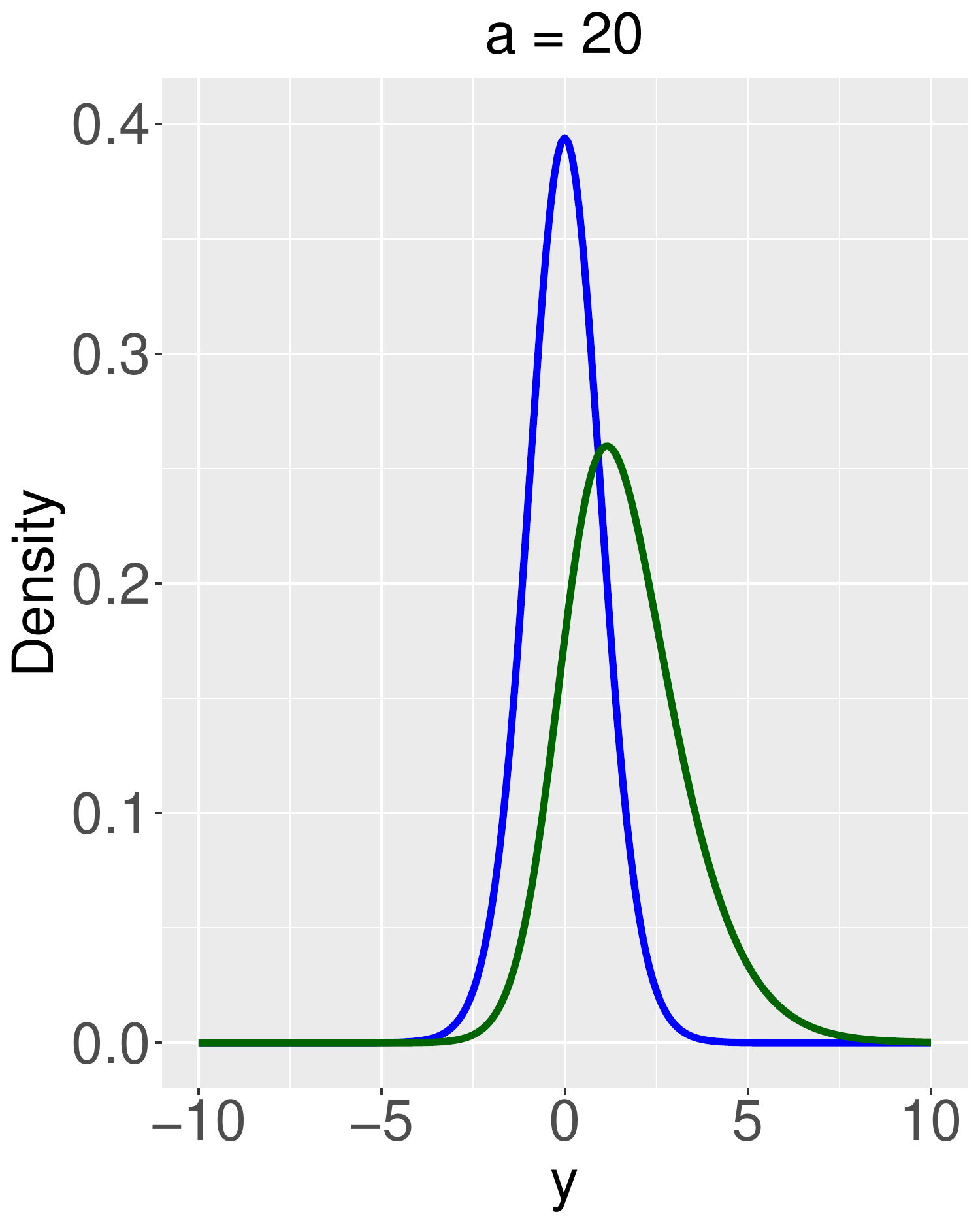}
\vspace{-2mm}
\caption{Univariate skew-$t$ density functions for different choices of the parameters. We fix the location parameter to $\mu = 0$ and scale to $b= 1$. }
\label{skewt_desnity}
\end{figure}

	
In Figure \ref{skewt_desnity}, the univariate skew-$t$ density functions for different choices of the model parameters are illustrated. The right-skewed nature of the distribution for $\lambda = 2$ compared to the symmetric nature for $\lambda = 0$ illustrates how the hierarchically defined skew-$t$ distribution extends the normal distribution to a larger class of models that are capable of modeling asymmetry. Increasing $a=0.5$ to $a=20$, the heaviness of the tails reduces towards that of a normal distribution and hence, by varying $a$, the skew-$t$ distribution extends the normal distribution to a class of models with flexible tail-heaviness. For more details, see \cite{azzalini2014skew}.
    
\section*{Appendix D: GEV-log transformation}
\label{transform}
In extreme value analysis, a block maximum $Y$ is assumed to be distributed as a generalized extreme value (GEV) distribution with CDF-- $\textrm{Pr}(Y \leq y) = \exp[-t(y)]$, where
\[
  \text{$t(y)$} =
  \left\{
  \begin{aligned}
     & \left[ 1 + \frac{\xi_y}{\sigma_y} (y - \mu_y) \right]^{- \frac{1}{\xi_y}} & \text{if $\xi_y \neq 0$}, \\
    & \exp\left[ -(y - \mu_y) / \sigma_y \right] & \text{if $\xi_y = 0$}.
  \end{aligned}
  \right.
\]

The location, scale, and shape parameters  are $\mu_y \in \mathbb{R}, \sigma_y>0$ and $ \xi_y \in \mathbb{R}$ respectively. These three parameters jointly determine the support of $Y$-- $(\mu_y - \sigma_y / \xi_y, \infty)$ if $\xi_y > 0$, $\mathbb{R}$ if $\xi_y = 0$ and $(-\infty, \mu_y - \sigma_y / \xi_y)$ if $\xi_y < 0$. We denote $Y \sim \textrm{GEV}(\mu_y, \sigma_y, \xi_y)$. 

The flexible tools of Bayesian nonparametrics (BNP) are easier to implement if the support of $Y$ is assumed to be the whole real line. Thus, we consider a transformation so that we can easily implement the BNP tools along with allowing the observations to have a more generalized support. Suppose, we have $\xi_y \neq 0$. By GEV transformation, we consider $\tilde{Y} =  \left \lbrace 1 + \xi_y \sigma_y^{-1} (Y - \mu_y)\right \rbrace^{\xi_y^{-1}}$ so that $\tilde{Y} \sim \textrm{GEV}(1, 1, 1)$, the unit-Fr\'echet distribution. The support of $\tilde{Y}$ is $(0, \infty)$. Further, we consider a log transformation to obtain $Y^* = \log(\tilde{Y})$ so that $Y^*$ follows the standard Gumbel distribution which has support over the whole real line. If $\xi_y = 0$, the transformation is defined in a limiting sense ($\xi_y \rightarrow 0$), and $Y^*$ is only a location-scale transformation of $Y$ in that case; both $Y$ and $Y^*$ are supported over the whole real line.  Finally, we apply the BNP tools over $Y^*$. Additional to flexible support, the transformation can add skewness for simple models like Gaussian processes (GPs). While the marginals become more flexible, the extremal dependence $\chi$ is invariant of the GEV-log transformation (or any other monotonically increasing transformation), and hence, an inverse GEV-log transformed GP would still remain asymptotically independent like a simple GP. Thus, location/scale mixtures of GPs are still required to model spatial data that exhibits spatial extremal dependence.


\begin{figure}[ht]
\centering
\includegraphics[height = 0.35\linewidth]{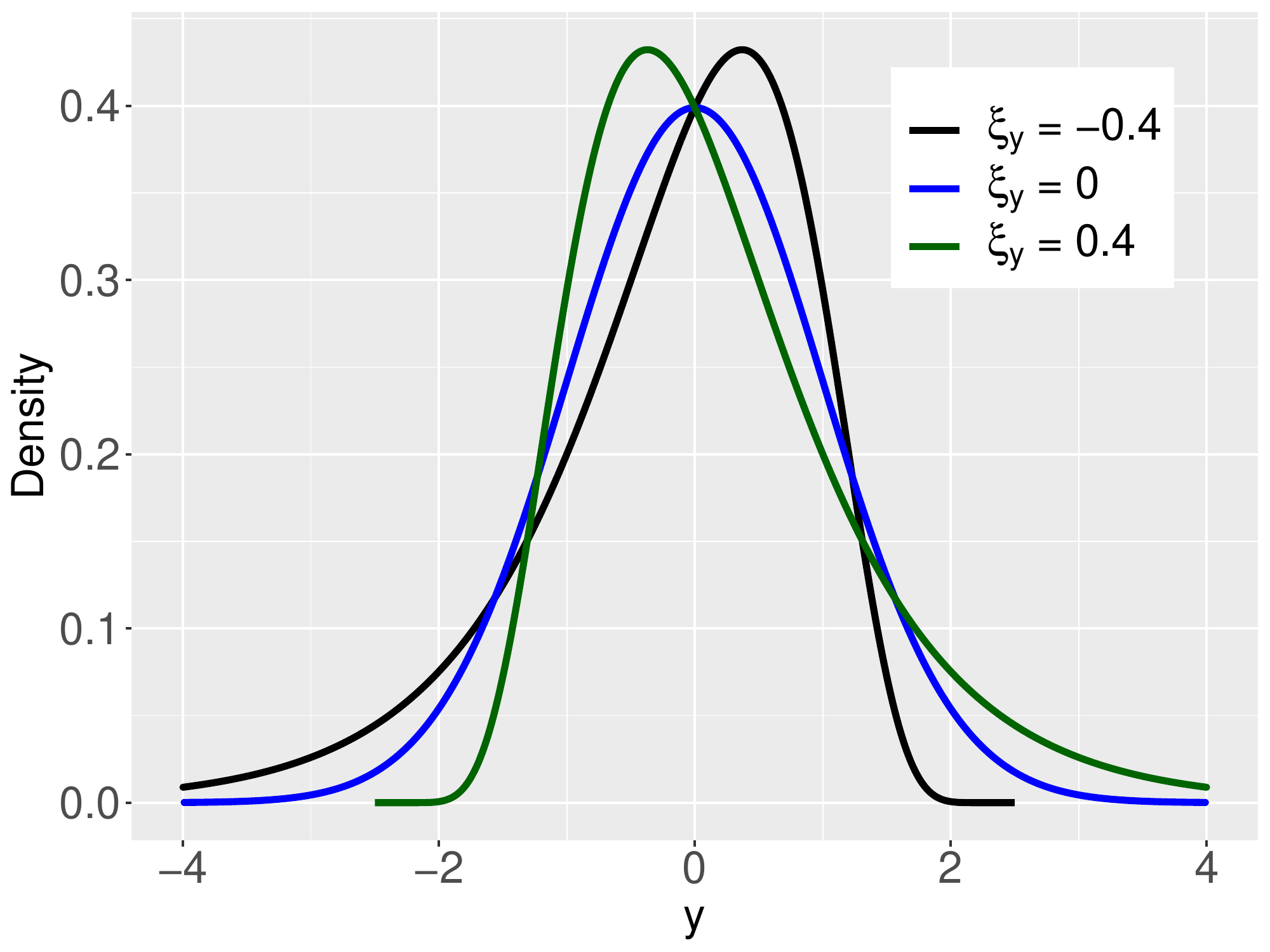}
\vspace{-2mm}
\caption{Density of the inverse GEV-log transformed standard normal random variable. Here $\mu_y = 0$, $\sigma_y = 1$ and $\xi_y = -0.4, 0, 0.4$.}
\label{transformation}
\vspace{-2mm}
\end{figure}

As an illustration, we consider three transformations of a standard normal random variable in Figure \ref{transformation}. Here we consider $Y^* \sim \textrm{Normal}(0, 1)$ and transform $Y^*$ to obtain $Y$. We fix the location and scale parameters to $\mu_y = 0$ and $\sigma_y = 1$. When $\xi_y = 0$, the density of $Y$ and $Y^*$ coincide. When $\xi_y = 0.4$, the density of $Y$ is bounded below and have heavier right tail compared to $Y^*$. Alternatively, when $\xi_y = -0.4$, the density of $Y$ is bounded above and have heavier left tail compared to $Y^*$. 

In case we consider $Y^* \sim \textrm{Normal}(\mu, \sigma^2)$, all the parameters $(\mu_y, \sigma_y, \xi_y, \mu, \sigma^2)$ are not individually identifiable unlike the support of the distribution. For the proposed STP-DPM model, the parameters involved within the distribution of $Y^*$ and $(\mu_y, \sigma_y, \xi_y)$ are also not identifiable; although, the support of the distribution is identifiable. Considering our main interest lies in the spatial maps of the high quantiles, non-identifiability of the individual parameters is not a major concern here.


\section*{Appendix E: Proof of Remark 2}
\label{proofs_meanvar}


By setting $\lambda_k = 0$, $a_k = \infty$ and $b_k = 0$ for each $k$ (thus, $Y_t^*(\bm{\cdot})$ has discrete support $\mu_k(\cdot)$'s), the proposed STP-DPM model has $\textrm{E}\left[Y_t^*(\bm{s}) \right] = \sum_{k=1}^{K} \pi_k \mu_{k}(\bm{s}) = \bar{\mu}(\bm{s})$ and $\textrm{Cov}\left[Y_t^*(\bm{s}), Y_t^*(\bm{s}') \right] = \sum_{k=1}^{K} \pi_k [\mu_{k}(\bm{s}) - \bar{\mu}(\bm{s})] [\mu_{k}(\bm{s}') -  \bar{\mu}(\bm{s}')]$ given the cluster parameters $\Theta_k$ and the mixture probabilities $\pi_k$. Specifically, consider a generic process $Y^*(\bm{s})$ with continuous mean function $\mu_0(\bm{s})$ and continuous covariance function $C_0(\bm{s}, \bm{s}')$. We will show that there exists a $K$ and model components $\mu_k(\cdot)$, $\pi_k$ such that: 
\begin{eqnarray}
\begin{cases}
& \sum_{k=1}^{K} \pi_k {\mu}_{k}(\bm{s}) = {\mu}_0(\bm{s}) \label{eqn1}  \\
& \sum_{k=1}^{K} \pi_k \left\{{\mu}_{k}(\bm{s}) - {\mu}_0(\bm{s}) \right\} \left\{ {\mu}_{k}(\bm{s}') - {\mu}_0 (\bm{s}')\right\} = C_0(\bm{s}, \bm{s}'), \label{eqn2} 
\end{cases}
\end{eqnarray}
for all $\bm{s}, \bm{s}'\in \mathcal{D}.$

From Mercer's theorem, the spectral decomposition of $C_0$ is $C_0(\bm{s}, \bm{s}') = \sum_{\ell\geq 1} \lambda_\ell \psi_\ell(\bm{s}) \psi_\ell(\bm{s}')$, where $\lambda_1\geq \lambda_2\ldots \geq0 $ and $\{ \psi_\ell(\bm{s})\}_{\ell\geq 1}$ forms an orthogonal basis in $L^2[\mathcal{D}]$. Let $L>1$ be sufficiently large such that $\lambda_L>0$ and $\lambda_{L+1}=0$ so that $C_0(\bm{s},\bm{s}')= \sum_{\ell=1}^{L} \lambda_\ell\psi_\ell(\bm{s}) \psi_\ell(\bm{s}')$. The proof still holds when $L=\infty $ and $\lambda_k\neq 0$ for any $k\geq 1$. In this case, we would have operations with infinite dimensional vectors and matrices.

Now select $K=L+1$ and consider $\pi_1, \ldots \pi_{L+1}$ such that $\sum_{k=1}^{L+1} \pi_k=1$; without loss of generality assume that $\pi_{L+1}\neq0$. It is sufficient to prove that there exists $\mu^*_1(\bm{s}), \ldots \mu^*_L(\bm{s})$ such that 
\begin{eqnarray}
\sum_{k=1}^L \pi_k {\mu}^*_{k}(\bm{s}) {\mu}^{*}_{k}(\bm{s}') + \pi_{L+1}^{-1} \sum_{k=1}^L \sum_{k'}^{L}\pi_k \pi_{k'}{\mu}^*_{k}(\bm{s})  {\mu}^*_{k'} (\bm{s}') 
=\sum_{\ell=1}^{L} \lambda_\ell\psi_\ell(\bm{s}) \psi_\ell(\bm{s}');    \ \label{eqn:red} 
\end{eqnarray}
simple algebra shows that the above system of equations is satisfied by choosing $ \mu_k(\bm{s}) = \mu_0(\bm{s}) + \mu^*_k(\bm{s})$ for $k=1, \ldots, L$ and $\mu_{L+1}(\bm{s}) = \mu_0(\bm{s}) - \pi_{L+1}^{-1} \sum_{k=1}^L \pi_k \mu^*_k(\bm{s})$.

Consider now the left hand side of the equality (\ref{eqn:red}) and rewrite it as \\
$\sum_{k=1}^L \sum_{k'}^{L}  {\mu}^*_{k}(\bm{s}) {\mu}^{*}_{k'}(\bm{s}') M_{kk'} $, where $M_{kk'} $ is the $(k,k')$ element of the $L\times L$ matrix $\bm{M}$ defined by 
\begin{eqnarray}
\bm{M}=\begin{pmatrix}
\pi_1 & 0 &\ldots& 0\\
0&\pi_2&\ldots &0\\
\ldots &\ldots &\ldots &\ldots\\
0&0&\ldots&\pi_L
\end{pmatrix}+ \frac{1}{\pi_{L+1}} 
\begin{pmatrix}
\pi^2_1 & \pi_1\pi_2 &\ldots& \pi_1\pi_L\\
 \pi_1\pi_2 &\pi^2_2&\ldots & \pi_2\pi_L \\
\ldots &\ldots &\ldots &\ldots\\
\pi_L\pi_1&\pi_L\pi_2&\ldots&\pi^2_L
\end{pmatrix}.	
\end{eqnarray}
This matrix is symmetric and is clearly positive definite as $a^T \bm{M} a = \sum_{\ell =1}^{L} \pi_\ell a^2_\ell + \sum_{\ell=1}^{L} (a^T \bm\pi)^2/\pi_{L+1}>0$ for any $a=(a_1, \ldots, a_L)'\neq 0$, where $\bm{\pi} = (\pi_1, \ldots, \pi_L)'$; thus it is also non-singular. It follows that  (\ref{eqn:red}) can be re-expressed as 
$\sum_{k=1}^L \sum_{k'}^{L}  {\mu}^*_{k}(\bm{s}) {\mu}^{*}_{k'}(\bm{s}') M_{kk'} =\sum_{\ell=1}^{L} \lambda_\ell\psi_\ell(\bm{s}) \psi_\ell(\bm{s}') $, which has a simple solution. Specifically, denote by $\bm{M}^{-1/2}$ the inverse square root of the matrix $\bm{M}$, and by $[{M}^{-1/2}]_{\ell k}$ the $(\ell,k)$ element of $\bm{M}^{-1/2}$. Then a solution of (\ref{eqn:red}) is 
\begin{eqnarray}
\mu_k^*(\bm{s}) = \sum_{\ell=1}^{L} [{M}^{-1/2}]_{\ell k} \lambda^{1/2}_\ell \psi_\ell(\bm{s}).
\end{eqnarray}

\section*{Appendix F: Derivation of spatial extremal dependence}
\label{proofs_chi}



We drop the time series structure (also the subscript $t$) and calculate the spatial extremal dependence. While some of the following results can be found in the literature (\cite{bortot2010tail}, for example), proofs are non-trivial, particularly for the skew-$t$ process and the mixture of skew-$t$ processes, and not published elsewhere as of authors' knowledge. Hence, we provide the detailed derivations. 
Before considering the mixture as well as the skewed structure, we derive the $\chi$-measure for a spatial Student's $t$-process. 
    
Suppose $R_k(\cdot)$ is a spatial Student's $t$-process constructed by random scaling of a Gaussian process with zero mean and correlation function (possibly nonstationary) $r_k(\cdot, \cdot)$. For two spatial locations $\bm{s}_1$ and $\bm{s}_2$, by an abuse of notation (only for the derivation and not elsewhere), we denote the realizations as $R_{k1} = R_k(\bm{s}_1)$ and $R_{k2} = R_k(\bm{s}_2)$, respectively, and also denote the correlation of the underlying Gaussian random variables as $r_k = r_k(\bm{s}_1, \bm{s}_2)$. The conditional (on $\sigma^2$) distribution of the bivariate observation $(R_{k1}, R_{k2})'$ is
	\begin{eqnarray}
	\nonumber &&
	\left(
	\begin{array}{c}
	R_{k1}\\
	R_{k2}
	\end{array}
	\right) \bigg\rvert \sigma^2 \sim \textrm{Normal}_2\left( \left(
	\begin{array}{c}
	0\\
	0
	\end{array}
	\right), \sigma^2 \left(
	\begin{array}{c}
	1 ~~ r_k \\
	r_k ~~ 1
	\end{array}
	\right) \right).
	\end{eqnarray}

The random scale term $\sigma^2 \sim \textrm{Inverse-Gamma}\left(a_k/2, a_k b_k/2\right)$. Thus, the conditional distribution of $R_{k1}$ given $R_{k2} = y_2$ and $\sigma^2$ is $R_{k1} | \lbrace R_{k2} = y_2, \sigma^2 \rbrace \sim \textrm{Normal}\left(r_k y_2,  (1 - r_k^2) \sigma^2 \right)$.
After marginalizing the random scale, we have
\begin{eqnarray}
	\nonumber && \pi_{R_{k1}}(y_1 | R_{k2} = y_2) = \int_{0}^{\infty} \pi_{R_{k1}}(y_1 | \sigma^2, R_{k2} = y_2) \pi(\sigma^2 | R_{k2} = y_2) d \sigma^2.
	\end{eqnarray}
By Bayes' theorem, we have $\sigma^2| \lbrace R_{k2} = y_2 \rbrace \sim \textrm{Inverse-Gamma}\left( \frac{a_k + 1}{2}, \frac{a_k b_k + y_2^2}{2} \right)$.
Thus, by marginalizing through the posterior distribution of $\sigma^2$, we have
	\begin{eqnarray}
	\nonumber \pi_{R_{k1}}(y_1 | R_{k2} = y_2) &=&  \frac{1}{\sqrt{\pi (a_k b_k + y_2^2)}} \frac{\Gamma(\frac{a_k + 2}{2})}{\Gamma(\frac{a_k + 1}{2})}  \left(1 + \frac{(y_1 - r_k y_2)^2}{(1 - r_k^2) (a_k b_k + y_2^2)}\right)^{- \frac{a_k + 2}{2}},
	\end{eqnarray}
	which implies that, conditional on $R_{k2} = y_2$,
	\begin{equation}
	\nonumber \sqrt{\frac{(a_k + 1)}{a_k b_k + y_2^2}} \left[\frac{R_{k1} - r_k y_2}{\sqrt{1 - r_k^2}} \right] \sim F_T(\cdot~; a_k + 1).
	\end{equation}
Hence, $\textrm{Pr}(R_{k1} \leq y | R_{k2} = y) = F_T\left( \sqrt{(a_k + 1)} \sqrt{\frac{1 - r_k}{1 + r_k}} \frac{y}{\sqrt{a_k b_k + y^2}}; a_k + 1 \right) $.


Thus, the extremal dependence between $R_{k1}$ and $R_{k2}$ is
\begin{eqnarray}
	\nonumber \chi_{R_k}\left(\bm{s}_1, \bm{s}_2 \right) &=& \lim_{y \rightarrow \infty} \textrm{Pr}\left( R_{k1} > y | R_{k2} > y \right) \\
\nonumber &=& 2 - 2\lim_{y \rightarrow \infty} \textrm{Pr}( R_{k1} \leq y |  R_{k2} = y)~~\textrm{[follows~from~L'Hospital's rule]} \\
	\nonumber &=& 2 - 2 F_T\left( \sqrt{(a_k + 1)} \sqrt{\frac{1 - r_k}{1 + r_k}}; a_k + 1 \right).
	\end{eqnarray}
    
Further, we consider a spatial skew-$t$ process $\tilde{R}_k(\cdot)$. Additional to the process $R_k(\cdot)$, here we consider a non-zero skewness term and we have $\tilde{R}_k(\cdot) = R_k(\cdot) + \lambda_k |z|$ where $z \sim \textrm{Normal}(0, \sigma^2)$ with same random $\sigma^2$ term as in $R_k(\cdot)$. By an abuse of notation, we denote the realizations of $\tilde{R}_k(\cdot)$ at $\bm{s}_1$ and $\bm{s}_2$ as $\tilde{R}_{k1} = \tilde{R}_k(\bm{s}_1)$ and $\tilde{R}_{k2} = \tilde{R}_k(\bm{s}_2)$, respectively. Following \cite{bortot2010tail}, the limit of the joint exceedance probability of $\tilde{\bm{R}} = (\tilde{R}_{k1}, \tilde{R}_{k2} )'$ is
\begin{eqnarray}
\nonumber && \lim_{y \rightarrow \infty} \textrm{Pr}(\tilde{R}_{k1} > y, \tilde{R}_{k2} > y) \\
\nonumber &=& \lim_{y \rightarrow \infty} \int_{y}^{\infty} \int_{y}^{\infty} \frac{2}{b_k(1 + 2\lambda_k^2/ (1 + r_k))} f_{T_2}\left(\tilde{\bm{R}}^*; \bm{\Sigma_{\tilde{R}}}, a_k \right) \\
\nonumber && ~~~~~~~~~~~\times F_T\left( \lambda_k \bm{1}_2' \bm{\Sigma}^{-1} \tilde{\bm{R}}^* \sqrt{\frac{a_k + 2}{a_k + \tilde{\bm{R}}^{*'}\bm{\Sigma_{\tilde{R}}}^{-1} \tilde{\bm{R}}^*}}; a_k + 2 \right) d\tilde{R}_{k1} d\tilde{R}_{k2} \\
\nonumber &\sim& \lim_{y \rightarrow \infty} F_T\left( \lambda_k y \frac{2 / (1 +r_k )}{\sqrt{b_k(1 + 2\lambda_k^2/ (1 + r_k))}} \sqrt{\frac{a_k + 2}{a_k + \frac{2/ (1 + r_k)}{b_k(1 + 2\lambda_k^2/ (1 + r_k))} y^2}}; a_k + 2 \right) \\
\nonumber && \times  \lim_{y \rightarrow \infty} \int_{y}^{\infty} \int_{y}^{\infty} \frac{2}{b_k(1 + 2\lambda_k^2/ (1 + r_k))} f_{T_2}\left(\tilde{\bm{R}}^*; \bm{\Sigma_{\tilde{R}}}, a_k \right) d\tilde{R}_{k1} d\tilde{R}_{k2} \\
\nonumber &=& F_T\left( \lambda_k \sqrt{a_k+2} \sqrt{2 / (1 +r_k )}; a_k + 2 \right) \\
\nonumber && \times  \lim_{y \rightarrow \infty} \int_{y}^{\infty} \int_{y}^{\infty}
\frac{2}{\pi a_k b_k(1 + 2\lambda_k^2/ (1 + r_k)) \lvert \bm{\Sigma_{\tilde{R}}} \rvert^{1/2} }
 \frac{\Gamma(\frac{a_k + 2}{2})}{\Gamma(\frac{a_k}{2})} \\
 \nonumber &&~~~~~~~~~~~~~~~~~~~~~~~~ \times \left(1 + \frac{1}{a_k b_k} \frac{\tilde{\bm{R}}'\bm{\Sigma_{\tilde{R}}}^{-1}\tilde{\bm{R}}}{1 + 2\lambda_k^2/ (1 + r_k)} \right)^{- \frac{a_k + 2}{2}} d\tilde{R}_{k1} d\tilde{R}_{k2},
\end{eqnarray}
where $\tilde{\bm{R}}^* = \tilde{\bm{R}} / \sqrt{b_k(1 + 2\lambda_k^2/ (1 + r_k))}$. The matrix $\bm{\Sigma_{\tilde{R}}}$ is given by $\bm{\Sigma_{\tilde{R}}} = (1 + 2\lambda_k^2/ (1 + r_k) )^{-1} \left( \bm{\Sigma} + \lambda_k^2 \bm{1}_2 \bm{1}_2' \right)$. Here $\bm{\Sigma}$ denotes the $2\times 2$-dimensional correlation matrix with the off-diagonal element $r_k$.


For the spatial location $\bm{s}$, the limiting univariate exceedance probability is
\begin{eqnarray}
\nonumber && \lim_{y \rightarrow \infty} \textrm{Pr}(\tilde{R}_{k2} > y) \\
\nonumber &=& \lim_{y \rightarrow \infty} \int_{y}^{\infty} \frac{2}{\sqrt{b_k (1 + \lambda_k^2)}} f_T\left(\frac{\tilde{R}_{k2}}{\sqrt{b_k (1 + \lambda_k^2)}}; a_k \right) \\
\nonumber && ~~~~~~~~~~~~~ \times F_T\left( \lambda_k \frac{\tilde{R}_{k2}}{\sqrt{b_k (1 + \lambda_k^2)}} \sqrt{\frac{a_k + 1}{a_k + \frac{\tilde{R}_{k2}^2}{b_k (1 + \lambda_k^2)}}}; a_k + 1 \right) d\tilde{R}_{k2} \\
\nonumber &\sim& \lim_{y \rightarrow \infty} F_T\left( \lambda_k \frac{y}{\sqrt{b_k (1 + \lambda_k^2)}} \sqrt{\frac{a_k + 1}{a_k + \frac{y^2}{b_k (1 + \lambda_k^2)}}}; a_k + 1 \right) \\
\nonumber &&~~~~~~~~ \times \lim_{y \rightarrow \infty} \int_{y}^{\infty} 2 \frac{1}{\sqrt{b_k (1 + \lambda_k^2)}} f_T\left(\frac{\tilde{R}_{k2}}{\sqrt{b_k (1 + \lambda_k^2)}}; a_k \right) d\tilde{R}_{k2} \\
\nonumber &=& F_T\left( \lambda_k \sqrt{a_k + 1}; a_k + 1 \right) \times \lim_{y \rightarrow \infty} \left( 1 - F_T\left(\frac{y}{\sqrt{b_k (1 + \lambda_k^2)}}; a_k \right) \right).
\end{eqnarray}

We define $\tilde{R}^*_{k1} = \sqrt{b_k(1 + \lambda_k^2)} \tilde{\epsilon}_{k1} $ and $\tilde{R}^*_{k2} = \sqrt{b_k(1 + \lambda_k^2)} \tilde{\epsilon}_{k2} $ with 
\begin{eqnarray}
\nonumber &&
\left(
\begin{array}{c}
\tilde{\epsilon}_{k1}\\
\tilde{\epsilon}_{k2}
\end{array}
\right) \bigg\vert \sigma^2 \sim \textrm{Normal}_2\left( \left(
\begin{array}{c}
0\\
0
\end{array}
\right), \sigma^2 \left(
\begin{array}{c}
1 ~~~~~~~~~~~ \frac{r_k + \lambda_k^2}{1 + \lambda_k^2}  \\
\frac{r_k + \lambda_k^2}{1 + \lambda_k^2} ~~~~~~~~~~~ 1
\end{array}
\right) \right).
\end{eqnarray}

Let us denote the correlation matrix by $\bm{\tilde{\Sigma}}$ and the vector $( \tilde{R}^*_{k1}, \tilde{R}^*_{k2})'$ by $\tilde{\bm{R}}^*$. Then,
\begin{eqnarray}
\nonumber && \lim_{y \rightarrow \infty} \textrm{Pr}(\tilde{R}^*_{k1} > y, \tilde{R}^*_{k2} > y)  \\
\nonumber &=& \lim_{y \rightarrow \infty} \int_{y}^{\infty} \int_{y}^{\infty}
\frac{\Gamma(\frac{a_k + 2}{2})}{\Gamma(\frac{a_k}{2})} \frac{1}{\pi a_k b_k (1 + \lambda_k^2) \lvert \bm{\tilde{\Sigma}} \rvert^{1/2}} \left(1 + \frac{1}{a_k b_k} \frac{\tilde{\bm{R}}^{*'} \bm{\tilde{\Sigma}}^{-1}  \tilde{\bm{R}}^*}{1 + \lambda_k^2} \right)^{- \frac{a_k + 2}{2}} d\tilde{R}^*_{k1} d\tilde{R}^*_{k2} \\
\nonumber &=& \lim_{y \rightarrow \infty} \int_{y}^{\infty} \int_{y}^{\infty}
\frac{2}{\pi a_k b_k(1 + 2\lambda_k^2/ (1 + r_k)) \lvert \bm{\Sigma_{\tilde{R}}} \rvert^{1/2} }
 \frac{\Gamma(\frac{a_k + 2}{2})}{\Gamma(\frac{a_k}{2})} \\
 \nonumber &&~~~~~~~~~~~~~~~~~~~~~~~~~~~~ \times \left(1 + \frac{1}{a_k b_k} \frac{\tilde{\bm{R}}'\bm{\Sigma_{\tilde{R}}}^{-1}\tilde{\bm{R}}}{1 + 2\lambda_k^2/ (1 + r_k)} \right)^{- \frac{a_k + 2}{2}}d\tilde{R}_{k1} d\tilde{R}_{k2}.
\end{eqnarray}

follows from the facts that 
\begin{eqnarray}
\nonumber && (1 + \lambda_k^2) \bm{\tilde{\Sigma}} = (1 + 2\lambda_k^2/ (1 + r_k)) \bm{\Sigma_{\tilde{R}}}, \\
\nonumber && (1 + \lambda_k^2) \lvert \bm{\tilde{\Sigma}} \rvert^{1/2} = \sqrt{1 + 2\lambda_k^2/ (1 + r_k)} \lvert \bm{\Sigma} \rvert^{1/2} = (1 + 2\lambda_k^2/ (1 + r_k)) \lvert \bm{\Sigma_{\tilde{R}}} \rvert^{1/2}.
\end{eqnarray}

Thus, we can write
\begin{eqnarray}
\nonumber && \lim_{y \rightarrow \infty} \textrm{Pr}(\tilde{R}_{k1} > y, \tilde{R}_{k2} > y) \\
\nonumber &=& 2F_T\left( \lambda_k \sqrt{a_k+2} \sqrt{2 / (1 +r_k )}; a_k + 2 \right) \times \lim_{y \rightarrow \infty} \textrm{Pr}(\tilde{R}^*_{k1} > y, \tilde{R}^*_{k2} > y),
\end{eqnarray}
and
\begin{eqnarray}
\nonumber && \lim_{y \rightarrow \infty} \textrm{Pr}(\tilde{R}_{k2} > y) = 2F_T\left( \lambda_k \sqrt{a_k + 1}; a_k + 1 \right) \times \lim_{y \rightarrow \infty} \textrm{Pr}(\tilde{R}^*_{k2} > y^*).
\end{eqnarray}
Thus, the extremal dependence between $\tilde{R}_{k1}$ and $\tilde{R}_{k2}$ is
\begin{eqnarray}
\nonumber \chi_{\tilde{R}_k}\left(\bm{s}_1, \bm{s}_2\right) &=& \frac{F_T\left( \lambda_k \sqrt{a_k+2} \sqrt{2 / (1 +r_k )}; a_k + 2 \right)}{F_T\left( \lambda_k \sqrt{a_k + 1}; a_k + 1 \right)} \times  \lim_{y \rightarrow \infty} \frac{\textrm{Pr}(\tilde{R}^*_{k1} > y, \tilde{R}^*_{k2} > y)}{\textrm{Pr}(\tilde{R}^*_{k2} > y)} \\
\nonumber &=& \frac{F_T\left( \lambda_k \sqrt{a_k+2} \sqrt{2 / (1 +r_k )}; a_k + 2 \right)}{F_T\left( \lambda_k \sqrt{a_k + 1}; a_k + 1 \right)} \times \left[ 2 - 2 F_T \left( \sqrt{\frac{(a_k + 1)(1 - r_k)}{1 + r_k + 2 \lambda_k^2}}; a_k + 1 \right) \right].
\end{eqnarray}

Instead of a zero-mean Gaussian process assumption on $R_k(\cdot)$ conditioned on random scale, considering a spatially-varying mean surface $\mu_k(\cdot)$, suppose we have $R^{\mu}_k(\cdot) = R_k(\cdot) + \mu_k(\cdot)$ and $\tilde{R}^{\mu}_k(\cdot) = \tilde{R}_k(\cdot) + \mu_k(\cdot)$. Considering two spatial locations $\bm{s}_1$ and $\bm{s}_2$, the marginal distributions are not same for $\tilde{R}^{\mu}_k(\bm{s}_1)$ and $\tilde{R}^{\mu}_k(\bm{s}_2)$ and suppose the CDFs are $F_{\tilde{R}^{\mu}_k(\bm{s}_1)}$ and $F_{\tilde{R}^{\mu}_k(\bm{s}_2)}$ respectively. Because of the location-shift, for any $u \in (0, 1)$, $F_{\tilde{R}^{\mu}_k(\bm{s}_1)}^{-1}(u) = \mu_k(\bm{s}_1) + F_{\tilde{R}_k}^{-1}(u)$ and $F_{\tilde{R}^{\mu}_k(\bm{s}_2)}^{-1}(u) = \mu_k(\bm{s}_2) + F_{\tilde{R}_k}^{-1}(u)$ where $F_{\tilde{R}_k}$ is the marginal CDF at any spatial location for $\tilde{R}_k(\cdot)$. Thus, the sets $\lbrace \tilde{R}^{\mu}_k(\bm{s}_i) > F_{\tilde{R}^{\mu}_k(\bm{s}_i)}^{-1}(u) \rbrace$ and $\lbrace \tilde{R}_k(\bm{s}_i) > F_{\tilde{R}_k}^{-1}(u) \rbrace$ are equal for each $i=1,2$. Thus, the extremal dependence between $\tilde{R}^{\mu}_k(\bm{s}_1)$ and $\tilde{R}^{\mu}_k(\bm{s}_2)$ is
	\begin{eqnarray}
	\nonumber \chi_{\tilde{R}^{\mu}_k}\left(\bm{s}_1, \bm{s}_2 \right) &=& \lim_{u \rightarrow 1} \textrm{Pr}\left( \tilde{R}^{\mu}_k(\bm{s}_1) > F_{\tilde{R}^{\mu}_k(\bm{s}_1)}^{-1}(u) | \tilde{R}^{\mu}_k(\bm{s}_2) > F_{\tilde{R}^{\mu}_k(\bm{s}_2)}^{-1}(u) \right) \\
    \nonumber &=&  \lim_{u \rightarrow 1} \textrm{Pr}\left( \tilde{R}_k(\bm{s}_1) > F_{\tilde{R}_k}^{-1}(u) | \tilde{R}_k(\bm{s}_2) > F_{\tilde{R}_k}^{-1}(u) \right) \\
    \nonumber &=&  \chi_{\tilde{R}_k}\left(\bm{s}_1, \bm{s}_2\right).
	\end{eqnarray}


Finally, we consider a spatial process $Y^*(\cdot)$ that follows a mixture of skew-$t$ processes $\tilde{R}^{\mu}_k(\cdot); k=1, \ldots, K$ where $K$ can be infinite. As described in the main article, suppose the latent variable denoting the cluster index is $g$ with the mixture probabilities are $\textrm{Pr}(g = k) = \pi_k$. Considering two spatial locations $\bm{s}_1$ and $\bm{s}_2$, the marginal distributions (conditioned on the mixture probabilities and cluster-specific parameters) are not same for $Y^*(\bm{s}_1)$ and $Y^*(\bm{s}_2)$ and suppose the CDFs are $F_{Y^*(\bm{s}_1)}$ and $F_{Y^*(\bm{s}_2)}$ respectively. Here $F_{Y^*(\bm{s}_i)} = \sum_{k=1}^{K} \pi_k F_{\tilde{R}^{\mu}_k(\bm{s}_i)}$ for each $i=1,2$. 

The $\chi$-measure between $Y^*(\bm{s}_1)$ and $Y^*(\bm{s}_2)$ is 
\begin{eqnarray}
\nonumber \chi_{*}\left(\bm{s}_1, \bm{s}_2\right) &=& \lim_{u \rightarrow 1} \textrm{Pr}(Y^*(\bm{s}_1) > F^{-1}_{Y^*(\bm{s}_1)}(u) | Y^*(\bm{s}_2) > F^{-1}_{Y^*(\bm{s}_2)}(u)) \\
\nonumber &=& \lim_{u \rightarrow 1} \sum_{k=1}^{K} \left[ \textrm{Pr}\left( Y^*(\bm{s}_1) > F^{-1}_{Y^*(\bm{s}_1)}(u) | \lbrace Y^*(\bm{s}_2) > F^{-1}_{Y^*(\bm{s}_2)}(u), g = k \rbrace \right) \right. \\
\nonumber &&  \left.~~~~~~~~~~~ \times \textrm{Pr}\left(g = k | Y^*(\bm{s}_2) > F^{-1}_{Y^*(\bm{s}_2)}(u) \right) \right] \\
\nonumber &\overset{\textrm{MCT}}{=}& \sum_{k=1}^{K} \left\lbrace \lim_{u \rightarrow 1} \left[  \textrm{Pr}\left( Y^*(\bm{s}_1) > F^{-1}_{Y^*(\bm{s}_1)}(u) | \lbrace Y^*(\bm{s}_2) > F^{-1}_{Y^*(\bm{s}_2)}(u), g = k \rbrace \right) \right] \right. \\
\nonumber &&  ~~~~~~~~~~~ \times \left. \lim_{u \rightarrow 1} \left[  \textrm{Pr}\left(g = k | Y^*(\bm{s}_2) > F^{-1}_{Y^*(\bm{s}_2)}(u) \right) \right] \right\rbrace.
\end{eqnarray}

The second term
\begin{eqnarray}
\nonumber && \lim_{u \rightarrow 1} \textrm{Pr}(g = k| Y^*(\bm{s}_2) > F^{-1}_{Y^*(\bm{s}_2)}(u)) \\
\nonumber &=& \lim_{y^* \rightarrow \infty} \textrm{Pr}(g = k| Y^*(\bm{s}_2) > y^*) \\
\nonumber &=&  \lim_{y^* \rightarrow \infty} \frac{\pi_k \textrm{Pr}(\tilde{R}^{\mu}_k(\bm{s}_2) > y^*) }{\sum_{l=1}^{K} \pi_l \textrm{Pr}(\tilde{R}^{\mu}_l(\bm{s}_2) > y^*)} \\
\nonumber &=& \lim_{y^* \rightarrow \infty} \frac{\pi_k \textrm{Pr}\left( \frac{y^*}{y^* - \bm{\mu}_k(\bm{s}_2)} \tilde{R}_k(\bm{s}_2) > y^* \right) }{ \sum_{l=1}^{K} \pi_l \textrm{Pr}\left( \frac{y^*}{y^* - \bm{\mu}_l(\bm{s}_2)} \tilde{R}_l(\bm{s}_2) > y^* \right)} \\
\nonumber &=& \lim_{y^* \rightarrow \infty} \frac{\pi_k \textrm{Pr} \left(\tilde{R}_k(\bm{s}_2) > y^* \right) }{ \sum_{l=1}^{K} \pi_l \textrm{Pr} \left(\tilde{R}_l(\bm{s}_2) > y^* \right)} ~~\textrm{[follows from Slutsky's theorem]} \\
\nonumber &=& \lim_{y^* \rightarrow \infty} \frac{\pi_{k} f_{\tilde{R}_k(\bm{s}_2)}(y^*)}{\sum_{l=1}^{K} \pi_{l} f_{\tilde{R}_l(\bm{s}_2)}(y^*)} ~~\textrm{[follows~from~L'Hospital's rule]} \\
\nonumber &=& \lim_{y^* \rightarrow \infty} \frac{\pi_{k} \frac{2}{\sqrt{b_k (1 + \lambda_k^2)}} f_T\left(\frac{y^*}{\sqrt{b_k (1 + \lambda_k^2)}}; a_k \right) F_T\left( \lambda_k \frac{y^*}{\sqrt{b_k (1 + \lambda_k^2)}} \sqrt{\frac{a_k + 1}{a_k + \frac{y^{*2}}{b_k (1 + \lambda_k^2)}}}; a_k + 1 \right)}{\sum_{l=1}^{K} \pi_{l}  \frac{2}{\sqrt{b_l (1 + \lambda_l^2)}} f_T\left(\frac{y^*}{\sqrt{b_l (1 + \lambda_l^2)}}; a_l \right) F_T\left( \lambda_l \frac{y^*}{\sqrt{b_l (1 + \lambda_l^2)}} \sqrt{\frac{a_l + 1}{a_l + \frac{y^{*2}}{b_l (1 + \lambda_l^2)}}}; a_l + 1 \right)}  \\
\nonumber &=& \begin{cases}
1 & ~~\textrm{if $k = \arg \min \left \lbrace a_l \right \rbrace$}\\
0 & ~~\textrm{otherwise.}\\
\end{cases}
\end{eqnarray}

Considering $u > 1 - \pi_k$, where $k = \arg \min \left \lbrace a_l \right \rbrace$, suppose the $u$-th quantile of $F_{Y^*(\bm{s})}$ is $y^*_u$.
\begin{eqnarray}
\nonumber \textrm{Pr}(Y^*(\bm{s}) > y^*_u) &=& \pi_k \textrm{Pr}(\tilde{R}^{\mu}_k(\bm{s}) > y^*_u) \times \left( 1 + \frac{\sum_{l\neq k}^{} \pi_l \textrm{Pr}(\tilde{R}^{\mu}_l(\bm{s}) > y^*_u)}{\pi_k \textrm{Pr}(\tilde{R}^{\mu}_k(\bm{s}) > y^*_u)} \right) \\
\nonumber &\sim& \pi_k \textrm{Pr}(\tilde{R}^{\mu}_k(\bm{s}) > y^*_u) ~~~~\textrm{[follows from the derivation of the second term]}.
\end{eqnarray}
Thus, the $u$-th quantile of $F_{Y^*(\bm{s})}$ and $\left(1 - \frac{1 - u}{\pi_k} \right)$-th quantile of $\tilde{R}^{\mu}_k(\bm{s})$ are equal in the limiting sense. Hence, $F^{-1}_{Y^*(\bm{s}_1)}(u) \sim F^{-1}_{\tilde{R}^{\mu}_k(\bm{s}_1)}\left(1 - \frac{1 - u}{\pi_k} \right)$ and $F^{-1}_{Y^*(\bm{s}_2)}(u) \sim F^{-1}_{\tilde{R}^{\mu}_k(\bm{s}_2)}\left(1 - \frac{1 - u}{\pi_k} \right)$.


The first term 
\begin{eqnarray}
\nonumber && \lim_{u \rightarrow 1}  \textrm{Pr}\left( Y^*(\bm{s}_1) > F^{-1}_{Y^*(\bm{s}_1)}(u) | \lbrace Y^*(\bm{s}_2) > F^{-1}_{Y^*(\bm{s}_2)}(u), g = k \rbrace \right) \\
\nonumber &=& \lim_{u \rightarrow 1} \textrm{Pr}\left( \tilde{R}^{\mu}_k(\bm{s}_1) > F^{-1}_{Y^*(\bm{s}_1)}(u) | \tilde{R}^{\mu}_k(\bm{s}_2) > F^{-1}_{Y^*(\bm{s}_2)}(u) \right) \\
\nonumber &=& \lim_{u \rightarrow 1}  \textrm{Pr}\left( \tilde{R}^{\mu}_k(\bm{s}_1) > F^{-1}_{\tilde{R}^{\mu}_k(\bm{s}_1)}\left(1 - \frac{1 - u}{\pi_k} \right) \bigg\vert \tilde{R}^{\mu}_k(\bm{s}_2) > F^{-1}_{\tilde{R}^{\mu}_k(\bm{s}_2)}\left(1 - \frac{1 - u}{\pi_k} \right) \right)  \\
\nonumber &=& \lim_{y^* \rightarrow \infty}  \textrm{Pr}\left( \tilde{R}_k(\bm{s}_1) > y^* | \tilde{R}_k(\bm{s}_2) > y^* \right) \\
\nonumber &=& \chi_{\tilde{R}_k}\left(\bm{s}_1, \bm{s}_2\right).
\end{eqnarray}


The observed process $Y(\cdot)$ is obtained from $Y^*(\cdot)$ using inverse GEV-log transformation, i.e., $Y(\cdot) = \mu_y + \frac{\sigma_y}{\xi_y}\left[ \exp(\xi_y Y^*(\cdot)) - 1 \right]$ if $\xi_y \neq 0$ and $Y(\cdot) = \mu_y + \sigma_y Y^*(\cdot)$ if $\xi_y = 0$. It is easy to verify that the GEV-log transformation is a strictly increasing transformation (and hence, the inverse transformation as well). As the $\chi$-measure is invariant of any strictly increasing transformation, the $\chi$-measure between $Y(\bm{s}_1)$ and $Y(\bm{s}_2)$, $\chi(\bm{s}_1, \bm{s}_2)$ equals $\chi_{*}\left(\bm{s}_1, \bm{s}_2\right)$.
Thus, for $k = \arg \min \left \lbrace a_l \right \rbrace$,
$$\chi(\bm{s}_1, \bm{s}_2) = \frac{F_T\left( \lambda_k \sqrt{a_k+2} \sqrt{2 / (1 +r_k )}; a_k + 2 \right)}{F_T\left( \lambda_k \sqrt{a_k + 1}; a_k + 1 \right)} \times \left[ 2 - 2 F_T \left( \sqrt{\frac{(a_k + 1)(1 - r_k)}{1 + r_k + 2 \lambda_k^2}}; a_k + 1 \right) \right].$$


\section*{Appendix G: Sub-asymptotic conditional exceedance probabilities}

While the extremal dependence between two random variables $Y_1$ and $Y_2$ (with CDFs $F_1$ and $F_2$, respectively) is often characterized by $\chi = \lim_{u \rightarrow 1} \chi_u$, where
\begin{eqnarray} 
\nonumber \chi_u = \textrm{Pr}\left[Y_1 > F_1^{-1}(u) | Y_2 > F_2^{-1}(u)\right],
\end{eqnarray}
in this section, we study the variation of $\chi_u$ as a function of $u \in (0, 1)$, i.e., the sub-asymptotic dependence, for different parameter choices of the bivariate skew-$t$ distribution defined in model (3.1) of the main paper. To match notations, suppose $Y_1 \equiv Y(\bm{s}_1)$ and $Y_2 \equiv Y(\bm{s}_2)$. We assume $\lambda(\bm{s}_1) = \lambda(\bm{s}_2) = \lambda$; under this assumption, $\chi_u$ is invariant to  $\mu(\bm{s}_1)$, $\mu(\bm{s}_2)$, and $b$. We vary the parameters $a$, $\lambda$ and $r$ (here $r$ corresponds to $r(\cdot, \cdot)$ of the main paper), and study how $\chi_u$ varies with $u$ under different settings.

For $a = \infty$ and $\lambda = 0$, the bivariate skew-$t$ model reduces to the bivariate normal model and as $u \rightarrow 1$, the probability goes to zero indicating that a GP is questionable for spatial prediction of extremes. Setting $a = \infty$ and $\lambda = 1$ gives the skew-normal model and again the limiting probability is zero indicating that a skew-normal process has no spatial extremal dependence. For the third and fourth panels with $a = 5$, the limiting probabilities are non-zero with stronger extremal dependence for the skew-$t$ case with $\lambda = 1$ compared to the symmetric-$t$ case with $\lambda = 0$.

\begin{figure}[ht]
\centering
\adjincludegraphics[height = 0.3\linewidth, width = 0.24\linewidth, trim = {{.0\width} {.0\width} 0 {.0\width}}, clip]{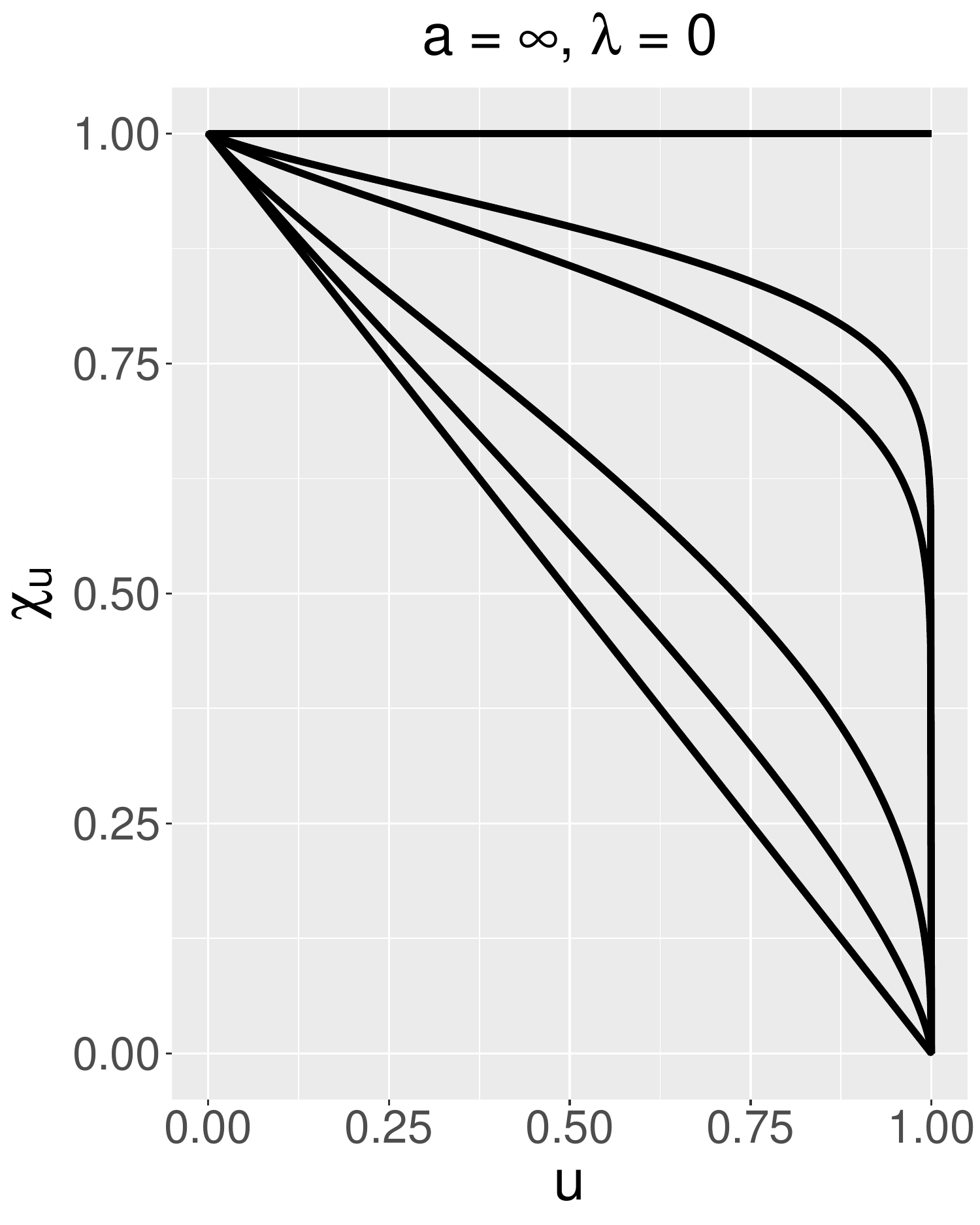}
\adjincludegraphics[height = 0.3\linewidth, width = 0.24\linewidth, trim = {{.0\width} {.0\width} 0 {.0\width}}, clip]{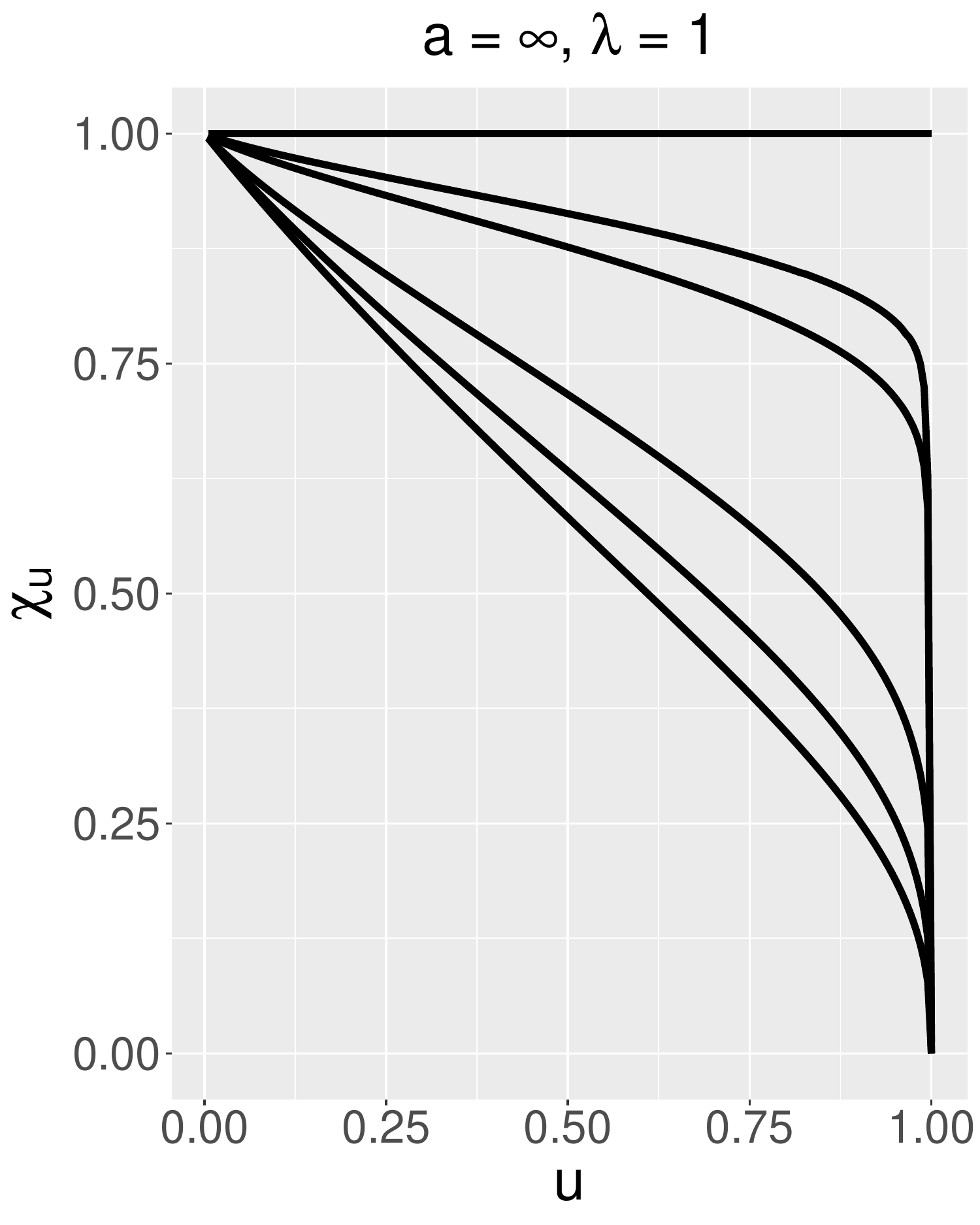}
\adjincludegraphics[height = 0.3\linewidth, width = 0.24\linewidth, trim = {{.0\width} {.0\width} 0 {.0\width}}, clip]{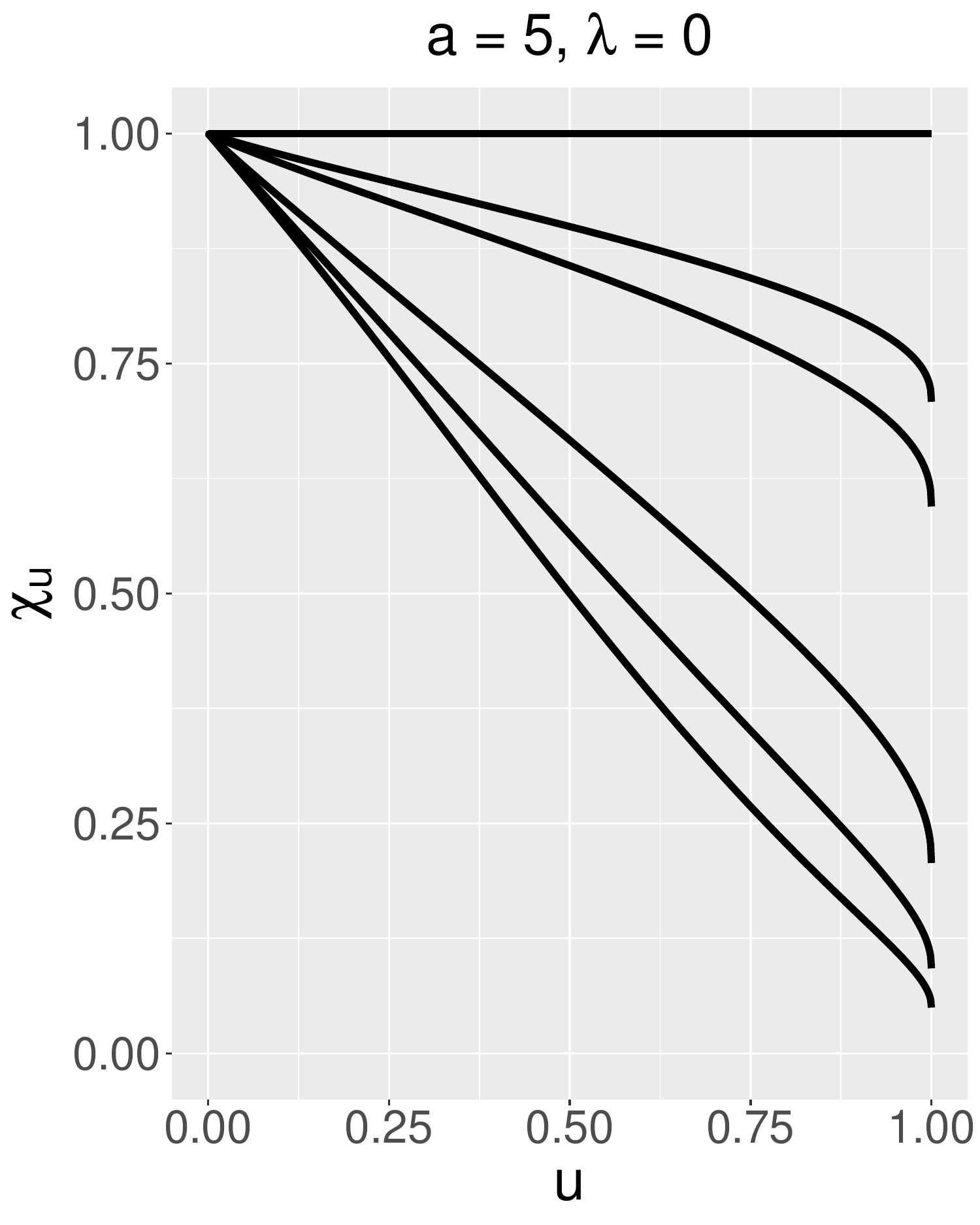}
\adjincludegraphics[height = 0.3\linewidth, width = 0.24\linewidth, trim = {{.0\width} {.0\width} 0 {.0\width}}, clip]{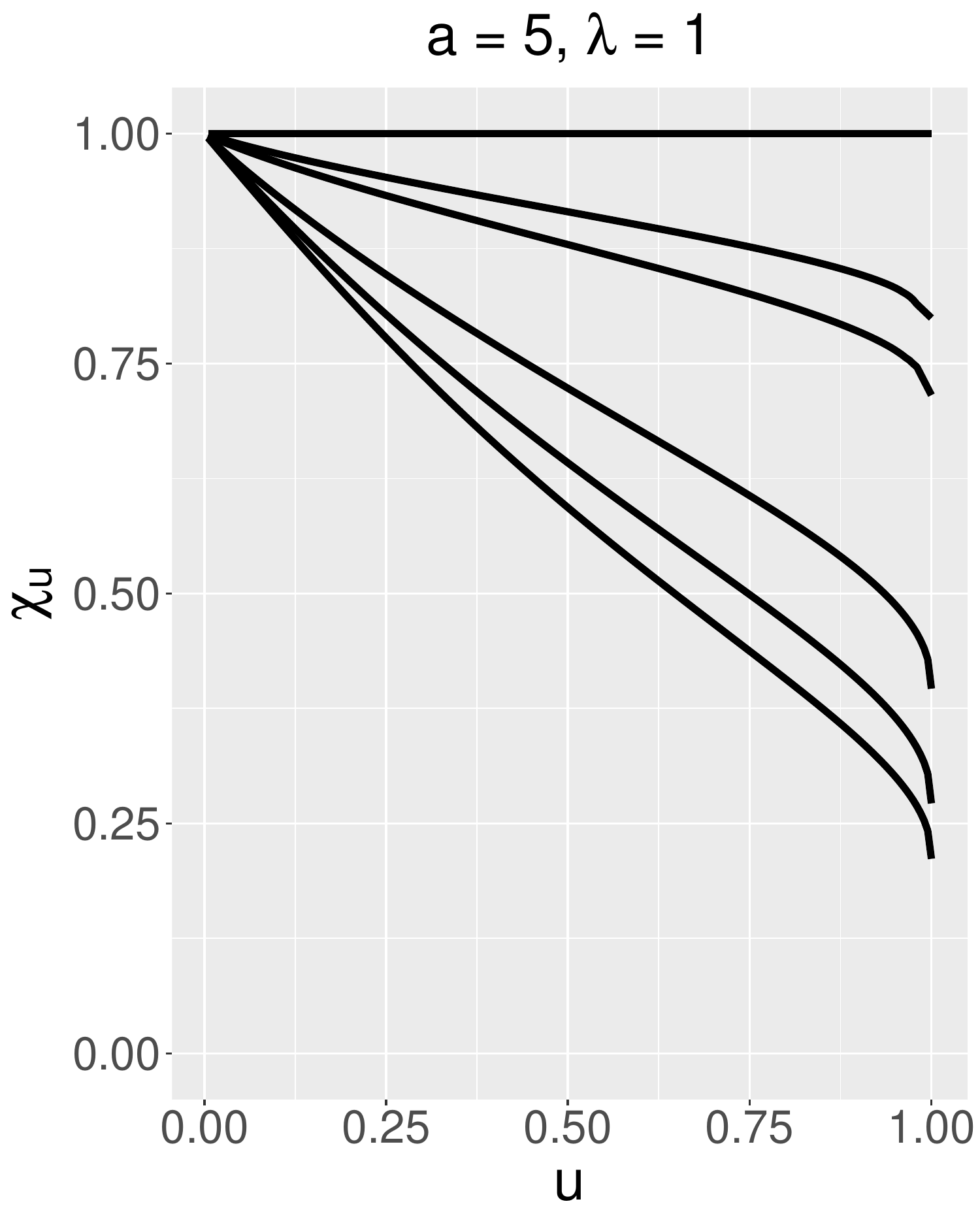}
\vspace{-2mm}
\caption{Conditional exceedance probabilities $\chi_u = \textrm{Pr}\left[Y_1 > F_1^{-1}(u) | Y_2 > F_2^{-1}(u)\right]$ for bivariate skew-$t$ distributions with different choices of the parameters $a$, $\lambda$, and $r$. The terms $F_{Y_1}$ and $F_{Y_2}$ are the marginal CDFs of $Y_1$ and $Y_2$, respectively. We assume same skewness parameter $\lambda$ for each component; $\chi_u$ is invariant of the location and scale parameters under this assumption. From bottom to top, the correlation term $r = 0, 0.2, 0.5, 0.9, 0.95$ and 1.}
\label{fig5}
\vspace{-3mm}
\end{figure}

\section*{Appendix H: Simulation Studies}
\label{simulation}


We perform a simulation study to assess the performance of our proposed model in spatial prediction of marginal quantiles and in estimation of extremal dependence. We compare the performances of GP, TP, STP, GP-DPM and TP-DPM with the proposed STP-DPM model. For simplicity and reduced computational burden, here we ignore the zone-specific divisions, and consider same skewness parameters for all the data locations (for STP and STP-DPM models), consider isotropic spatial correlation (Euclidean distance in (3.3) of the main paper, instead of Mahalanobis distance), and also ignore the temporally dependent models; we simulate data from temporally independent models and fit purely spatial models-- this is reasonable as our main goal is in spatial prediction. The STP is considered to be of the form discussed in Section 3.1 of the main paper, with $\lambda(\bm{s}) = \lambda$ for all $\bm{s} \in \mathcal{D}$, while the TP and the GP are sub-models of STP with $\lambda = 0$ for TP, and $\lambda = 0$ and $a = \infty$ both for GP. 


\subsection*{Simulation design}
We generate 100 datasets from each of the 6 designs: (1) GP, (2) TP, (3) STP, (4) Mixture of GPs, (5) Mixture of TPs and (6) Mixture of STPs. In each case, data are generated at $n = 60$ sites and $T = 100$ time points. The sites are generated uniformly on the unit square. For Design (6), we consider a three-component mixture of STPs as in (3.2) of the main paper, with the parameters in Table \ref{stpchoices}. For simulation from Design (5), we consider same parameters as in Table \ref{stpchoices} except that we set the $\lambda_k$'s to zero. Additionally we set $a_k$'s to infinity in case of Design (4), i.e., the components have fixed variance equal to $b_k$'s. For Design (3), we consider the parameters of the third component in Table \ref{stpchoices} except that we set the Mat\'ern parameters $\gamma = 0.8$, $\nu = 0.5$ and $\rho = 1$. Design (2) has same parameters as in Design (3) except $\lambda = 0$ and additionally $a = \infty$ for Design (1). For each case, we transform the simulated data $Y_t^*(\bm{s})$ to $Y_t(\bm{s})$ using inverse GEV-log transformation with $\mu_y = 10$, $\sigma_y = 2$ and $\xi_y = 0.2$. We choose the parameters arbitrarily except ensuring that the true marginal distributions are multimodal (based on visual inspection), and the component with the smallest $a_k$, the extremal cluster, has small $\pi_k = 0.25$. 


 \begin{table}  
\caption{\label{stpchoices} The parameter choices for the skew-$t$ process mixture components (Design 6). Here $\bm{s} = (s_1, s_2)$.}
\centering
 \begin{tabular}{rllllllll}
   \hline
 $k$ & $\pi_k$ & $\mu_k(\bm{s})$ & $\lambda_k$ & $a_k$ & $b_k$ & $\gamma_k$ & $\nu_k$ &  $\rho_k$ \\ 
   \hline
 1 & 0.25  & $0.5 -\sqrt{s_1}$  & 1 & 2 & $0.5^2$ & 0.9 & 0.5 & 1 \\ 
   2 & 0.25  & $-(0.5 + \sqrt{s_2})$ & -0.5 & 4 & $0.4^2$ & 0.5 & 0.1 & 0.1 \\ 
   3 & 0.5 & $1 + 2 \sqrt{s_1 s_2}$ & 1 & 6 & 1 & 0.1 & 2 & 0.5 \\ 
    \hline
 \end{tabular}
 \end{table}


We use the priors discussed in Section 4 of the main paper. We run each MCMC chain for 20,000 iterations, discard first 10,000 iterations as burn-in and out of the post-burn-in samples, we perform thinning by keeping one in each five samples. Here we are not interested in estimating specific parameters which is complicated due to label switching throughout the MCMC. Rather, we are interested in marginal quantiles of the posterior predictive distribution and the extremal dependence. We monitor the convergence for a set of quantiles and the MCMC converges well for these quantities.

\begin{figure}[h]
\includegraphics[width=0.32\linewidth]{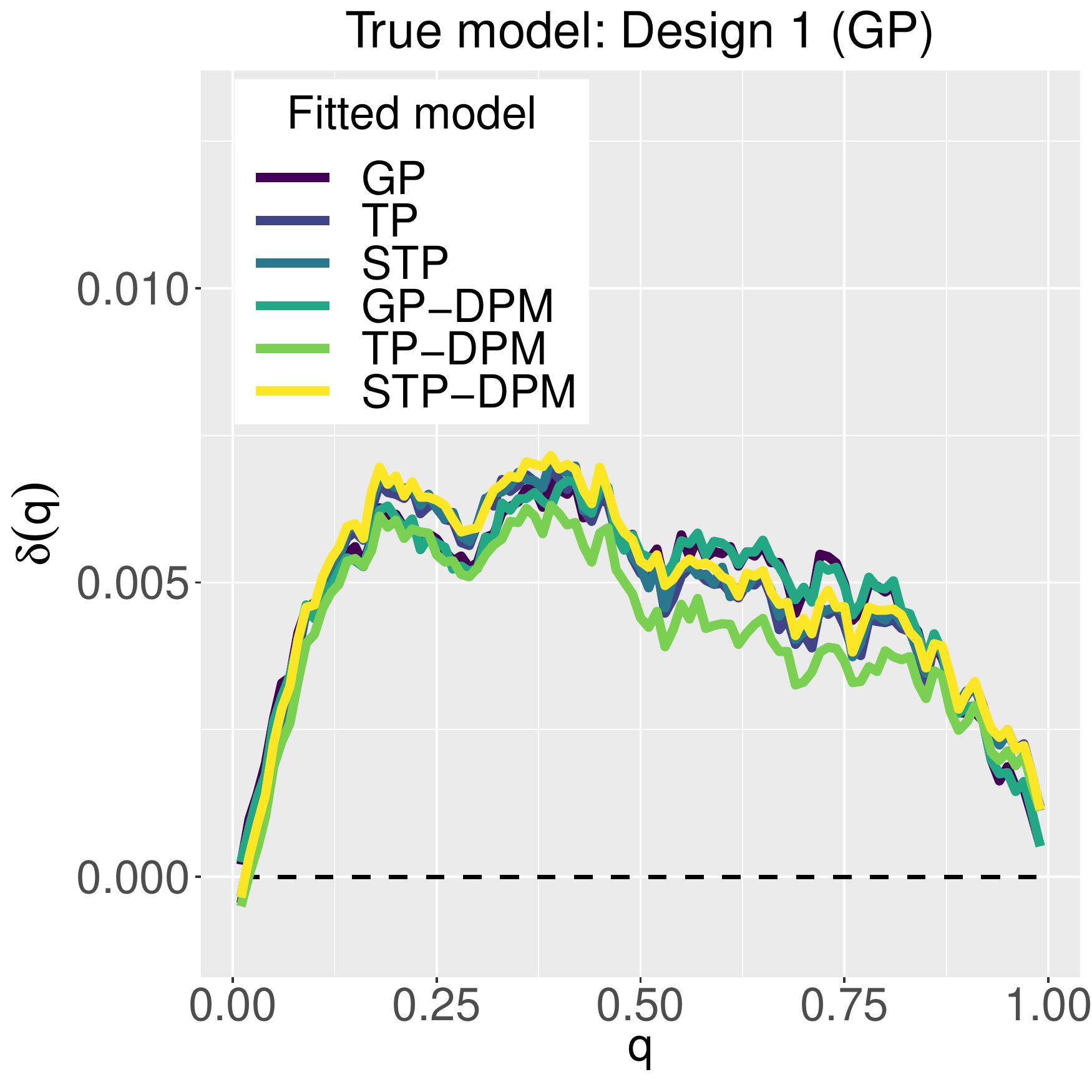} 
\includegraphics[width=0.32\linewidth]{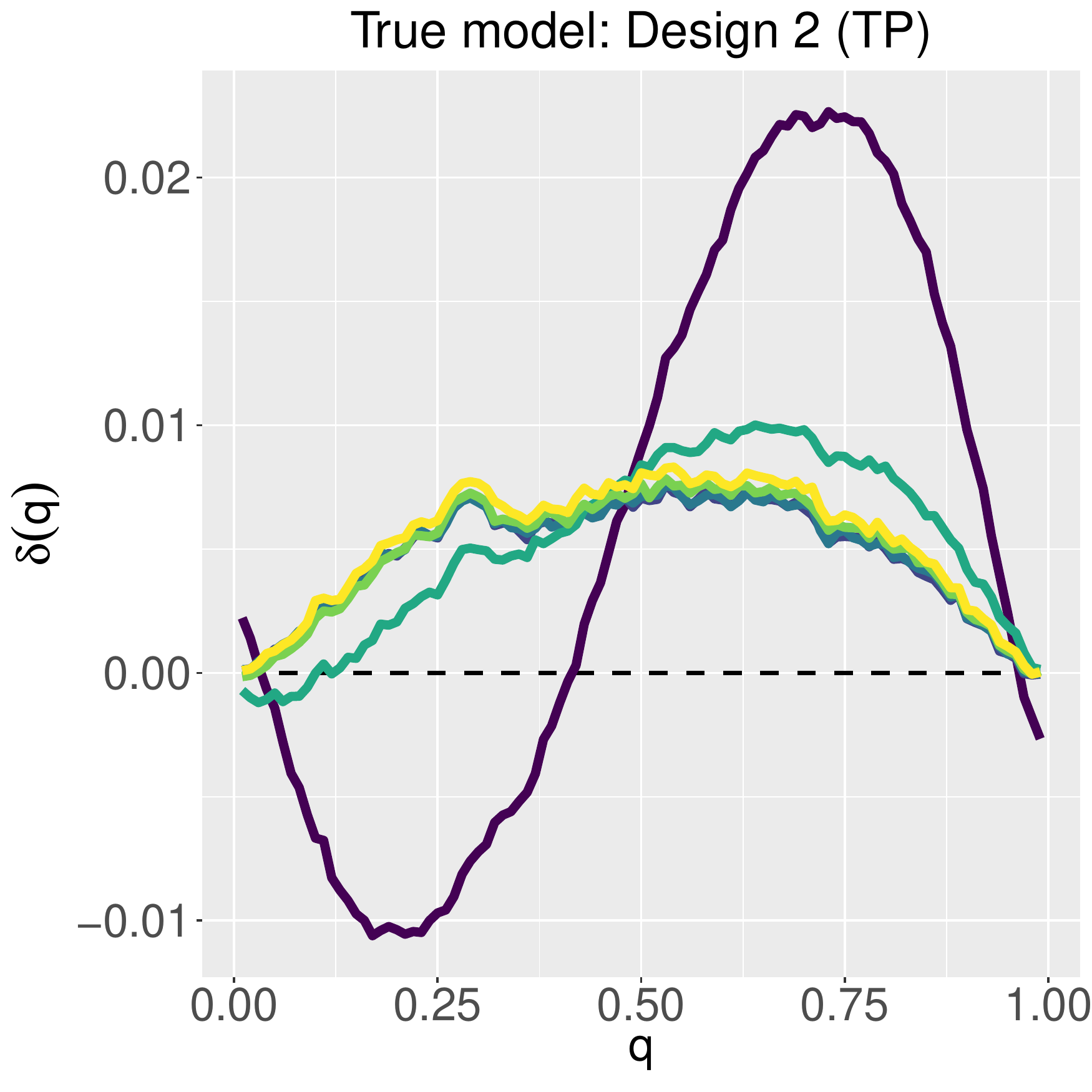}
\includegraphics[width=0.32\linewidth]{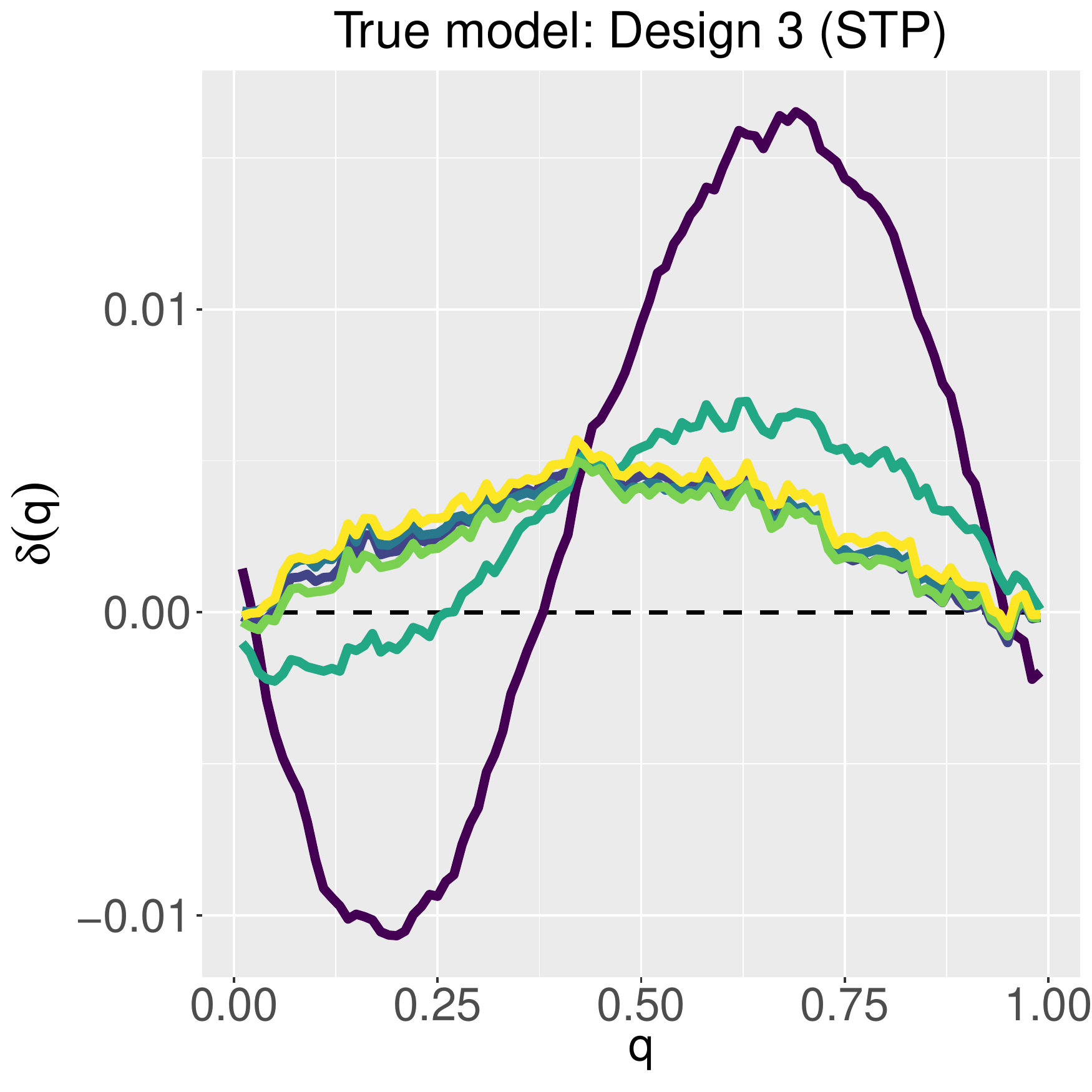} \\
\includegraphics[width=0.32\linewidth]{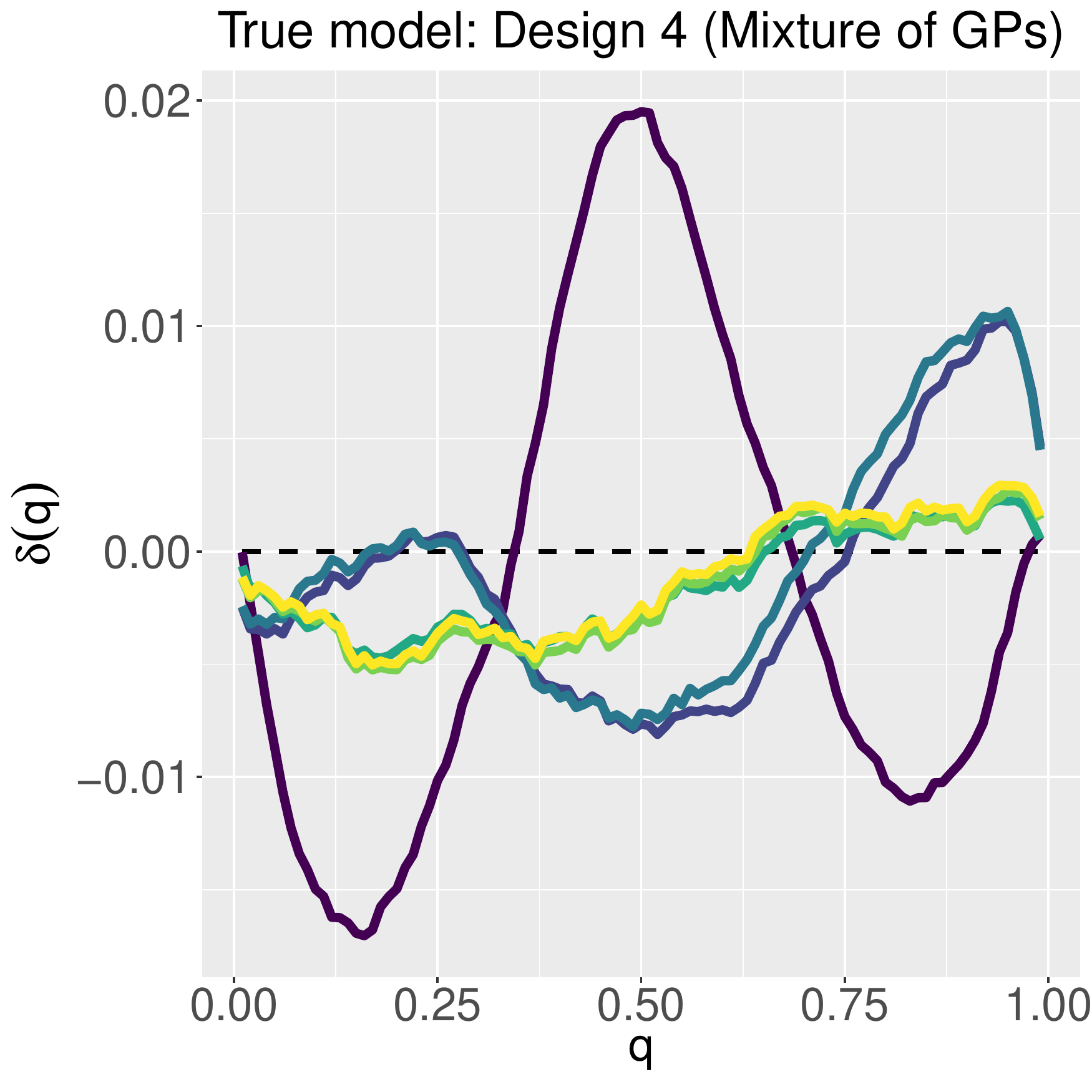}
\includegraphics[width=0.32\linewidth]{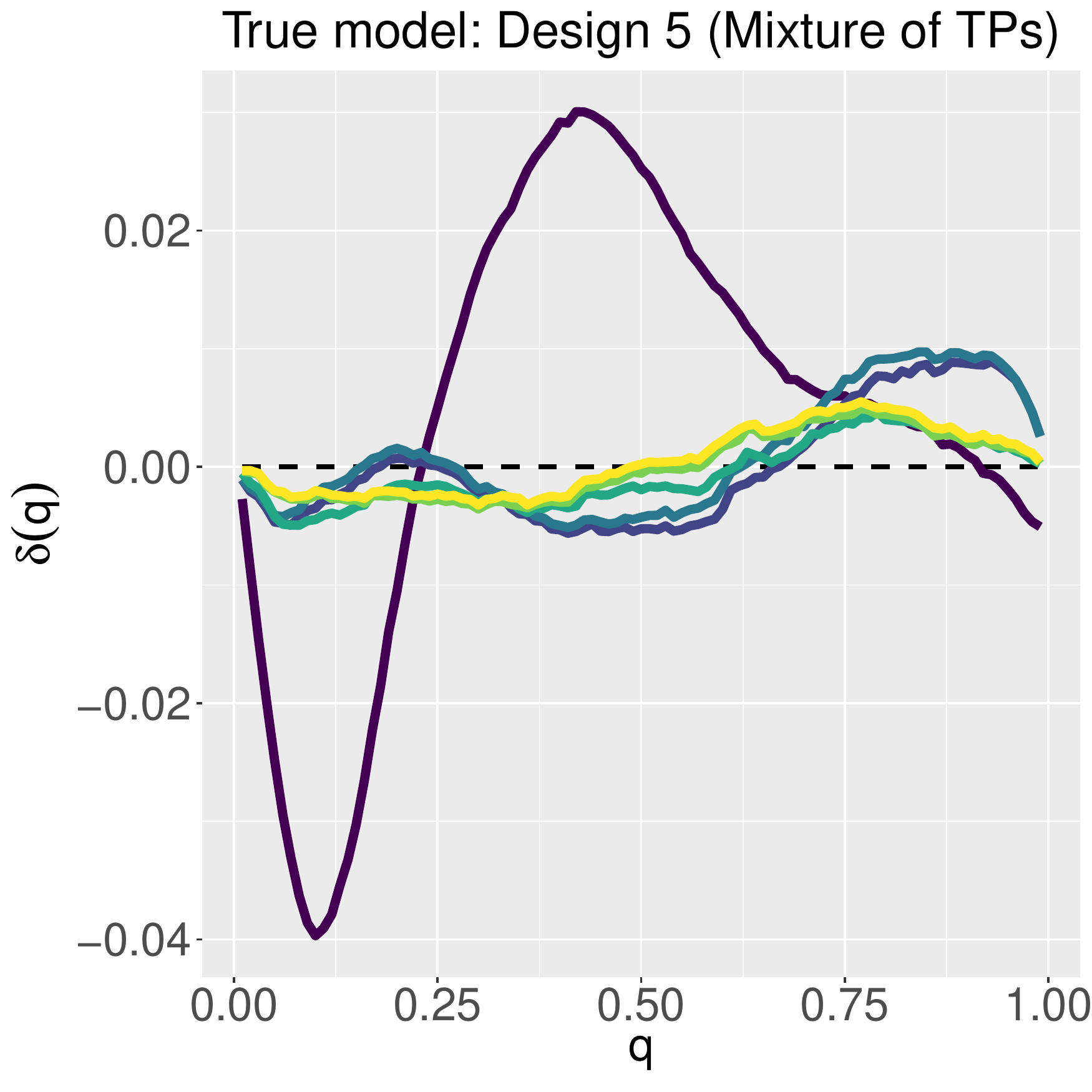}
\includegraphics[width=0.32\linewidth]{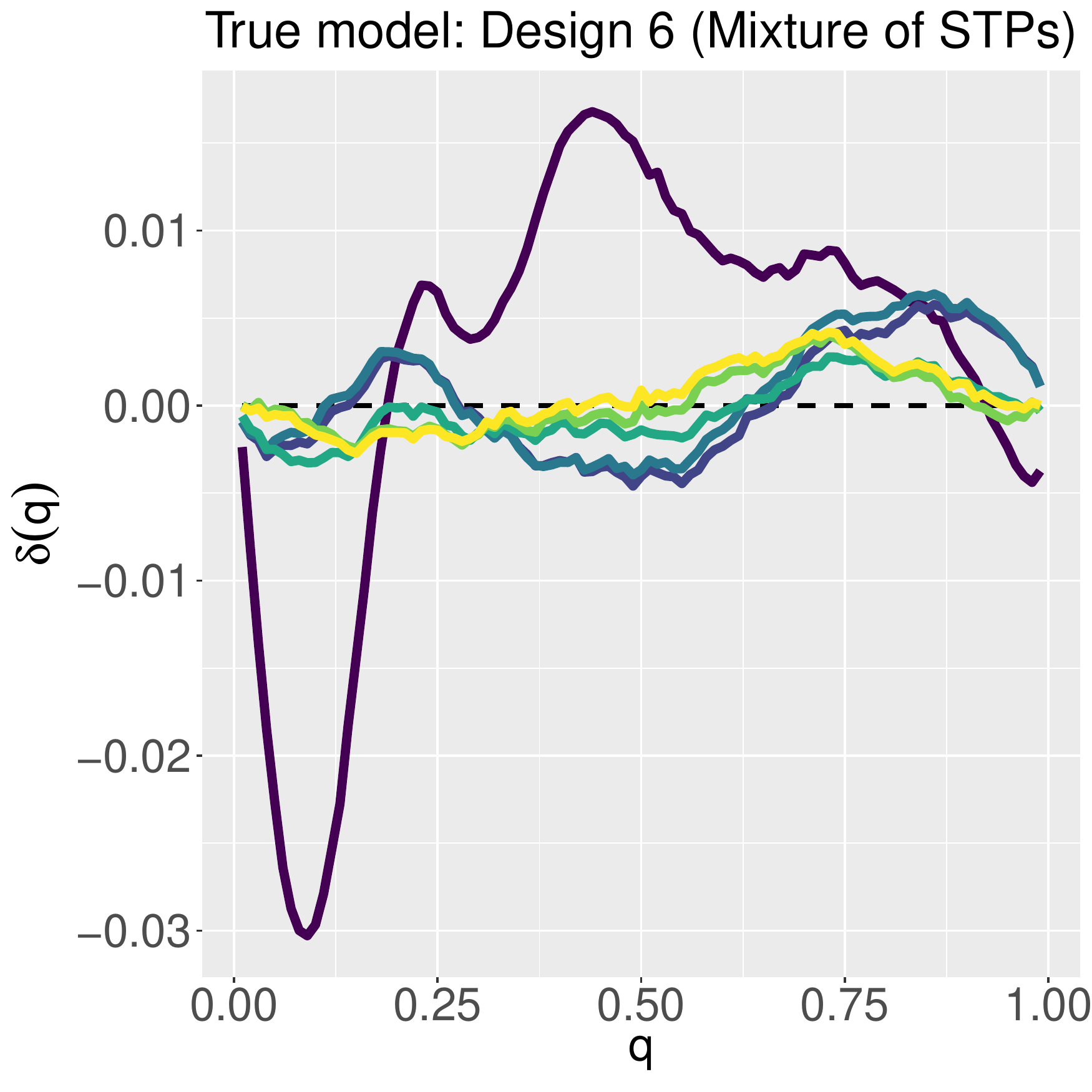}
\vspace{-2mm}
\caption{Comparison of models GP, TP, STP, GP-DPM, TP-DPM and STP-DPM based on the performance in spatial prediction of true quantiles when the data are generated from Designs 1-6. A model with $\delta(q)$ closer to zero is preferred for that $q \in (0,1)$. A positive (negative) value of $\delta(q)$ indicates overestimation (underestimation) of the true quantile.}
\label{fig8}
\end{figure}

We fit the models to 50 sites and predict the true quantiles for 10 additional test sites. For a test site $\bm{s}_P$, let the true marginal distribution function (CDF) and the posterior predictive CDF be denoted by $F_{P}$ and $\tilde{F}_{P}$, respectively. For $q \in (0,1)$, models are judged based on the difference $\delta(q) = F_{P}[\tilde{F}_{P}^{-1}(q)] - q$; a model with $\delta(q)$ close to zero is preferred for that $q$. A positive (negative) value of $\delta(q)$ indicates overestimation (underestimation) of the true quantile. For each of the 100 datasets generated from six designs, we fit all the models, and plot $q$ versus $\delta(q)$ for $q=0.01, \ldots, 0.99$, averaged across the test sites and the datasets, in Figure \ref{fig8}. While our main interest lies in the inference about the tails, the proposed method can model the full support flexibly and hence we consider low-through-high values of $q$. 
While $\delta(q)$ indicates the bias in prediction, the root mean squared errors (RMSE) of prediction for the 0.95$^{th}$ quantile are provided in Table \ref{table2}.  

For evaluating the performance in estimating the true spatial extremal dependence, we compare the true $\chi$-measure versus the estimates based on the six models, GP through STP-DPM. For models GP and GP-DPM, spatial extremal dependence is always zero. The plots of the estimated $\chi(h)$ as a function of the spatial Euclidean distance $h$, based on models TP, STP, TP-DPM and STP-DPM, are provided in Figure \ref{fig_estimated_chi}. 

%

\begin{table}
 \caption{\label{table2} Comparison of models GP, TP, STP, GP-DPM, TP-DPM and STP-DPM based on the prediction RMSE of the 0.95$^{th}$ quantile when the data are generated from the Simulation Designs 1-6. A smaller value indicates better performance. The standard errors are provided in the parentheses.}
\centering
\begin{tabular}{ccccccc}
  \hline
 Design & GP & TP & STP & GP-DPM & TP-DPM & STP-DPM \\ 
  \hline
1 & 0.45 (0.02) & 0.52 (0.02) & 0.54 (0.03) & \textbf{0.43} (0.02) & 0.50 (0.02) & 0.56 (0.03) \\ 
  2 & 0.83 (0.05) & \textbf{0.58} (0.03) & 0.67 (0.04) & 0.90 (0.05) & 0.60 (0.03) & 0.74 (0.04) \\ 
  3 & 1.75 (0.18) & 1.76 (0.09) & 1.56 (0.11) & 1.61 (0.10) & 1.76 (0.09) & \textbf{1.47} (0.09) \\ 
  4 & 1.96 (0.03) & 2.94 (0.05) & 4.29 (0.12) & \textbf{0.44} (0.02) & 0.49 (0.02) & 0.49 (0.02) \\ 
  5 & 1.75 (0.04) & 3.64 (0.11) & 11.66 (0.47) & 0.66 (0.03) & \textbf{0.63} (0.02) & \textbf{0.63} (0.03) \\ 
  6 & 2.61 (0.11) & 4.93 (0.24) & 20.52 (0.86) & 1.40 (0.08) & 1.53 (0.08) & \textbf{1.20} (0.08) \\
   \hline
\end{tabular}

\end{table}


\subsection*{Results}
First, we compare the models based on Figure \ref{fig8} and Table \ref{table2}. When the data are generated from Design 1, all models perform well (with the highest $\vert \delta(q) \vert $ is approximately 0.0075). 
For Design 2, GP and GP-DPM perform worse than the other models, particularly for $q$ between 0.5 and 0.9. In case of Design 3, GP and GP-DPM again perform worse than the other models.
In all three cases, the STP-DPM model has only slightly higher RMSE than the models with smallest RMSE values. When the data are generated from the Designs 4-6, the DPM models perform equally well and better than the parametric models both in terms of prediction bias and prediction RMSE.
For Design 4, the prediction RMSE is slightly smaller for GP-DPM.
For Design 5, TP-DPM and STP-DPM perform better than GP-DPM in terms of prediction RMSE, and for Design 6, STP-DPM perform the best followed by GP-DPM. Similar ordering is observed for the prediction RMSE of the 0.98$^{th}$ and 0.99$^{th}$ quantiles as well. 



\begin{figure}[h]
\includegraphics[width=0.32\linewidth]{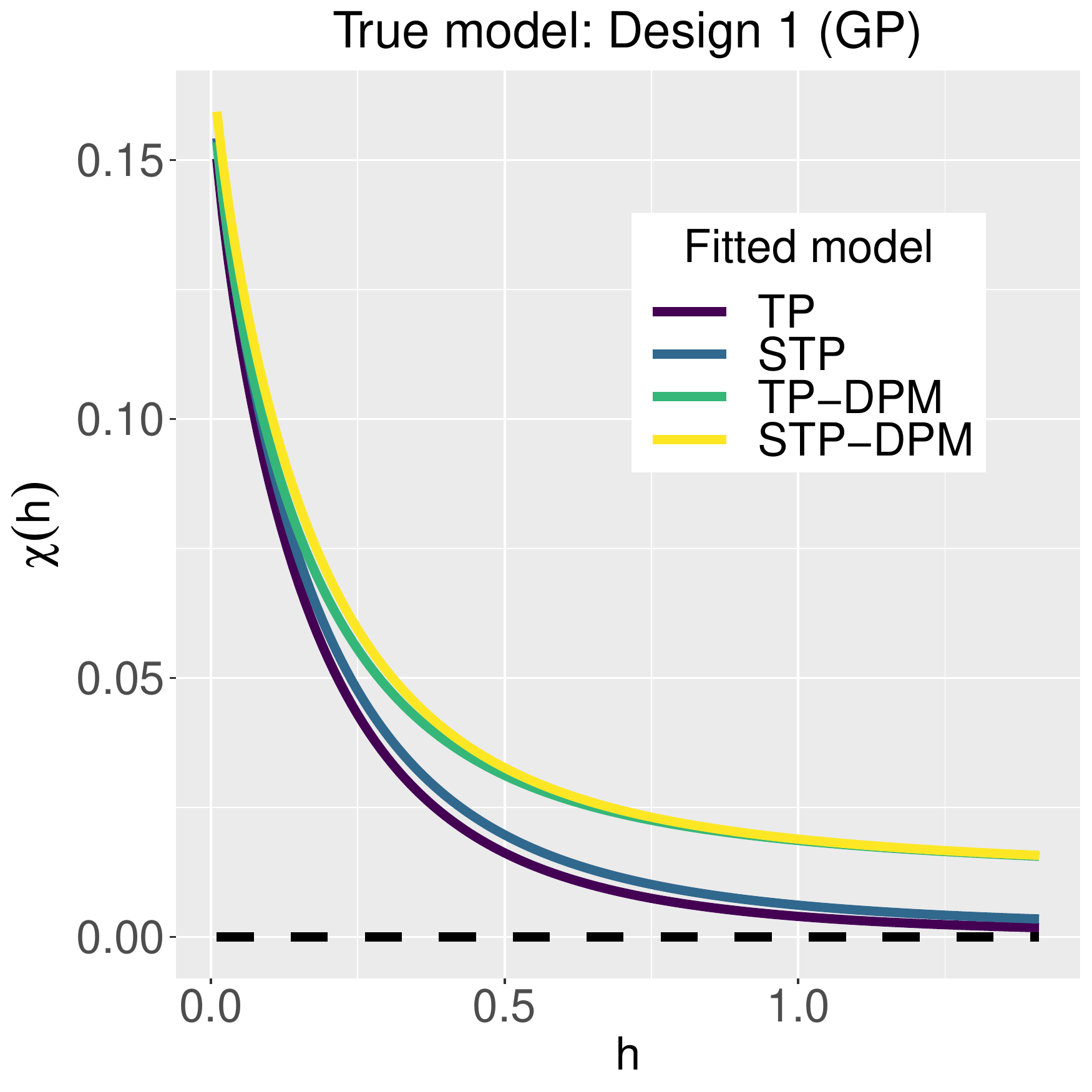} 
\includegraphics[width=0.32\linewidth]{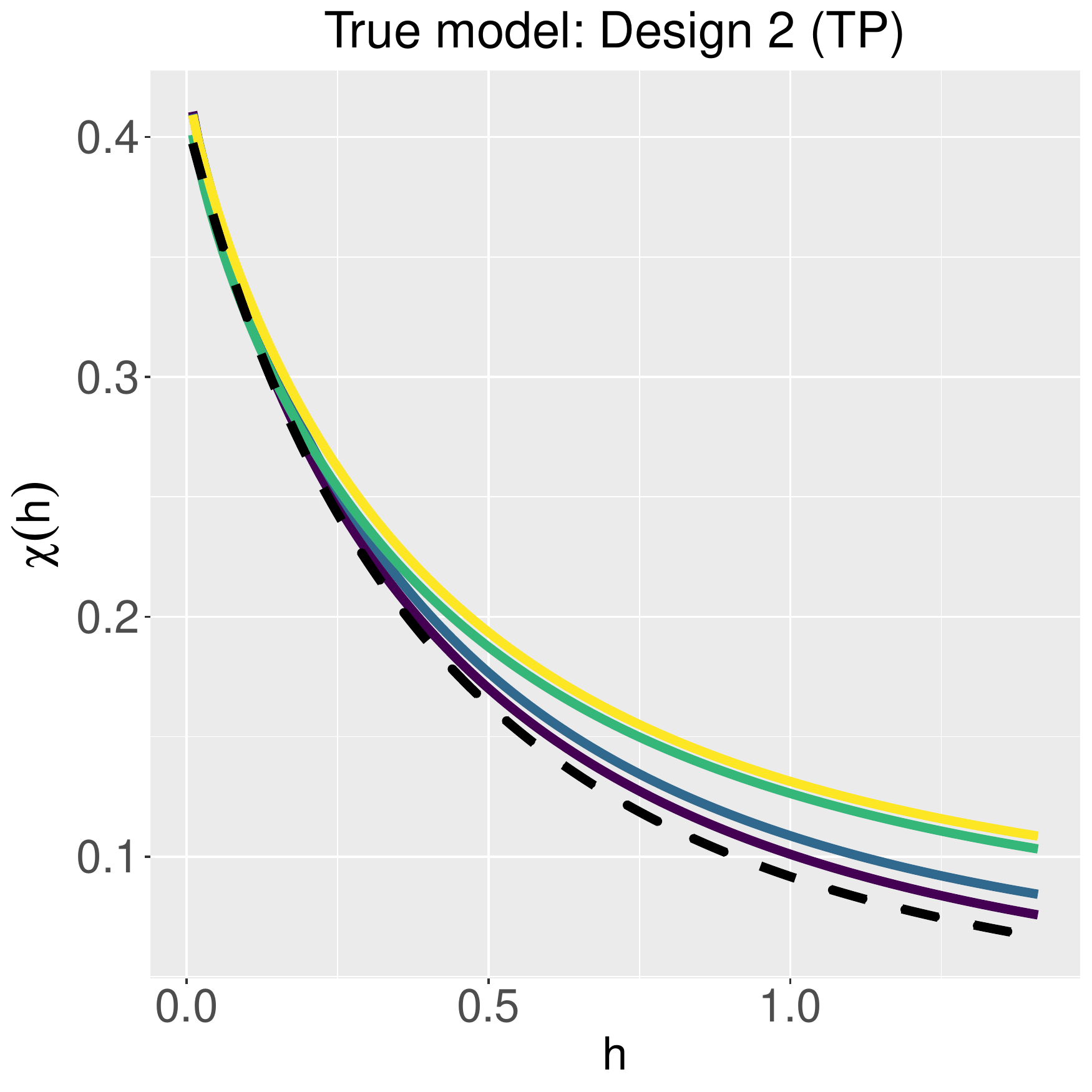}
\includegraphics[width=0.32\linewidth]{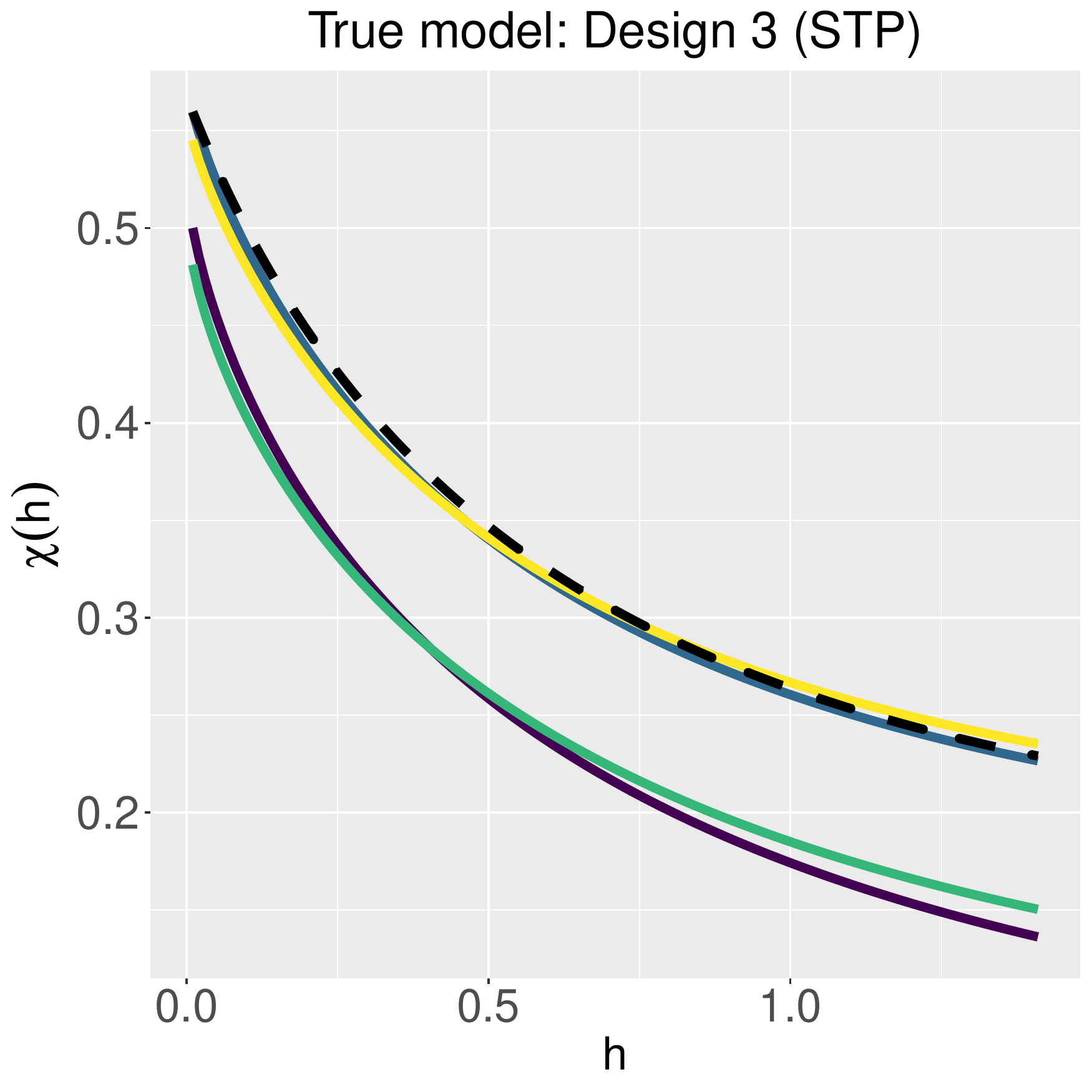} \\
\includegraphics[width=0.32\linewidth]{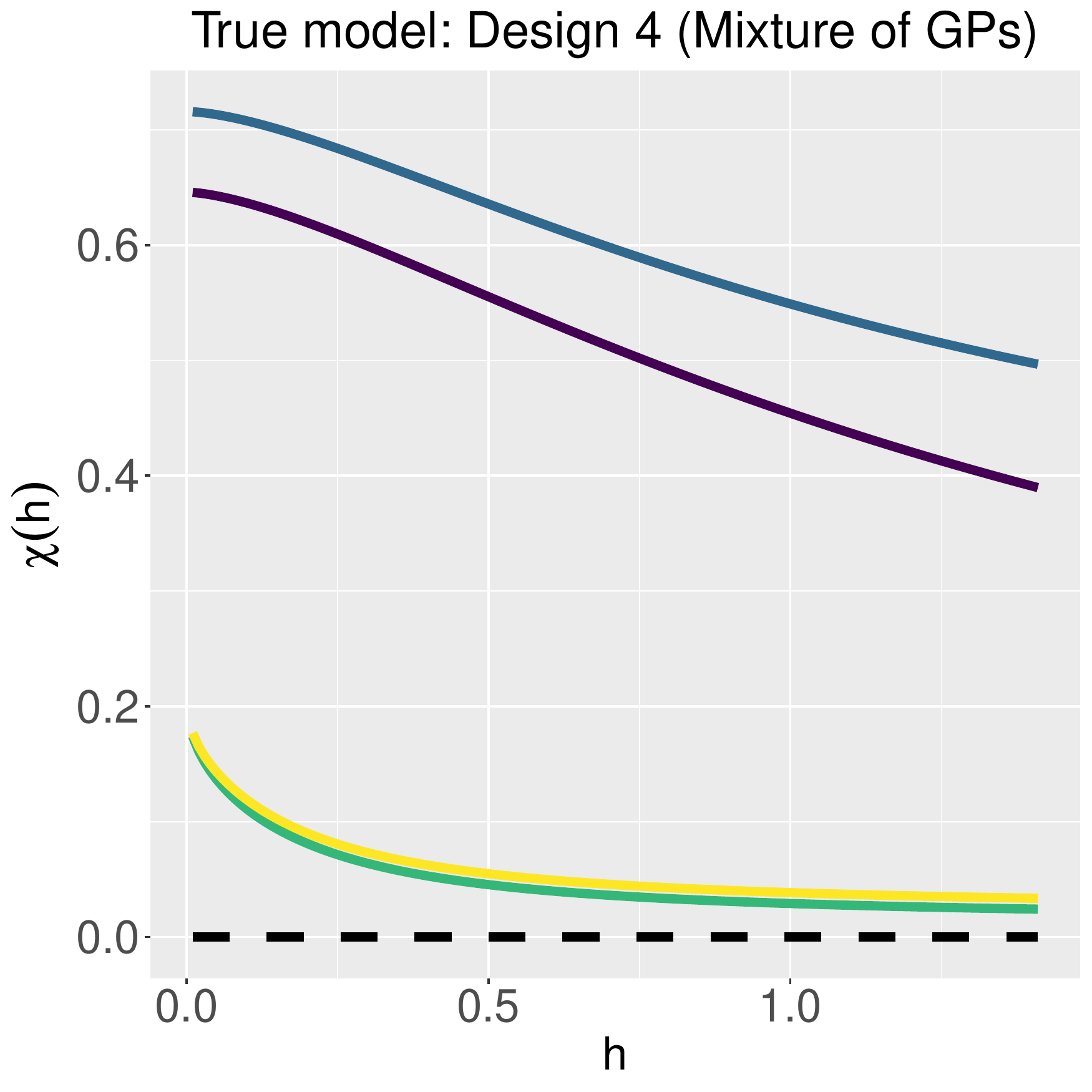}
\includegraphics[width=0.32\linewidth]{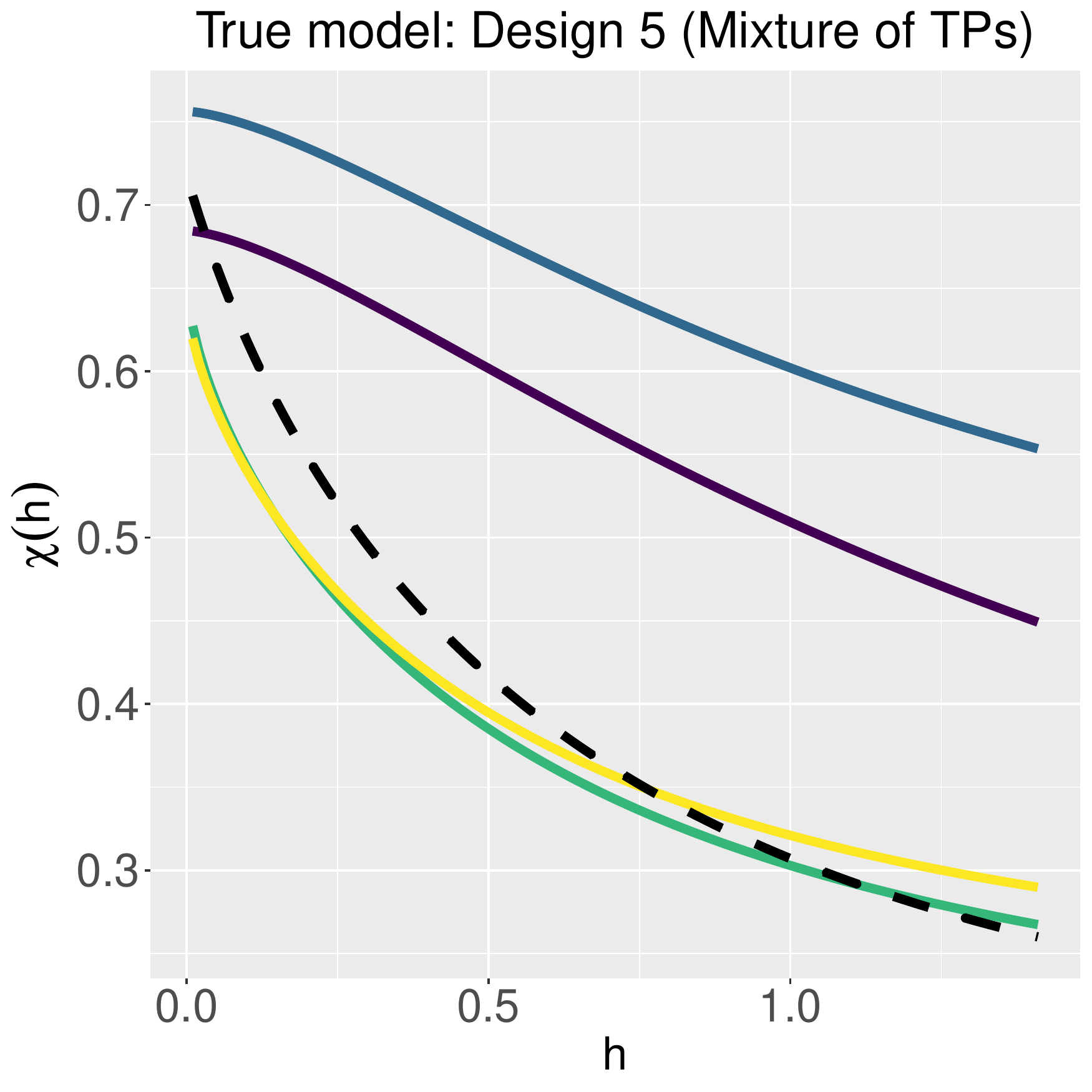}
\includegraphics[width=0.32\linewidth]{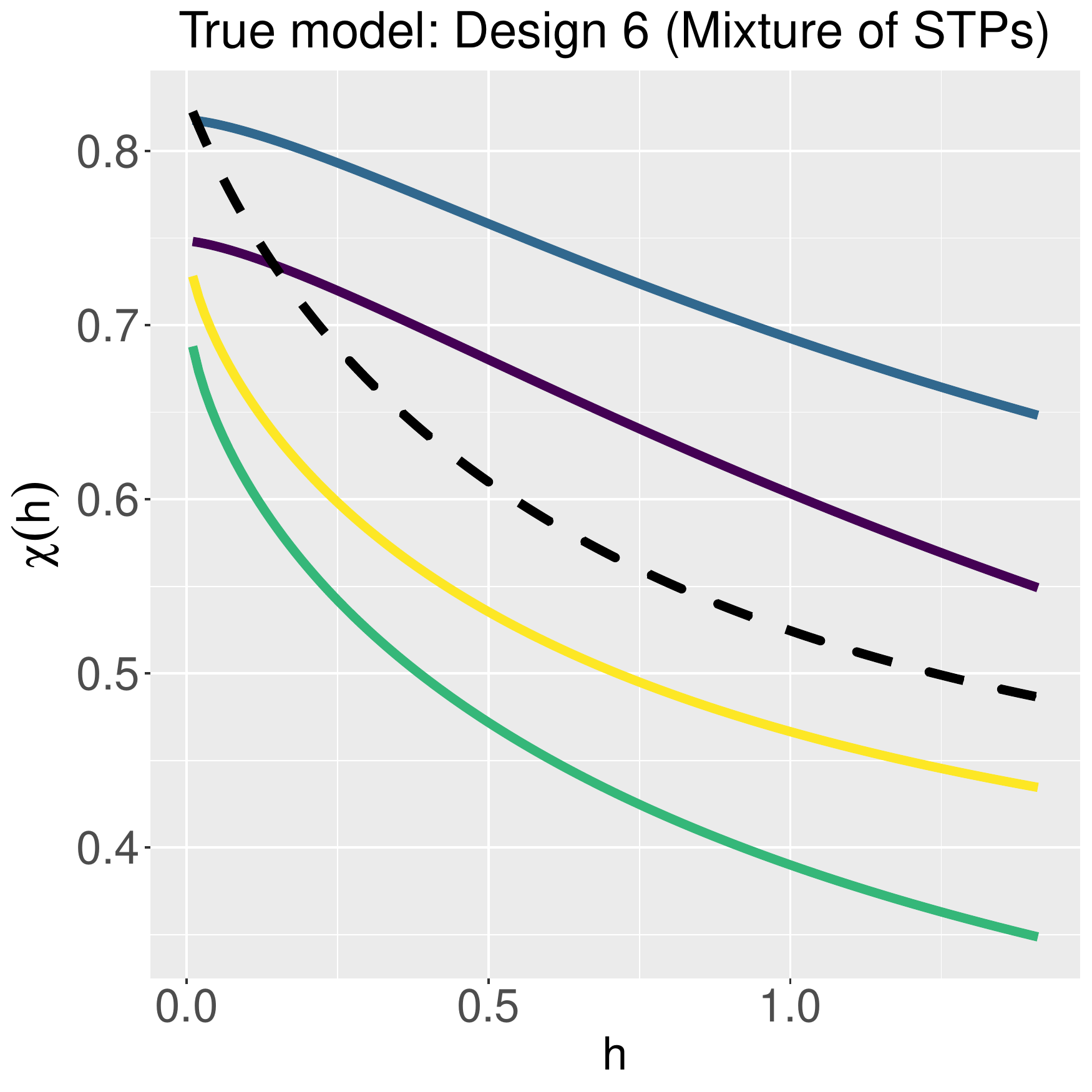}
\vspace{-2mm}
\caption{Comparison of models GP, TP, STP, GP-DPM, TP-DPM and STP-DPM based on the performance in estimation of the true spatial extremal dependence (dashed line) when the data are generated from Designs 1-6. For models GP and GP-DPM, extremal dependence is zero and hence not presented.}
\label{fig_estimated_chi}
\end{figure}

Next, we compare the models based on Figure \ref{fig_estimated_chi}. The estimates are obtained from the posterior samples based on Theorem 3.1. When the data are generated from Design 1, all the models TP, STP, TP-DPM and STP-DPM perform comparably in estimating $\chi(h)$, and the estimated $\chi(h)$ sharply drops as $h$ increases for all the models; the parametric models have less estimation bias for large values of $h$. For Design 2, again the four models estimate $\chi(h)$ well, and the parametric models have less bias. For Design 3, the models STP and STP-DPM perform better than the models TP and TP-DPM-- this indicates the necessity of the skewness component in the model. These three cases indicate that the parametric models perform well in estimation of $\chi(h)$ when the data generating model is parametric. For Design 4, the semiparametric models, TP-DPM and STP-DPM, perform well while the parametric models TP and STP lead to inaccurate estimates. For Design 5, again we notice the similar pattern that the nonparametric models estimate $\chi(h)$ with higher accuracy, while parametric models highly overestimate $\chi(h)$. For Design 6, among the nonparametric models, STP-DPM estimates $\chi(h)$ more accurately. While the estimated $\chi(h)$ based on TP is also close to true $\chi(h)$, the STP-DPM captures the true shape of $\chi(h)$ as a function of $h$, while TP fails. Although the pointwise biases are high for large $h$, a pointwise 95\% confidence interval (wide due to sparse tail) covers the true $\chi(h)$ for STP-DPM. 
Overall, STP-DPM performs equally well or better than other models, both in terms of prediction bias and prediction RMSE, as well as in the estimation of spatial extremal dependence, irrespective of the data generating model.

\section*{Appendix I: Temporal dependence}
Here we provide the proof of Theorem 3.2 in the main paper. 
    
\vspace{-2mm}
    
\begin{proof} 
We drop the spatial structure (also the notation $\bm{s}$) and calculate the temporal extremal dependence. For two time points $t$ and $t+h$, suppose the observations are $Y_{t}$ and $Y_{t+ h}$. After the GEV-log transformation, let the observations be $Y^*_{t}$ and $Y^*_{t+ h}$. Because of the stationary time series construction, the marginal distributions of the observations at two time points $t$ and $t+h$ are same and let the common CDF be $F$. Given the cluster index $g_t = k$, consider the notation $R_t = Y^*_t - \bm{\mu}_k = \lambda_k \lvert z_t \rvert + \sigma_t \epsilon_t$ where $\epsilon_t \overset{\textrm{iid}}{\sim} \textrm{Normal}(0, 1)$, $z_t|\sigma_t^2 \overset{\textrm{indep}}{\sim} \textrm{Normal}(0, \sigma_t^2)$ and $\sigma_t^2 \overset{\textrm{indep}}{\sim} \textrm{Inverse-Gamma}(a_k/2, a_k b_k / 2)$. Thus, $R_t \sim \textrm{Skew-}t(0, \lambda_k, a_k, b_k)$. The extremal dependence between $Y_{t}$ and $Y_{t+ h}$, $\nonumber \chi(t, t+h)$ is equal to the extremal dependence between $Y^*_{t}$ and $Y^*_{t+ h}$, follows from invariance of the $\chi$-measure under strictly increasing GEV-log transformation. Further, note that the sets $\lbrace Y^*_t > y^* \vert g_t = k \rbrace$ and $\lbrace \frac{y^*}{y^* - \mu_k} R_t > y^* \vert g_t = k \rbrace$ are equal. Under the limiting condition $y^* \rightarrow \infty$, $\frac{y^*}{y^* - \mu_k} \rightarrow 1$, and $\lim_{y^* \rightarrow \infty} \textrm{Pr} ( \frac{y^*}{y^* - \mu_k} R_t > y^* \vert g_t = k )$ equals $\lim_{y^* \rightarrow \infty} \textrm{Pr}(R_t > y^* \vert g_t = k)$, follows from Slutsky's theorem. Thus,
\begin{eqnarray}
	\nonumber \chi(t, t+h) &=& \lim_{y^* \rightarrow \infty} \textrm{Pr}\left(Y^*_{t+ h} > y^* | Y^*_t > y^* \right) \\
	\nonumber &=& \lim_{y^* \rightarrow \infty} \frac{\sum_{k=1}^{K} \sum_{l=1}^{K} \textrm{Pr}(Y^*_{t+ h} > y^*, Y^*_t > y^*, g_{t+h} = l, g_t = k) }{\sum_{k=1}^{K} \textrm{Pr}(Y^*_t > y^*, g_t = k) } \\
	\nonumber &=& \lim_{y^* \rightarrow \infty} \frac{ \splitfrac{\sum_{k=1}^{K} \sum_{l=1}^{K} \left\lbrace \textrm{Pr}(Y^*_{t+ h} > y^*| g_{t+h} = l) \right.}{ ~~~~~~~~~~~~~~~~~~ \times \left.\textrm{Pr}(Y^*_t > y^* | g_t = k) \textrm{Pr}(g_{t+h} = l, g_t = k) \right\rbrace }}{\sum_{k=1}^{K} \textrm{Pr}(Y^*_t > y^*| g_t = k) \textrm{Pr}(g_t = k) } \\
	\nonumber &=& \lim_{y^* \rightarrow \infty} \frac{\splitfrac{\sum_{k=1}^{K} \sum_{l=1}^{K} \left\lbrace \textrm{Pr}\left(R_{t+h} > y^*| g_{t+h} = l\right) \right.}{  ~~~~~~~~~~~~~~~~~~ \times \left. \textrm{Pr}\left(R_{t} > y^*| g_t = k\right) \textrm{Pr}(g_{t+h} = l, g_t = k) \right\rbrace }    }{\sum_{k=1}^{K} \textrm{Pr}\left(R_{t} > y^* | g_t = k\right) \textrm{Pr}(g_t = k)}.
\end{eqnarray}    

	Suppose $i = \arg \min \left \lbrace a_l \right \rbrace$ is unique. Then, for any $l \neq i$,
	\begin{eqnarray}
	\nonumber && \lim_{y^* \rightarrow \infty} \frac{\textrm{Pr}\left(R_{t+h} > y^*| g_{t+h} = l\right)}{\textrm{Pr}\left(R_{t+h} > y^*| g_{t+h} = i\right)} \\
	\nonumber &=& \lim_{y^* \rightarrow \infty} \frac{ f_{R_{t+h}}(y^* | g_{t+h} = l)}{ f_{R_{t+h}}(y^* | g_{t+h} = i)} ~~\textrm{[follows~from~L'Hospital's rule]} \\
	\nonumber &=& \lim_{y^* \rightarrow \infty} \frac{ \frac{2}{\sqrt{b_l (1 + \lambda_l^2)}} f_T\left(\frac{y^*}{\sqrt{b_l (1 + \lambda_l^2)}}; a_l \right) F_T\left( \lambda_l \frac{y^*}{\sqrt{b_l (1 + \lambda_l^2)}} \sqrt{\frac{a_l + 1}{a_l + \frac{y^{*2}}{b_l (1 + \lambda_l^2)}}}; a_l + 1 \right)}{ \frac{2}{\sqrt{b_i (1 + \lambda_i^2)}} f_T\left(\frac{y^*}{\sqrt{b_i (1 + \lambda_i^2)}}; a_i \right) F_T\left( \lambda_i \frac{y^*}{\sqrt{b_i (1 + \lambda_i^2)}} \sqrt{\frac{a_i + 1}{a_i + \frac{y^{*2}}{b_i (1 + \lambda_i^2)}}}; a_i + 1 \right)}  \\
	\nonumber &=& 0. 
	\end{eqnarray}
Thus, in the limit, $\textrm{Pr}\left(R_{t+h} > y^*| g_{t+h} = i\right) > \textrm{Pr}\left(R_{t+h} > y^*| g_{t+h} = l\right)$ for all $l \neq i$. Hence,
	\begin{eqnarray}
		\nonumber \chi(t, t+h) &\leq& \lim_{y^* \rightarrow \infty} \frac{\splitfrac{\sum_{k=1}^{K} \sum_{l=1}^{K} \left\lbrace \textrm{Pr}\left(R_{t+h} > y^*| g_{t+h} = i\right) \right.}{  ~~~~~~~~~~~~~~~~~~ \times \left. \textrm{Pr}\left(R_{t} > y^*| g_t = k\right) \textrm{Pr}(g_{t+h} = l, g_t = k) \right\rbrace }    }{\sum_{k=1}^{K} \textrm{Pr}\left(R_{t} > y^* | g_t = k\right) \textrm{Pr}(g_t = k)} \\
	\nonumber &=& \lim_{y^* \rightarrow \infty} \textrm{Pr}\left(R_{t+h} > y^* | g_{t+h} = i \right) \\
	\nonumber && ~~~~~~~~ \times \lim_{y^* \rightarrow \infty} \frac{\sum_{k=1}^{\infty}  \textrm{Pr}\left(R_{t} > y | g_{t} = k\right) \sum_{l=1}^{\infty} \textrm{Pr}(g_{t+h} = l, g_t = k) }{\sum_{k=1}^{\infty} \textrm{Pr}\left(R_{t} > y\right| g_{t} = k) \textrm{Pr}(g_t = k)} \\
	\nonumber &=& \lim_{y^* \rightarrow \infty} \textrm{Pr}\left(R_{t+h} > y* | g_{t+h} = i\right) \\
	\nonumber &=& 0 ~~\textrm{[as $R_{t+h}$ is a valid random variable with $\textrm{Pr}(R_{t+h} < \infty | g_{t+h} = i) = 1$]}.
	\end{eqnarray}
	Also, by definition, $\chi_t(h) \geq 0$. Hence, $\chi_t(h) = 0$ for any $t, h \geq 1$. In case $b_k = 0$, we obtain the trivial case of exact dependence.
\end{proof}

\section*{Appendix J: MCMC details}

Posterior inference about the model parameters is drawn using Markov chain Monte Carlo (MCMC) procedure implemented in R (\url{http://www.r-project.org}). In case it is possible to consider a conjugate prior, we select it. For some parameters, existences of conjugate priors are unknown. We use random walk Metropolis-Hastings (M-H) steps to update such parameters individually. We tune the candidate distributions in M-H steps during the burn-in period so that the acceptance rate is in between 0.3 and 0.5.



For the purpose of computation, we fix an upper limit of the number of components in the stick-breaking model, say, $K$. Suppose the observations are available at $T$ time points. The design matrix corresponding to the prior mean of the spatially-varying Dirichlet process atoms-- $\bm{\mu}_k, k=1, \ldots, K$ be denoted by $\bm{B}$, where the $i$-th row of $\bm{B}$ is $\bm{B}(\bm{s}_i)$. For updating the censored as well as the missing data faster, we consider a further hierarchical construction of the Dirichlet process mixture components. Suppose, $\bm{\Sigma}_k$ is the $n \times n$-dimensional correlation matrix evaluated at the data locations and obtained from the parameters $\Omega_k = \lbrace \rho_k, \tilde{\rho}_k, \nu_k, \psi_k, \gamma_k \rbrace$, and $\tilde{\bm{\Sigma}}_k$ is the correlation matrix obtained from the parameters $\tilde{\Omega}_k = \lbrace \rho_k, \tilde{\rho}_k, \nu_k, \psi_k \rbrace$ ignoring the nugget component. Then, following Equation 3.3 of the main paper, $\bm{\Sigma}_k = \gamma_k \tilde{\bm{\Sigma}}_k + (1 - \gamma_k) \bm{I}_n$. While $\bm{Y}_t^* \sim \textrm{Normal}_n(\bm\mu_k + \bm{A} \bm{\lambda}_k \lvert z_t \rvert, \sigma_t^2 \bm{\Sigma}_k)$, we can define the distribution of $\bm{Y}_t^*$ hierarchically as $\bm{Y}_t^* \vert \bm{X}_t \sim \textrm{Normal}_n(\bm\mu_k + \bm{X}_t + \bm{A} \bm{\lambda}_k \lvert z_t \rvert, (1 - \gamma_k) \sigma_t^2 \bm{I}_n)$ with $\bm{X}_t \sim \textrm{Normal}_n(\bm{0}_n, \gamma_k \sigma_t^2 \tilde{\bm{\Sigma}}_k)$. While we define $z_t \sim \textrm{Normal}(0, 1)$ in the main paper, here we re-define $(\sigma_t z_t)$ as $z_t$; hence, $z_t | \sigma_t^2 \sim \textrm{Normal}(0, \sigma_t^2)$; in case of the temporally independent model, this choice allows Gibbs sampling for $z_t$ and $\sigma_t^2$. Thus, although the model remains the same, the reparametrization is motivated by computational gain. We denote the set of anisotropic Mat\'ern parameters for the prior of $\mu_k(\cdot)$ as $\Omega_{\mu} = \lbrace \rho_{\mu}, \tilde{\rho}_{\mu}, \nu_{\mu}, \psi_{\mu}, \gamma_{\mu} \rbrace$. Additionally, we denote the spatial correlation matrix obtained from $\Omega_{\mu}$ by $\bm{\Sigma}_{\mu}$. The set of parameters and hyper-parameters in the model is
\begin{eqnarray}
\nonumber \Theta &=& \left\lbrace \mu_y, \sigma_y^* = \log(\sigma_y), \xi_y, \lbrace\bm\mu_k \rbrace_{k=1}^{K}, \lbrace\bm\lambda_k \rbrace_{k=1}^{K}, \lbrace \bm{X}_t \rbrace_{t=1}^{T}, \lbrace \sigma_t^2 \rbrace_{t=1}^{T}, \lbrace z_t \rbrace_{t=1}^{T},  \lbrace g_t \rbrace_{t=1}^{T}, \right. \\
\nonumber && \left. \lbrace a_k \rbrace_{k=1}^K,  \lbrace b_k \rbrace_{k=1}^K, \lbrace \Omega_k \rbrace_{k=1}^K, \lbrace \pi_k \rbrace_{k=1}^K, \phi_z, \phi_\sigma, \bm{\beta}, \sigma_\mu^2, \Omega_{\mu}, \delta \right\rbrace.
\end{eqnarray}
The MCMC steps for updating the parameters in $\Theta$ are as follows. Corresponding to a parameter (or a set of parameters), by $rest$, we mean the data, all the parameters and hyperparameters in $\Theta$ except that parameter. The notations used in the full posteriors are as follows:




\begin{itemize}
\item $\Phi$ - Standard normal distribution function
\vspace{-1mm}
    \item $f_{\textrm{Normal}}(\cdot; \mu, \sigma^2)$ - Univariate normal density with mean $\mu$ and variance $\sigma^2$
    \vspace{-1mm}
    \item $f_{\textrm{Normal}_n}(\cdot; \bm{\mu}, \bm{\Sigma})$ - $n$-variate normal density with mean $\bm{\mu}$ and covariance $\bm{\Sigma}$
    \vspace{-1mm}
    \item $f_{\textrm{G}}(\cdot; a, b)$ - Gamma density with shape $a$ and rate $b$
    \vspace{-1mm}
    \item $F_{\textrm{IG}}(\cdot; a, b)$ - Inverse-gamma distribution function with shape $a$ and scale $b$
    \vspace{-1mm}
    \item $f_{\textrm{IG}}(\cdot; a, b)$ - Inverse-gamma density with shape $a$ and scale $b$
    \vspace{-1mm}
    \item $F_{\textrm{HN}}$ - Half-normal distribution function with scale $\sigma^2$
    \vspace{-1mm}
    \item $f_{\textrm{HN}}$ - Half-normal density with scale $\sigma^2$
\end{itemize}

	\noindent \underline{\textbf{$\mu_y, \sigma^*_y, \xi_y | rest$}} \\
The parameters $\mu_y$, $\sigma^*_y$ and $\xi_y$ are updated one-at-a-time using M-H algorithm. We show the step only for $\mu_y$; the other two parameters can be similarly updated. First, we generate a candidate using the random walk Gaussian candidate distribution-- $\mu_y^{(c)} \sim \textrm{Normal}(\mu_y^{(m)}, s_{\mu_y}^2)$, where $\mu_y^{(m)}$ is the $m$-th MCMC sample from $\mu_y$, and $s_{\mu_y}$ is the standard deviation of the candidate normal distribution. Suppose $\sigma_y^{(m)}$ and $\xi_y^{(m)}$ denote the $m$-th MCMC samples from $\sigma_y$ and $\xi_y$, respectively. Let the GEV-log transformed $\bm{Y}_t(\bm{s}_i)$ based on the GEV parameters $\mu_y^{(m)}$, $\sigma_y^{(m)}$, and $\xi_y^{(m)}$ be denoted by $\bm{Y}_t^{*(m)}(\bm{s}_i)$ and $\bm{Y}_{t}^{*(m)} = [Y_t^{*(m)}(\bm{s}_1), \ldots, Y_t^{*(m)}(\bm{s}_n)]'$. Similarly, for the GEV parameters $\mu_y^{(c)}$, $\sigma_y^{(m)}$, and $\xi_y^{(m)}$, let the GEV-log transformed $\bm{Y}_t(\bm{s}_i)$ be denoted by $\bm{Y}_t^{*(c)}(\bm{s}_i)$ and $\bm{Y}_{t}^{*(c)} = [Y_t^{*(c)}(\bm{s}_1), \ldots, Y_t^{*(c)}(\bm{s}_n)]'$. Here, the acceptance ratio is
\begin{eqnarray}
\nonumber && R = \frac{\prod_{k=1}^{K} \prod_{t:g_t = k}^{} f_{\textrm{Normal}_n}\left( \bm{Y}_{t}^{*(c)}; \bm\mu_k + \bm{X}_t + \bm{A} \bm{\lambda}_k \lvert z_t \rvert, (1 - \gamma_k) \sigma_t^2 \bm{I}_n \right) }{\prod_{k=1}^{K} \prod_{t:g_t = k}^{} f_{\textrm{Normal}_n}\left( \bm{Y}_{t}^{*(m)}; \bm\mu_k + \bm{X}_t + \bm{A} \bm{\lambda}_k \lvert z_t \rvert, (1 - \gamma_k) \sigma_t^2 \bm{I}_n \right)}  \\
\nonumber && \times \frac{\prod_{t=1}^{T} \prod_{i = 1}^{n} \left(\sigma_y^{(m)} + \xi_y^{(m)} (\bm{Y}_t(\bm{s}_i) - \mu_y^{(m)} ) \right) }{\prod_{t=1}^{T} \prod_{i = 1}^{n} \left(\sigma_y^{(m)} + \xi_y^{(m)} (\bm{Y}_t(\bm{s}_i) - \mu_y^{(c)} ) \right)} \times \frac{f_{\mu_y}\left(\mu_y^{(c)}\right)}{f_{\mu_y}\left(\mu_y^{(m)}\right)},
\end{eqnarray}
where $f_{\mu_y}$ denotes the prior density of $\mu_y$. The candidate $\mu_y^{(c)}$ is accepted with probability $min \lbrace R,1 \rbrace$. If $(\sigma_y^{(m)} + \xi_y^{(m)} (\bm{Y}_t(\bm{s}_i) - \mu_y^{(c)} ))$ is negative for any $i$ and $t$, we set $R = 0$.

\noindent \underline{\textbf{$g_t | rest$}} \\
In case the temporal dependence is ignored, the posterior distribution of $g_t$ is as follows.
	\begin{eqnarray}
	\nonumber \textrm{Pr}(g_t = k|rest) &\propto& \pi_k \times f_{\textrm{Normal}_n}\left( \bm{Y}_{t}^*; \bm\mu_k + \bm{X}_t + \bm{A} \bm{\lambda}_k \lvert z_t \rvert, (1 - \gamma_k) \sigma_t^2 \bm{I}_n \right) \\
	\nonumber && \times f_{\textrm{IG}}(\sigma_t^2; a_k / 2, a_k b_k / 2).
	\end{eqnarray}
	
For models ignoring random scaling (GP-DPM), the third term is removed from the expression. The density of $z_t$ depends only on $\sigma_t^2$ and hence does not vary through $g_t$.

In case of temporal dependence, the posterior density of $g_t$ depends on $\lbrace g_t \rbrace_{t=1}^{T}$ through both $g_{t-1}$ and $g_{t+1}$. Given $\sigma_{t}^2$, $\sigma_{t-1}^2$, $g_{t-1}$, $\sigma_{t+1}^2$, $g_{t+1}$, $\lbrace a_k \rbrace_{k=1}^K$ and $\lbrace b_k \rbrace_{k=1}^K$, we consider the copula transformations as in the main article,
\begin{eqnarray}
\nonumber \sigma_{t-1}^{2*} &=& \Phi^{-1}[F_{\textrm{IG}}(\sigma_{t-1}^2; a_{g_{t-1}} / 2, a_{g_{t-1}} b_{g_{t-1}}/ 2)], \\
\nonumber \sigma_{t}^{2*(k)} &=& \Phi^{-1}[F_{\textrm{IG}}(\sigma_{t-1}^2; a_{k} / 2, a_{k} b_{k}/ 2)]; k=1, \ldots, K,\\
\nonumber \sigma_{t+1}^{2*} &=& \Phi^{-1}[F_{\textrm{IG}}(\sigma_{t+1}^2; a_{g_{t+1}} / 2, a_{g_{t+1}} b_{g_{t+1}}/ 2)].
\end{eqnarray}

After multiplying the Jacobian terms of the copula transformation, the posterior density of $g_t$ is 
\begin{eqnarray}
	\nonumber \textrm{Pr}(g_t = k|rest) &\propto& \pi_k \times f_{\textrm{Normal}_n}\left( \bm{Y}_{t}^*; \bm\mu_k + \bm{X}_t + \bm{A} \bm{\lambda}_k \lvert z_t \rvert, (1 - \gamma_k) \sigma_t^2 \bm{I}_n \right) \\
	\nonumber && \times f_{\textrm{IG}}(\sigma_t^2; a_k / 2, a_k b_k / 2)  \times f_{\textrm{Normal}}(\sigma_{t}^{2*(k)}; \phi_\sigma \sigma_{t-1}^{2*}, 1 - \phi_\sigma^2)\\
	\nonumber && \times f_{\textrm{Normal}}(\sigma_{t+1}^{2*}; \phi_\sigma \sigma_{t}^{2*(k)}, 1 - \phi_\sigma^2) / f_{\textrm{Normal}}(\sigma_{t}^{2*(k)}; 0, 1).
	\end{eqnarray}

For $t = 1$, the fourth and sixth terms are removed from the expression, while for $t = T$, the fifth term is removed from the expression.

   

\noindent \underline{\textbf{$\bm{\mu}_k | rest$}} \\
The prior distributions of $\bm{\mu}_k$'s are $\bm{\mu}_k \overset{\textrm{iid}}{\sim} \textrm{Normal}_n(\bm{B} \bm{\beta}, \sigma_{\mu}^2 \bm{\Sigma_{\mu}})$. The full conditional posterior distribution of $\bm{\mu}_k$ is $\bm\mu_k | rest \sim \textrm{Normal}_n(\bm{\mu_k}^\ast, \bm{\Sigma_k}^\ast)$, where
\begin{eqnarray}
\nonumber && \bm{\Sigma_k}^\ast = \left[ \frac{1}{1 - \gamma_k} \left( \sum_{t:g_t = k}^{} \sigma_t^{-2} \right) \bm{I}_n + \frac{1}{\sigma_{\mu}^2} \bm{\Sigma_{\mu}}^{-1}\right]^{-1}, \\
\nonumber && \bm{\mu_k}^\ast =  \bm{\Sigma_k}^\ast \left[\frac{1}{1 - \gamma_k} \left(\sum_{t:g_t = k}^{} \sigma_t^{-2} \left(\bm{Y}_t^* - \bm{X}_t - \bm{A} \bm{\lambda}_k \lvert z_t \rvert \right) \right) + \frac{1}{\sigma_{\mu}^2} \bm{\Sigma_{\mu}}^{-1} \bm{B} \bm{\beta} \right].
\end{eqnarray}



\noindent \underline{\textbf{$\bm{X}_t | rest$}} \\
The full conditional posterior distribution of $\bm{X}_t$ is $\bm{X}_t | rest \sim \textrm{Normal}_n(\bm{\mu_X}^\ast, \bm{\Sigma_X}^\ast)$, where
\begin{eqnarray}
\nonumber && \bm{\Sigma_X}^\ast = \left[ \frac{1}{(1 - \gamma_{g_t}) \sigma_t^2} \bm{I}_n + \frac{1}{\gamma_{g_t} \sigma_{t}^2} \tilde{\bm{\Sigma}}_{g_t}^{-1}\right]^{-1},~ \bm{\mu_X}^\ast =  \bm{\Sigma_X}^\ast \left[\frac{1}{(1 - \gamma_{g_t})\sigma_t^2} \left( \bm{Y}_t^* - \bm{\mu}_{g_t} - \bm{A} \bm{\lambda}_{g_t} \lvert z_t \rvert \right) \right].
\end{eqnarray}

    
\noindent \underline{\textbf{$\bm\lambda_k | rest$}} \\
For the prior $\bm\lambda_k \overset{\textrm{iid}}{\sim} \textrm{Normal}_q(\bm{0}, \bm{\Sigma_\lambda})$,  we have $\bm\lambda_k | rest \sim \textrm{Normal}_q(\bm{\mu_\lambda}^\ast, \bm{\Sigma_\lambda}^\ast)$, where
\begin{eqnarray}
\nonumber && \bm{\Sigma_\lambda}^\ast = \left[ \frac{1}{1 - \gamma_k} \left(\sum_{t:g_t = k}^{} \frac{z_t^2}{\sigma_t^2}\right) \bm{A}' \bm{A} + \bm{\Sigma_\lambda}^{-1}\right]^{-1}, \\
\nonumber && \bm{\mu_\lambda}^\ast =  \bm{\Sigma_\lambda}^\ast \left[\frac{1}{1 - \gamma_k}\bm{A}' \left( \sum_{t:g_t = k}^{} \frac{|z_t|}{\sigma_t^2}  \left(\bm{Y}_t^* - \bm\mu_k - \bm{X}_t \right) \right) \right].
\end{eqnarray}
    



\noindent \underline{\textbf{$\sigma_t^2 | rest$}} \\
	If $\sigma_t^2$ are independent over days, its full conditional posterior distribution is also inverse-gamma. Let $\bm{R}_t = \bm{Y}_{t}^* - \bm{\mu}_{g_t} - \bm{X}_t  - \bm{A} \bm{\lambda}_{g_t} \lvert z_t \rvert$. The distributions that involve $\sigma_t^2$ are $\bm{R}_t \sim \textrm{Normal}_n(\bm{0}_n, (1 - \gamma_{g_t}) \sigma_t^2 \bm{I}_n)$, $\bm{X}_t \sim \textrm{Normal}_n(\bm{0}_n, \gamma_{g_t} \sigma_t^2 \tilde{\bm{\Sigma}}_{g_t})$, and $z_t \sim \textrm{Normal}(0, \sigma_t^2)$. The conditional posterior distribution of $\sigma_t^2$ is $\sigma_t^2 | rest \sim \textrm{Inverse-Gamma}(a^*, b^*)$, where
	\begin{eqnarray}
	\nonumber a^* &=& 0.5\left(a_{g_t} + 2n + 1\right), \\
	\nonumber b^* &=& 0.5\left(a_{g_t} b_{g_t} + (1 - \gamma_{g_t})^{-1} \bm{R}_t'\bm{R}_t + \gamma_{g_t}^{-1} \bm{X}_t'\tilde{\bm{\Sigma}}_{g_t}^{-1} \bm{X}_t + z_t^2 \right).
	\end{eqnarray}


In case $\sigma_t^2$'s are dependent, the conditional posterior distribution of $\sigma_t^2$ has no closed form expression. Here we update $\sigma_t^2$ using M-H algorithm. For the $m$-th MCMC step, if $\sigma_t^{2(m)}$ denotes the sample from  $\sigma_t^{2}$, using the copula transformation, we obtain $\sigma_t^{2*(m)} = \Phi^{-1}(F_{\textrm{IG}}(\sigma_t^{2(m)}; a_{g_t} / 2, a_{g_t} b_{g_t} / 2))$. We generate a candidate for $\sigma_t^2$ using
\begin{eqnarray}
\nonumber \sigma_t^{2*(c)} \sim \textrm{Normal}(\sigma_t^{2*(m)}, s_\sigma^2);~~ \sigma_t^{2(c)} = F_{\textrm{IG}}^{-1}\left(\Phi\left(\sigma_t^{2*(c)}\right); a_{g_t} / 2, a_{g_t} b_{g_t} / 2 \right),
\end{eqnarray}
where $s_\sigma$ is the standard deviation of the candidate normal distribution.

Considering the Jacobian transformations regarding the candidate distribution, after a few steps of algebra, the acceptance ratio is
	\begin{eqnarray}
	\nonumber R &=& \frac{ f_{\textrm{Normal}_n}\left( \bm{Y}_{t}^{*}; \bm\mu_{g_t} + \bm{X}_t + \bm{A} \bm{\lambda}_{g_t} \lvert z_t \rvert, \sigma_t^{2(c)} (1 - \gamma_{g_t}) \bm{I}_n \right) }{ f_{\textrm{Normal}_n}\left( \bm{Y}_{t}^{*};  \bm\mu_{g_t} + \bm{X}_t + \bm{A} \bm{\lambda}_{g_t} \lvert z_t \rvert, \sigma_t^{2(m)} (1 - \gamma_{g_t}) \bm{I}_n \right)} \times \frac{f_{\textrm{HN}}(\lvert z_t \rvert; \sigma_t^{2(c)})}{f_{\textrm{HN}}(\lvert z_t \rvert; \sigma_t^{2(m)})} \\
	\nonumber && \times \frac{f_{\textrm{Normal}}\left(\sigma_t^{2*(c)}; \phi_\sigma \sigma_{t-1}^{2*}, 1 - \phi_\sigma^2\right)}{f_{\textrm{Normal}}\left(\sigma_t^{2*(m)}; \phi_\sigma \sigma_{t-1}^{2*}, 1 - \phi_\sigma^2 \right)} \times \frac{f_{\textrm{Normal}}\left(\sigma_{t+1}^{2*}; \phi_\sigma \sigma_t^{2*(c)}, 1 - \phi_\sigma^2\right)}{f_{\textrm{Normal}}\left(\sigma_{t+1}^{2*}; \phi_\sigma \sigma_t^{2*(m)}, 1 - \phi_\sigma^2 \right)}.
	\end{eqnarray} 
For $t=1$ and $t = T$, the third and fourth ratios in $R$ are dropped respectively. 

\noindent \underline{\textbf{$\lvert z_t \rvert | rest$}} \\
	As $z_t$ is not identifiable, we treat $\lvert z_t \rvert$ as a parameter and update $\lvert z_t \rvert$ within the MCMC steps. Here $\lvert z_t \rvert \sim \textrm{Half-Normal}(\sigma_t^2)$. If $\lvert z_t \rvert$ are independent over days, the full conditional posterior density of $\lvert z_t \rvert$ is $\lvert z_t \rvert | rest \sim \textrm{Normal}_{(0, \infty)}(\mu_z, \sigma_z^2)$, where
	\begin{eqnarray}
	\nonumber \sigma_z^2 &=& \sigma_t^2\left[1 + (1 - \gamma_{g_t})^{-1} \bm{\lambda}_{g_t}' \bm{A}'\bm{A} \bm{\lambda}_{g_t} \right]^{-1}, \\
	\nonumber \mu_z &=& \left[1 - \gamma_{g_t} + \bm{\lambda}_{g_t}' \bm{A}'\bm{A} \bm{\lambda}_{g_t} \right]^{-1} (\bm{\lambda}_{g_t}' \bm{A}' (\bm{Y}_{t}^* - \bm{\mu}_{g_t} - \bm{X}_t)).
	\end{eqnarray}

Here,  $\textrm{Normal}_{(0, \infty)}$ denotes the truncated normal distribution with support equals the positive real line.	

	If $\lvert z_t \rvert$ are dependent across days, the full conditional posterior distribution of $\lvert z_t \rvert$ has no closed form expression, and we update $\lvert z_t \rvert$ using M-H algorithm. For the $m$-th MCMC step, if $\lvert z_t \rvert^{(m)}$ denotes the sample from  $\lvert z_t \rvert$, using the copula transformation, we obtain $z_t^{*(m)} = \Phi^{-1}(F_{\textrm{HN}}(\lvert z_t \rvert^{(m)}; \sigma_t^2))$. We generate a candidate for $\lvert z_t \rvert$ using
\begin{eqnarray}
\nonumber z_t^{*(c)} \sim \textrm{Normal}\left( z_t^{*(m)}, s_z^2 \right);~~ \lvert z_t \rvert^{(c)} = F_{\textrm{HN}}^{-1}\left(\Phi\left( z_t^{*(c)} \right); \sigma_t^2 \right),
\end{eqnarray}
where $s_z$ is the standard deviation of the candidate normal distribution.

Considering the Jacobian transformations regarding the candidate distribution, after a few steps of algebra, the acceptance ratio is
\begin{eqnarray}
	\nonumber R &=& \frac{ f_{\textrm{Normal}_n}\left( \bm{Y}_{t}^{*}; \bm\mu_{g_t} + \bm{X}_t + \bm{A} \bm{\lambda}_{g_t} \lvert z_t \rvert^{(c)}, \sigma_t^2 (1 - \gamma_{g_t}) \bm{I}_n \right) }{ f_{\textrm{Normal}_n}\left( \bm{Y}_{t}^{*}; \bm\mu_{g_t} + \bm{X}_t + \bm{A} \bm{\lambda}_{g_t} \lvert z_t \rvert^{(m)}, \sigma_t^2 (1 - \gamma_{g_t}) \bm{I}_n \right)} \\
	\nonumber && \times \frac{f_{\textrm{Normal}}(z_t^{*(c)}; \phi_z z_{t-1}^{*}, 1 - \phi_z^2)}{f_{\textrm{Normal}}(z_t^{*(m)}; \phi_z z_{t-1}^{*}, 1 - \phi_z^2)} \times \frac{f_{\textrm{Normal}}(z_{t+1}^{*}; \phi_z z_t^{*(c)}, 1 - \phi_z^2)}{f_{\textrm{Normal}}(z_{t+1}^{*}; \phi_z z_t^{*(m)}, 1 - \phi_z^2)}.
	\end{eqnarray}
For $t=1$ and $t = T$, the second and third ratios in $R$ are dropped, respectively. 
	
	\noindent \underline{\textbf{$a_k | rest$}} \\
	First, we consider the discrete prior case-- $a_k \overset{\textrm{iid}}{\sim} \textrm{Discrete-Uniform}\{0.5, 0.6, \ldots, 39.9, 40.0\}$. If $\sigma_t^2$ are independent over days, the full conditional posterior distribution of $a_k$ is
	\begin{eqnarray}
	\nonumber \textrm{Pr}(a_k = a^\ast | rest) \propto \prod_{t:g_t = k}^{} f_{\textrm{IG}}(\sigma_t^2, a^\ast / 2, a^\ast \beta_k / 2),
	\end{eqnarray}
	and we draw a random sample from the discrete support $\{0.5, 0.6, \ldots, 39.9, 40.0\}$, with probabilities proportional to $\textrm{Pr}(a_k = a^\ast | rest)$. In case $\sigma_t^2$ are dependent over days, the conditional posterior distribution of $a_k$ is proportional to the product of the terms involving $a_k$ in the expression of the joint density of $\sigma_t^2;t=1, \ldots,T$,
	\begin{eqnarray}
	\nonumber f(\sigma_1^{2}, \ldots, \sigma_{T}^{2}) &=& \prod_{t = 1}^{T} f_{\textrm{IG}}(\sigma_t^{2}; a_{g_t} / 2, a_{g_t} b_{g_t} / 2) \times \prod_{t = 2}^{T} \frac{f_{\textrm{Normal}}\left(\sigma_t^{2*}; \phi_\sigma \sigma_{t-1}^{2*}, 1 - \phi_\sigma^2 \right)}{f_{\textrm{Normal}}\left(\sigma_t^{2*}; 0, 1 \right)}.
	\end{eqnarray}

This Sampling step is slow and hence, we also consider updating $a_k$ using M-H algorithm for the very flexible models like STP-DPM. We consider $a_k \overset{\textrm{iid}}{\sim} \textrm{Uniform}(0.5, 40)$. Suppose $a_k^{(m)}$ denotes the MCMC sample from $a_k$ at the $m$-th iteration. Considering a logit transformation, we obtain $a_k^{*(m)} \in \mathbb{R}$, and generate a sample $a_k^{*(c)} \sim \textrm{Normal}( a_k^{*(m)}, s_{a_k}^2 )$. Subsequently, using an inverse-logit transformation, we obtain $a_k^{(c)}$ from $a_k^{*(c)}$. Let $f_{a_k}(\sigma_1^{2}, \ldots, \sigma_{T}^{2})$ be the terms in $f(\sigma_1^{2}, \ldots, \sigma_{T}^{2})$ that involves $a_k$. Here the acceptance ratio is
\begin{eqnarray}
	\nonumber R = \frac{f_{a_k^{(c)}}(\sigma_1^{2}, \ldots, \sigma_{T}^{2})}{f_{a_k^{(m)}}(\sigma_1^{2}, \ldots, \sigma_{T}^{2})} \times \frac{\left(a_k^{(c)} - 0.5\right) \left(40 - a_k^{(c)}\right)}{\left(a_k^{(m)} - 0.5\right) \left(40 - a_k^{(m)}\right)}.
	\end{eqnarray}



	\noindent \underline{\textbf{$b_k | rest$}} \\
	We consider the priors-- $b_k \overset{\textrm{iid}}{\sim} \textrm{Gamma}(a^*, b^*)$. In case $\sigma_t^2$ are independent over days, the full conditional posterior distribution of $b_k$ is conjugate and given by
	\begin{eqnarray}
	\nonumber b_k | rest \propto \textrm{Gamma}\left(a^* +   \frac{a_k}{2} \sum_{t=1}^{T} \mathbb{I}(g_t = k), b^* +  \frac{a_k}{2} \sum_{t:g_t = k}^{} \sigma_t^{-2} \right).
	\end{eqnarray}
	If $\sigma_t^2$ are dependent over days, the conditional posterior distribution of $b_k$ is proportional to the product of the terms involving $b_k$ in the expression of the joint density of $\sigma_t^2;t=1, \ldots,T$ 
	multiplied with the prior density of $b_k$. The density has no closed form expression, and hence we update $b_k$ using M-H algorithm. Suppose $b_k^{(m)}$ denotes the MCMC sample from $b_k$ at the $m$-th iteration. Considering a log transformation, we obtain $b_k^{*(m)} \in \mathbb{R}$ and generate a sample $b_k^{*(c)} \sim \textrm{Normal}\left( b_k^{*(m)}, s_{b_k}^2 \right)$. Subsequently, by exponentiating, we obtain $b_k^{(c)}$ from $b_k^{*(c)}$. Let $f_{b_k}(\sigma_1^{2}, \ldots, \sigma_{T}^{2})$ be the terms in $f(\sigma_1^{2}, \ldots, \sigma_{T}^{2})$ that involves $b_k$. Here the acceptance ratio is
\begin{eqnarray}
	\nonumber R = \frac{f_{b_k^{(c)}}(\sigma_1^{2}, \ldots, \sigma_{T}^{2})}{f_{b_k^{(m)}}(\sigma_1^{2}, \ldots, \sigma_{T}^{2})} \times \frac{f_{\textrm{G}}(b_k^{(c)}; a^*, b^*)}{f_{\textrm{G}}(b_k^{(m)}; a^*, b^*)} \times \frac{b_k^{(c)}}{b_k^{(m)}}.
	\end{eqnarray}

\noindent \underline{\textbf{$\phi_\sigma | rest$}} \\
	We consider $\textrm{Uniform}(0, 1)$ prior for the parameter $\phi_\sigma$. Thus, the conditional posterior density is proportional to the joint density of $\sigma_t^{2*}, t=1, \ldots,{T}$. The M-H step is similar to the update of $a_k$.
	
	\noindent \underline{\textbf{$\phi_z | rest$}} \\
	We consider $\textrm{Uniform}(0, 1)$ prior for the parameter $\phi_z$. Thus, the conditional posterior density is proportional to the joint density of $z_t^*, t=1, \ldots,{T}$. The M-H step is similar to the update of $\phi_\sigma$.

\noindent \underline{\textbf{$\pi = [\pi_1, \ldots, \pi_K] | rest$}} \\
	Using the stick-breaking representation, there is a one-to-one correspondence between $\pi$ and $\bm{V} = [v_1, \ldots, v_K]$ as $\pi_k = v_k \Pi_{l=1}^{k-1} (1 - v_l)$. We set $v_K = 1$ and update $v_1, \ldots, v_{K-1}$. Here, $v_1, \ldots, v_{K-1} \overset{\textrm{iid}}{\sim} \textrm{Beta}(1, \delta)$. The posterior density of $v_k$ conditioned on $rest$ is given by
	\begin{equation}
	\nonumber v_k| rest \sim \textrm{Beta} \left( 1 + \sum_{t=1}^{T} \mathbb{I}(g_t = k), \delta + \sum_{t=1}^{T} \mathbb{I}(g_t > k) \right).
	\end{equation}
	
	\noindent \underline{\textbf{$\delta | rest$}} \\
	We consider the prior for $\delta$ to be $\delta \sim \textrm{Gamma}(a_\delta, b_\delta)$. Thus, the posterior density of $v_k$ conditioned on $rest$ is given by
	\begin{eqnarray}
	\nonumber && \delta | rest \sim \textrm{Gamma}\left(a_\delta + K - 1, b_\delta - \sum_{k=1}^{K-1} \log(1 - v_k) \right).
	\end{eqnarray}
	
\noindent \underline{$\tilde{\Omega}_k = \{ \rho_k, \tilde{\rho}_k, \nu_k, \psi_k \} | rest$} \\
These parameters are updated using M-H algorithm one-at-a-time. 
For $\rho_k$ and $\psi_k$, we draw candidate samples within MCMC similar to $a_k$. For $\tilde{\rho}_k$, we draw a candidate sample similar to $b_k$. We update $\nu_k$ in the log scale similar to $\sigma_y$. 


First we consider updating $\rho_k$. We consider the priors-- $\rho_k \overset{\textrm{iid}}{\sim} \textrm{Uniform}(0, 2.5\Delta)$, where $\Delta$ is the largest Euclidean distance between two data locations. At the $m$-th MCMC step, suppose $\tilde{\bm{\Sigma}}_k^{(m)}$ denotes the correlation matrix based on $\{\rho_k^{(m)}, \tilde{\rho}_k^{(m)}, \nu_k^{(m)}, \psi_k^{(m)} \}$, and $\tilde{\bm{\Sigma}}_k^{(c)}$ denotes the correlation matrix based on $\{\rho_k^{(c)}, \tilde{\rho}_k^{(m)}, \nu_k^{(m)}, \psi_k^{(m)}\}$, where $\rho_k^{(c)}$ is a candidate sample from $\rho_k$.
The acceptance ratio is 
\begin{eqnarray}
\nonumber R &=& \frac{\prod_{t:g_t = k}^{} f_{\textrm{Normal}_n}\left(\bm{X}_t; \bm{0}_n, \gamma_k \sigma_t^2 \tilde{\bm{\Sigma}}_{k}^{(c)} \right) }{\prod_{t:g_t = k}^{} f_{\textrm{Normal}_n}\left(\bm{X}_t; \bm{0}_n, \gamma_k \sigma_t^2 \tilde{\bm{\Sigma}}_{k}^{(m)} \right) } \times  \frac{\rho_k^{(c)}  \left(2.5\Delta - \rho_k^{(c)}\right)}{\rho_k^{(m)} \left(2.5\Delta - \rho_k^{(m)}\right)},
\end{eqnarray}
and the candidate is accepted with probability $min \lbrace R,1 \rbrace$.



Next, we consider updating $\tilde{\rho}_k$. We consider the priors-- $\tilde{\rho}_k \overset{\textrm{iid}}{\sim} \textrm{Gamma}(0.1, 0.1)$. At the $m$-th MCMC step, suppose $\tilde{\bm{\Sigma}}_k^{(m)}$ and $\tilde{\bm{\Sigma}}_k^{(c)}$ denote the correlation matrices based on $\{\rho_k^{(m)}, \tilde{\rho}_k^{(m)}, \nu_k^{(m)}, \psi_k^{(m)} \}$, and $\{\rho_k^{(m)}, \tilde{\rho}_k^{(c)}, \nu_k^{(m)}, \psi_k^{(m)}\}$, respectively, where $\tilde{\rho}_k^{(c)}$ is a candidate sample from $\tilde{\rho}_k$. The acceptance ratio is 
\begin{eqnarray}
\nonumber R &=& \frac{\prod_{t:g_t = k}^{} f_{\textrm{Normal}_n}\left(\bm{X}_t; \bm{0}_n, \gamma_k \sigma_t^2 \tilde{\bm{\Sigma}}_{k}^{(c)} \right) }{\prod_{t:g_t = k}^{} f_{\textrm{Normal}_n}\left(\bm{X}_t; \bm{0}_n, \gamma_k \sigma_t^2 \tilde{\bm{\Sigma}}_{k}^{(m)} \right) } \times \frac{f_{\textrm{G}}(\tilde{\rho}_k^{(c)}; 0.1, 0.1)}{f_{\textrm{G}}(\tilde{\rho}_k^{(m)}; 0.1, 0.1)} \times \frac{\tilde{\rho}_k^{(c)}}{\tilde{\rho}_k^{(m)}}.
\end{eqnarray}

Next, we consider updating $\nu_k^* = \log(\nu_k)$. We consider the priors-- $\nu_k^* \overset{\textrm{iid}}{\sim} \textrm{Normal}(-1.2, 1^2)$ with $\nu_k$ is truncated above at 40. At the $m$-th MCMC step, suppose $\tilde{\bm{\Sigma}}_k^{(m)}$ and $\tilde{\bm{\Sigma}}_k^{(c)}$ denote the correlation matrices based on $\{\rho_k^{(m)}, \tilde{\rho}_k^{(m)}, \nu_k^{(m)}, \psi_k^{(m)} \}$, and $\{\rho_k^{(m)}, \tilde{\rho}_k^{(m)}, \nu_k^{(c)}, \psi_k^{(m)} \}$, respectively, where $\nu_k^{(c)}$ is a candidate sample from $\nu_k$. The acceptance ratio is 
\begin{eqnarray}
\nonumber R &=& \frac{\prod_{t:g_t = k}^{} f_{\textrm{Normal}_n}\left(\bm{X}_t; \bm{0}_n, \gamma_k \sigma_t^2 \tilde{\bm{\Sigma}}_{k}^{(c)} \right) }{\prod_{t:g_t = k}^{} f_{\textrm{Normal}_n}\left(\bm{X}_t; \bm{0}_n, \gamma_k \sigma_t^2 \tilde{\bm{\Sigma}}_{k}^{(m)} \right) } \times \frac{f_{\textrm{Normal}}\left( \log(\nu_k^{(c)}); -1.2, 1^2 \right)}{f_{\textrm{Normal}}\left(\log(\nu_k^{(m)}); -1.2, 1^2 \right)},
\end{eqnarray}
when $\nu_k^{(c)} \leq 40$, and we set $R = 0$ if $\nu_k^{(c)} > 40$.

Further, we consider updating $\psi_k$. We consider the priors-- $\psi_k \overset{\textrm{iid}}{\sim} \textrm{Uniform}(0, \pi/2)$. At the $m$-th MCMC step, suppose $\tilde{\bm{\Sigma}}_k^{(m)}$ and $\tilde{\bm{\Sigma}}_k^{(c)}$ denote the correlation matrices based on $\{\rho_k^{(m)}, \tilde{\rho}_k^{(m)}, \nu_k^{(m)}, \psi_k^{(m)} \}$, and $\{\rho_k^{(m)}, \tilde{\rho}_k^{(m)}, \nu_k^{(m)}, \psi_k^{(c)} \}$, respectively, where $\psi_k^{(c)}$ is a candidate sample from $\psi_k$. The acceptance ratio is 
\begin{eqnarray}
\nonumber R &=& \frac{\prod_{t:g_t = k}^{} f_{\textrm{Normal}_n}\left(\bm{X}_t; \bm{0}_n, \gamma_k \sigma_t^2 \tilde{\bm{\Sigma}}_{k}^{(c)} \right) }{\prod_{t:g_t = k}^{} f_{\textrm{Normal}_n}\left(\bm{X}_t; \bm{0}_n, \gamma_k \sigma_t^2 \tilde{\bm{\Sigma}}_{k}^{(m)} \right) } \times \frac{\psi_k^{(c)}  \left(\pi/2 - \psi_k^{(c)}\right)}{\psi_k^{(m)} \left(\pi/2 - \psi_k^{(m)}\right)}.
\end{eqnarray}

\noindent \underline{$\gamma_k | rest$} \\
We consider the priors-- $\gamma_k \overset{\textrm{iid}}{\sim} \textrm{Uniform}(0, 1)$. We draw candidate samples within MCMC similar to $a_k$. Suppose $\gamma_k^{(m)}$ denotes the $m$-th MCMC sample from $\gamma_k$ and $\gamma_k^{(c)}$ is a candidate. Given $g_t = k$, let $\bm{R}_t = \bm{Y}_{t}^* - \bm\mu_k - \bm{X}_t - \bm{A} \bm{\lambda}_k \lvert z_t \rvert$. The acceptance ratio is 
\begin{eqnarray}
\nonumber R &=& \frac{ \prod_{t:g_t = k}^{} f_{\textrm{Normal}_n}\left(\bm{R}_t; \bm{0}_n, (1 - \gamma_k^{(c)}) \sigma_t^2 \bm{I}_n \right) \times f_{\textrm{Normal}_n}\left(\bm{X}_t; \bm{0}_n, \gamma_k^{(c)} \sigma_t^2 \tilde{\bm{\Sigma}}_{k} \right) }{ \prod_{t:g_t = k}^{} f_{\textrm{Normal}_n}\left(\bm{R}_t; \bm{0}_n, (1 - \gamma_k^{(m)}) \sigma_t^2 \bm{I}_n \right) \times f_{\textrm{Normal}_n}\left(\bm{X}_t; \bm{0}_n, \gamma_k^{(m)} \sigma_t^2 \tilde{\bm{\Sigma}}_{k} \right)} \\
\nonumber && \times \frac{\gamma_k^{(c)}  \left(1 - \gamma_k^{(c)}\right)}{\gamma_k^{(m)} \left(1 - \gamma_k^{(m)}\right)}.
\end{eqnarray}


\noindent \underline{\textbf{$\beta | rest$}} \\
We consider the conjugate noninformative prior $\bm{\beta} \sim \textrm{Normal}_p(0, 10^2 \bm{I}_p)$. The full conditional posterior distribution is $\beta | rest \sim \textrm{Normal}_p(\bm{\mu_{\beta}}, \bm{\Sigma_{\beta}})$, where
\begin{eqnarray}
\nonumber && \bm{\Sigma_{\beta}} = \left[ K \sigma_{\mu}^{-2} \bm{B}' \bm{\Sigma_{\mu}}^{-1} \bm{B} + 10^{-2} \bm{I}_p \right]^{-1},~~ \bm{\mu_{\beta}} = \bm{\Sigma_{\beta}} \left[ \sigma_{\mu}^{-2} \bm{B}' \bm{\Sigma_{\mu}}^{-1} \sum_{k=1}^{K} \bm{\mu}_k \right].
\end{eqnarray}

\noindent \underline{\textbf{$\sigma_\mu^2 | rest$}} \\
We consider the prior $\sigma_\mu^2 \sim \textrm{Inverse-Gamma}(a_\mu, b_\mu)$. The posterior of $\sigma_\mu^2$ given $rest$ is
\begin{eqnarray}
\nonumber && \sigma_\mu^2 | rest \sim \textrm{Inverse-Gamma}\left(a_\mu + \frac{nK}{2}, b_\mu + \frac{1}{2} \sum_{k=1}^{K} \bm{\mu}_k' \bm{\Sigma_{\mu}}^{-1} \bm{\mu}_k \right).
\end{eqnarray}
	
\noindent \underline{$ \Omega_{\mu} = \lbrace \rho_{\mu}, \tilde{\rho}_{\mu}, \nu_{\mu}, \psi_{\mu}, \gamma_{\mu} \rbrace | rest$} \\
The parameters $\rho_{\mu}, \tilde{\rho}_{\mu}, \nu_{\mu}$, and $\psi_{\mu}$ are updated similar to $\rho_k, \tilde{\rho}_k, \nu_k$, and $\psi_k$, respectively. We describe the case of $\gamma_{\mu}$ only. While updating $\gamma_\mu$, the candidates are generated similar to $a_k$.  At the $m$-th MCMC step, suppose $\bm{\Sigma_\mu}^{(m)}$ and $\bm{\Sigma_\mu}^{(c)}$ denote the correlation matrices based on $\{\rho_\mu^{(m)}, \tilde{\rho}_\mu^{(m)}, \nu_\mu^{(m)}, \psi_\mu^{(m)}, \gamma_\mu^{(m)}\}$ and $\{\rho_\mu^{(m)}, \tilde{\rho}_\mu^{(m)}, \nu_\mu^{(m)}, \psi_\mu^{(m)}, \gamma_\mu^{(c)}\}$, respectively, where $\gamma_\mu^{(c)}$ is a candidate from $\gamma_{\mu}$.
The acceptance ratio is 
\begin{eqnarray}
\nonumber R &=& \frac{\prod_{k=1}^{K} f_{\textrm{Normal}_n}\left( \bm{\mu}_k; \bm{B} \bm{\beta}, \sigma_\mu^2 \bm{\Sigma_\mu}^{(c)} \right) }{\prod_{k=1}^{K} f_{\textrm{Normal}_n}\left( \bm{\mu}_k; \bm{B} \bm{\beta}, \sigma_\mu^2 \bm{\Sigma_\mu}^{(m)} \right)} \times \frac{\gamma_{\mu}^{(c)}  \left(1 - \gamma_{\mu}^{(c)}\right)}{\gamma_{\mu}^{(m)} \left(1 - \gamma_k^{(m)}\right)}.
\end{eqnarray}

For updating the parameters $\rho_{\mu}, \tilde{\rho}_{\mu}, \nu_{\mu}$, and $\psi_{\mu}$, the acceptance ratios would be similarly defined; here the first terms of the acceptance ratios of $\rho_k, \tilde{\rho}_k, \nu_k$, and $\psi_k$ are replaced by the first term of the acceptance ratio for $\gamma_{\mu}$, along with properly redefined $\bm{\Sigma_\mu}^{(m)}$ and $\bm{\Sigma_\mu}^{(c)}$.

\subsection*{Prediction}
Our main objective is to generate spatial maps of FFWI at high quantiles over a fine grid across the spatial domain of interest, say, $\mathcal{S}_P = \lbrace \bm{s}_{P,1}, \ldots, \bm{s}_{P,m} \rbrace$ where the spatial process is unobserved. Similar to \cite{gelfand2005bayesian}, we are not interested in predicting $Y_t(\bm{s})$ at a $t \in \lbrace 1, \ldots, T \rbrace$ for some $\bm{s} \in \mathcal{S}_P$ but we want a new replication of the spatial process and hence we use the subscript $``0"$. Even if the observations are temporally dependent, we are interested in the marginal spatial process that is stationary across time. Let, $\bm{Y}^{(P)}_0 = [Y_0(\bm{s}_{P,1}), \ldots, Y_0(\bm{s}_{P,m})]'$ denotes the prediction at $\mathcal{S}_P$, $\bm{B}^{(P)}$ denotes the design matrix formed based on $\bm{B}(\bm{s}_{P,1}), \ldots, \bm{B}(\bm{s}_{P,m})$ and $\bm{A}^{(P)}$ denotes the design matrix formed based on region-specific indicators $\bm{A}(\bm{s}_{P,1}), \ldots, \bm{A}(\bm{s}_{P,m})$. After GEV-log transformation of $\bm{Y}^{(P)}_0$, $\bm{Y}^{*(P)}_0 = [Y_0^*(\bm{s}_{P,1}), \ldots, Y_0^*(\bm{s}_{P,m})]'$. Let $L$ be the number of posterior samples (after thinning of the post burn-in samples) we use for prediction.

Suppose we denote the $l$-th posterior sample from $\bm{\pi}$ by $\bm{\pi}^{(l)} = (\pi^{(l)}_1, \ldots, \pi^{(l)}_K)'$. First, we generate a sample $g_0^{(l)}$ from the distribution of cluster index $\textrm{Pr}(g_0 = k) = \pi_k^{(l)}$. Suppose, $g_0^{(l)} = k$. The samples $\bm{\beta}^{(l)}$, $\bm{\mu}_k^{(l)}$, $\sigma_{\mu}^{2(l)}$ and $\Omega_{\mu}^{(l)}$ are available from MCMC. Let the spatial correlation matrices for $\mathcal{S}$ and $\mathcal{S}_P$ be $\bm{\Sigma_{\mu}}^{(l)}$ and $\bm{\Sigma_{\mu}}^{(P,P)(l)}$, respectively and the cross-correlation matrix between $\mathcal{S}_P$ and $\mathcal{S}$ be $\bm{\Sigma_{\mu}}^{(P)(l)}$-- all three matrices are calculated based on $\Omega_{\mu}^{(l)}$. We draw a sample $\bm{\mu}_k^{(P)(l)}$ from $\bm{\mu}_k^{(P)}$ following 
\begin{eqnarray}
\nonumber && \bm{\mu}_k^{(P)} | \bm{\mu}_k = \bm{\mu}_k^{(l)} \sim \textrm{Normal}_{m} \left( \bm{B}^{(P)} \bm{\beta}^{(l)} + \bm{\Sigma_{\mu}}^{(P)(l)} \bm{\Sigma_{\mu}}^{(l)-1} \left(\bm{\mu}_k^{(l)} - \bm{B} \bm{\beta}^{(l)} \right), \right. \\
\nonumber &&~~~~~~~~~~~~~~~~~~~~~~~~~~~~~~~~~~~~~~ \left. \sigma_{\mu}^{2(l)} \left( \bm{\Sigma_{\mu}}^{(P,P)(l)} - \bm{\Sigma_{\mu}}^{(P)(l)} \bm{\Sigma_{\mu}}^{(l)-1} \bm{\Sigma_{\mu}}^{(P)(l)'}  \right)^{-1}  \right).
\end{eqnarray}


Let the random scaling term for time $``0"$ be $\sigma^{2}_0$. The posterior samples $a_k^{(l)}$ and $b_k^{(l)}$ are available from MCMC, and we draw a sample $\sigma^{2(l)}_0$ from $\sigma^{2}_0 \sim \textrm{Inverse-Gamma}(a_k / 2, a_k b_k / 2)$. Subsequently, we draw a sample $z_0^{(l)}$ from $z_0 \sim \textrm{Normal}(0, \sigma^{2(l)}_0)$. Based on the $l$-th posterior sample $\tilde{\Omega}_k^{(l)} = \{ \rho_k^{(l)}, \tilde{\rho}_k^{(l)}, \nu_k^{(l)}, \psi_k^{(l)} \}$, let the spatial correlation matrices (without nugget) corresponding to the $k$-th STP component for $\mathcal{S}$ and $\mathcal{S}_P$ be $\tilde{\bm{\Sigma}}_k^{(l)}$ and $\tilde{\bm{\Sigma}}_k^{(P,P)(l)}$, respectively, and the cross-correlation matrix between $\mathcal{S}_P$ and $\mathcal{S}$ be $\tilde{\bm{\Sigma}}_k^{(P)(l)}$. First, we generate a sample $\bm{X}^{(l)}_0$ following $\bm{X}_0 \sim \textrm{Normal}_n(\bm{0}_n, \gamma^{(l)}_k \sigma_0^{2(l)} \tilde{\bm{\Sigma}}^{(l)}_k)$. The conditional distribution of $\bm{X}^{(P)}_0$, given $\bm{X}_0= \bm{X}_0^{(l)}$ and other parameters, is
$$\bm{X}^{(P)}_0 | rest \sim \textrm{Normal}_{m} \left( \tilde{\bm{\Sigma}}_k^{(P)(l)} \tilde{\bm{\Sigma}}_k^{(l)-1} \bm{X}_0^{(l)}, \gamma^{(l)}_k \sigma_0^{2(l)} \left( \tilde{\bm{\Sigma}}_k^{(P,P)(l)} - \tilde{\bm{\Sigma}}_k^{(P)(l)} \tilde{\bm{\Sigma}}_k^{(l)-1} \tilde{\bm{\Sigma}}_k^{(P)(l)'}  \right)^{-1}  \right).$$

We draw a sample $\bm{X}^{(P)(l)}_0$ from $\bm{X}^{(P)}_0 | rest$. Subsequently, we draw a sample $\bm{Y}^{*(P)(l)}_0$ from the full conditional posterior distribution 
$$\bm{Y}^{*(P)}_0 | rest \sim \textrm{Normal}_m\left(\bm{\mu}_k^{(P)(l)} + \bm{X}^{(P)(l)}_0 + \bm{A}^{(P)} \bm{\lambda}_k^{(l)} \lvert z_0^{(l)} \rvert, (1 - \gamma_k^{(l)}) \sigma_0^{2(l)} \bm{I}_m\right)$$

Finally, a sample from $\bm{Y}^{(P)}_0$ is obtained following the inverse GEV-log transformation of $\bm{Y}^{*(P)(l)}_0$. We draw inference at new spatial locations $\mathcal{S}_P$ based on these samples.

\vspace{-1mm}

\section*{Appendix K: Model diagnostics}

Here we provide some model diagnostics, comparing the estimates based on the STP-DPM model and the empirical estimates. To compare the performance, as a reference, we also provide the results based on the highly flexible sub-model GP, where we fit a Gaussian process with unknown spatial mean surface and anisotropic Mat\'ern correlation function to the GEV-log transformed observations; all the parameters are assumed to be unknown and we fit a fully Bayesian model. 

The proposed STP-DPM model allows finite first and second order moments only when the degrees of freedom parameters for all the mixture components are larger than two. To allow heavier tail, we allow those parameters to be larger than 0.5; theoretically, the posterior probability of such parameters being less than two is nonzero. Hence, the posterior mean and variance do not exist for the proposed model with a nonzero probability. Considering this fact, we discuss results based on the median, along with low--through--high quantiles instead of the mean, the inter-quartile range (IQR) instead of the standard deviation, and the Spearman's rank correlation instead of the Pearson's correlation. While these assess the performance of the model in fitting the bulk of the data, we also discuss the estimated pairwise tail-dependence measure $\chi_u$ for $u=0.95$.

\begin{figure}[h]
\includegraphics[width=0.32\linewidth]{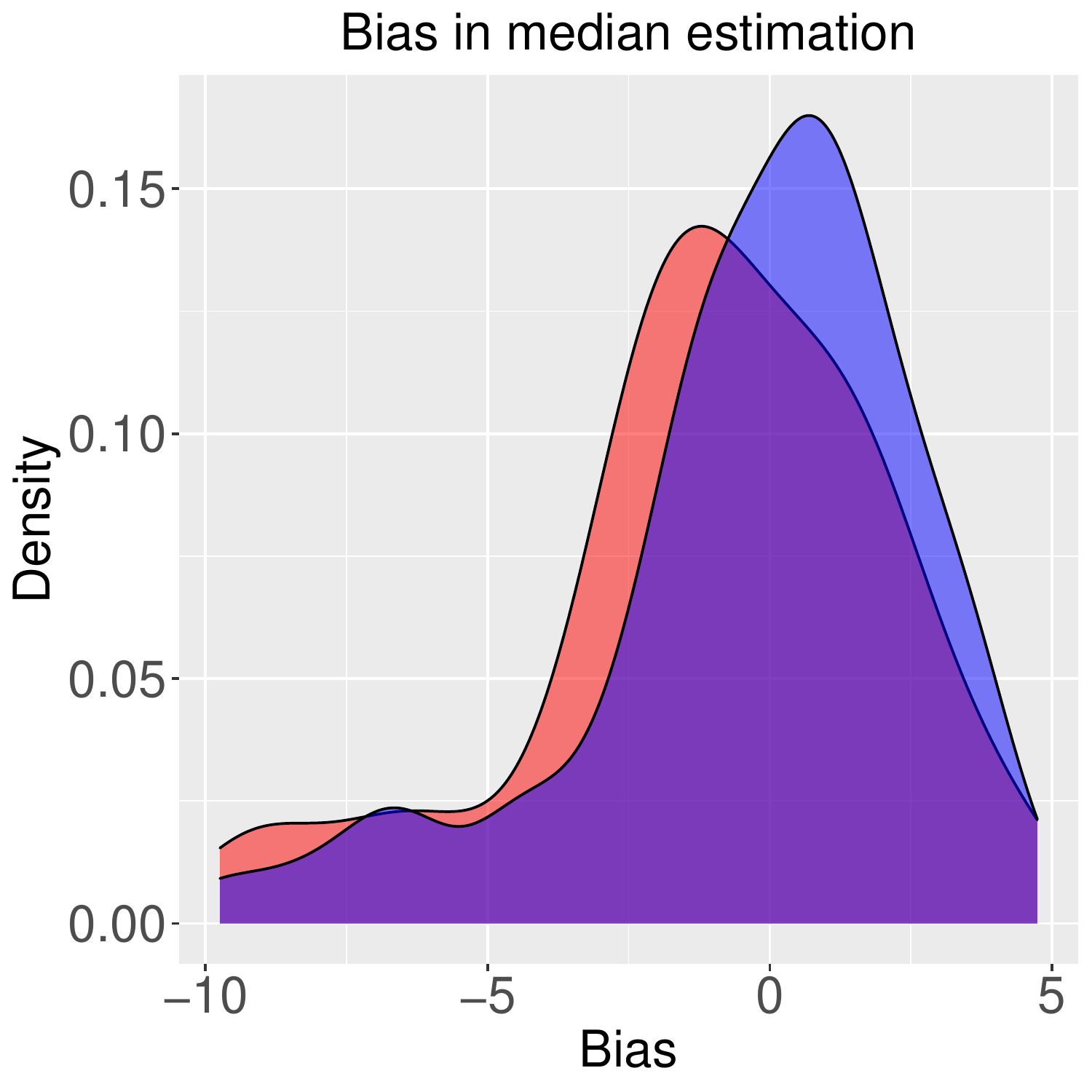} 
\includegraphics[width=0.32\linewidth]{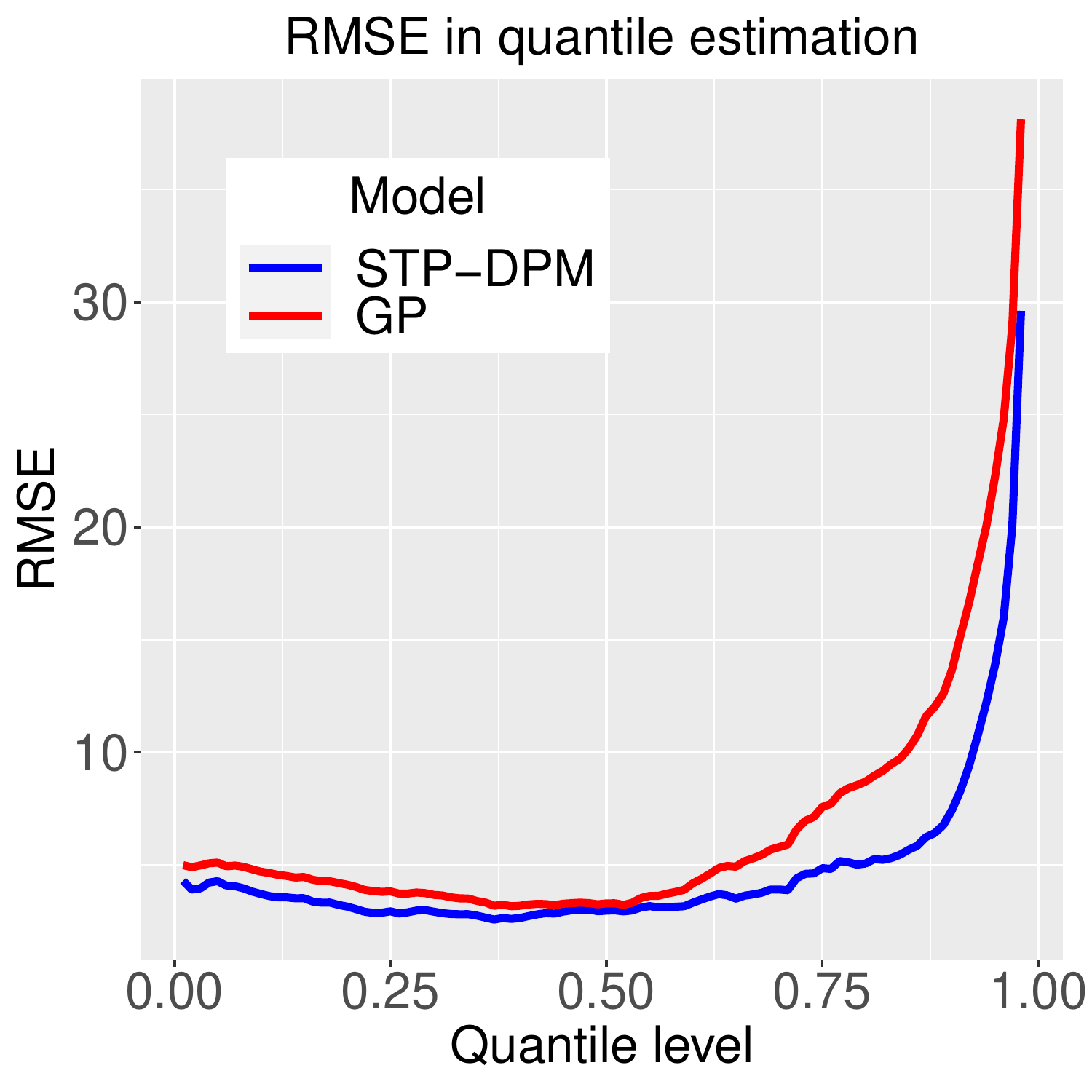}
\includegraphics[width=0.32\linewidth]{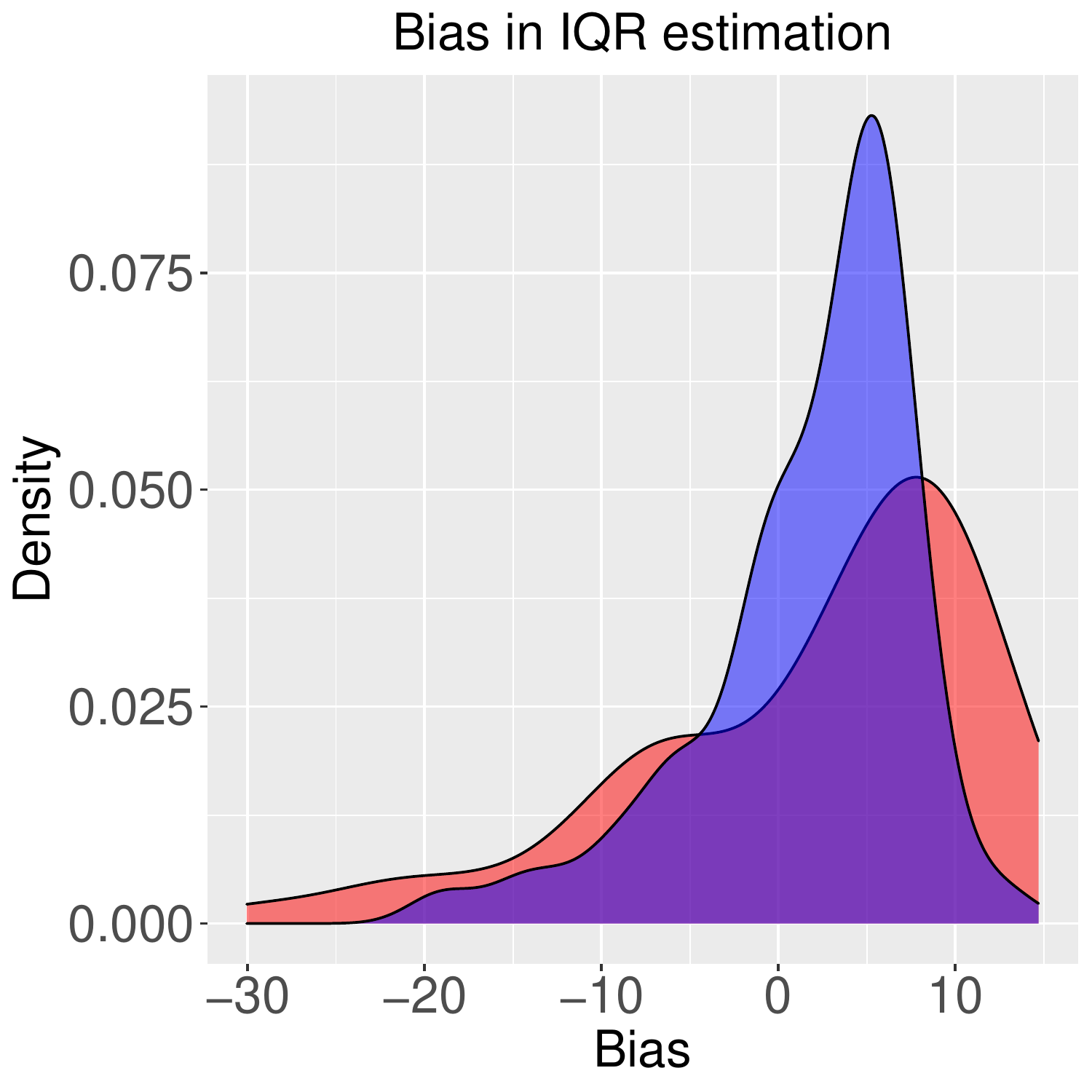}
\vspace{-2mm}
\caption{Comparison of biases or fitting RMSEs in estimating the station-wise medians (left), low--through--high quantiles (middle), and inter-quartile ranges (right), between the models STP-DPM (blue) and GP (red).}
\label{fig_diagnostics_marginal}
\end{figure}

The kernel densities of the ``biases'' in estimating the station-wise medians (estimates based on model fitting, minus the corresponding empirical estimates, henceforth) are presented in the left panel of Figure \ref{fig_diagnostics_marginal}. While both the models STP-DPM and GP underestimate the medians for some stations (thick left-tail for both the cases), the biases are closer to zero for most of the stations in case of the STP-DPM model, compared to the biases in case of fitting the GP model. The medians of the biases are 0.16 and -0.93 for the models STP-DPM and GP, respectively.

To compare the fitting at low--through--high quantile levels, for a model $M$, we calculate the fitting RMSE for a quantile $q$ as $\textrm{RMSE}^{fit}_{M} = \sqrt{n^{-1} \sum_{i=1}^{n}[\tilde{F}_i^{-1}(q) - {F}_i^{-1}(q)]^2}$, where $\tilde{F}_i$ and ${F}_i$ denote the CDF of the the posterior fitted distribution and the empirical CDF at site $i$ respectively. For $q = 0.01, \ldots, 0.98$, we present the $\textrm{RMSE}^{fit}_{M}$ for the models STP-DPM and GP in the middle panel of Figure \ref{fig_diagnostics_marginal}. Between the quantiles 0.01--0.75, the RMSE is smaller than 5 for the STP-DPM model. While the RMSE is higher for the very high quantiles, the STP-DPM model still performs better than GP.

The kernel densities of the biases in estimating the station-wise IQRs are presented in the right panel of Figure \ref{fig_diagnostics_marginal}. While both STP-DPM and GP underestimate the IQRs for some stations (thick left-tail for both the cases similar to that for the medians), the kernel density curve for the STP-DPM model has a sharper peak near zero. Out of the 61 stations, fitting STP-DPM leads to biases larger than 10 only for 4 stations, while GP leads to biases larger than 10 for 17 stations.


\begin{figure}
\includegraphics[width=0.32\linewidth]{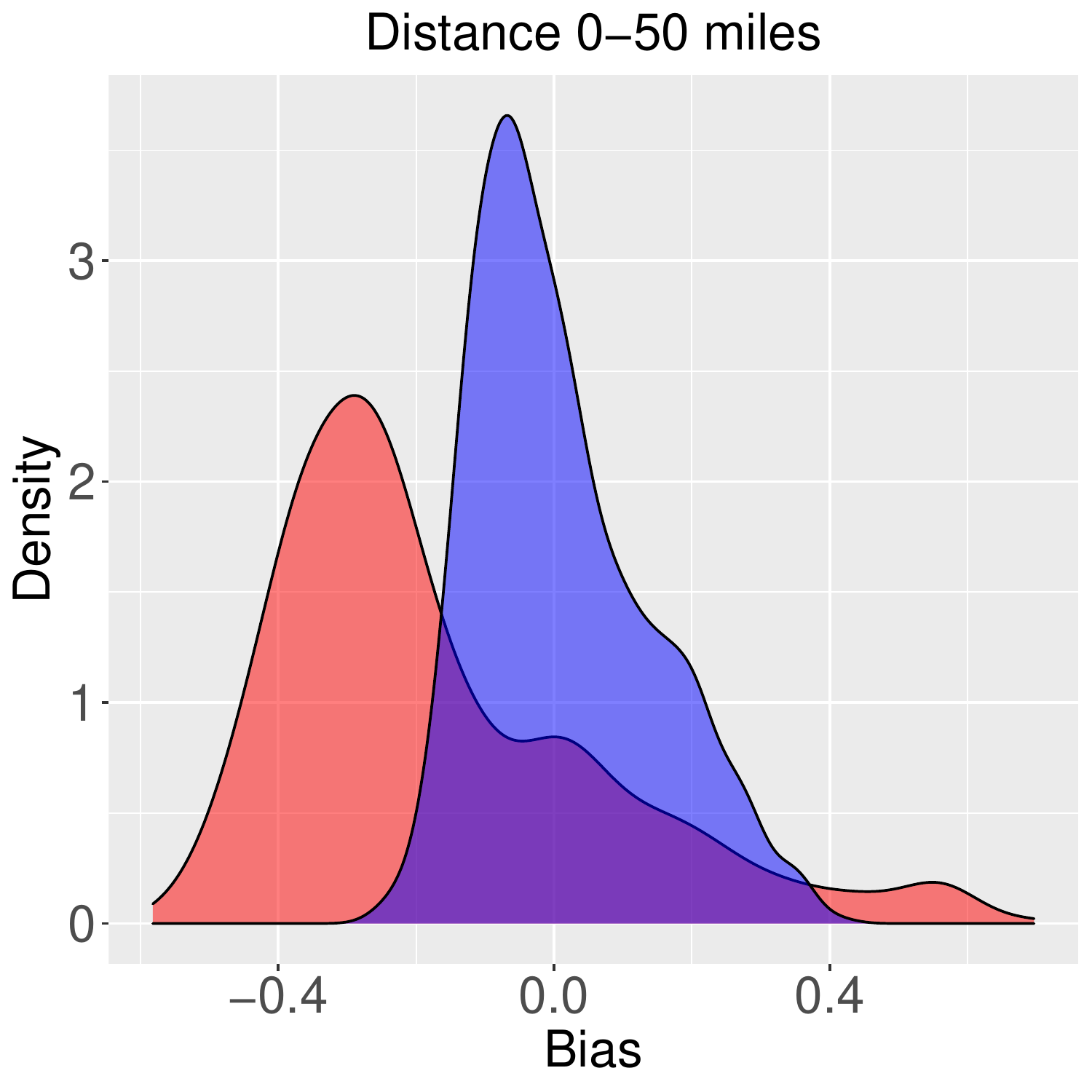} 
\includegraphics[width=0.32\linewidth]{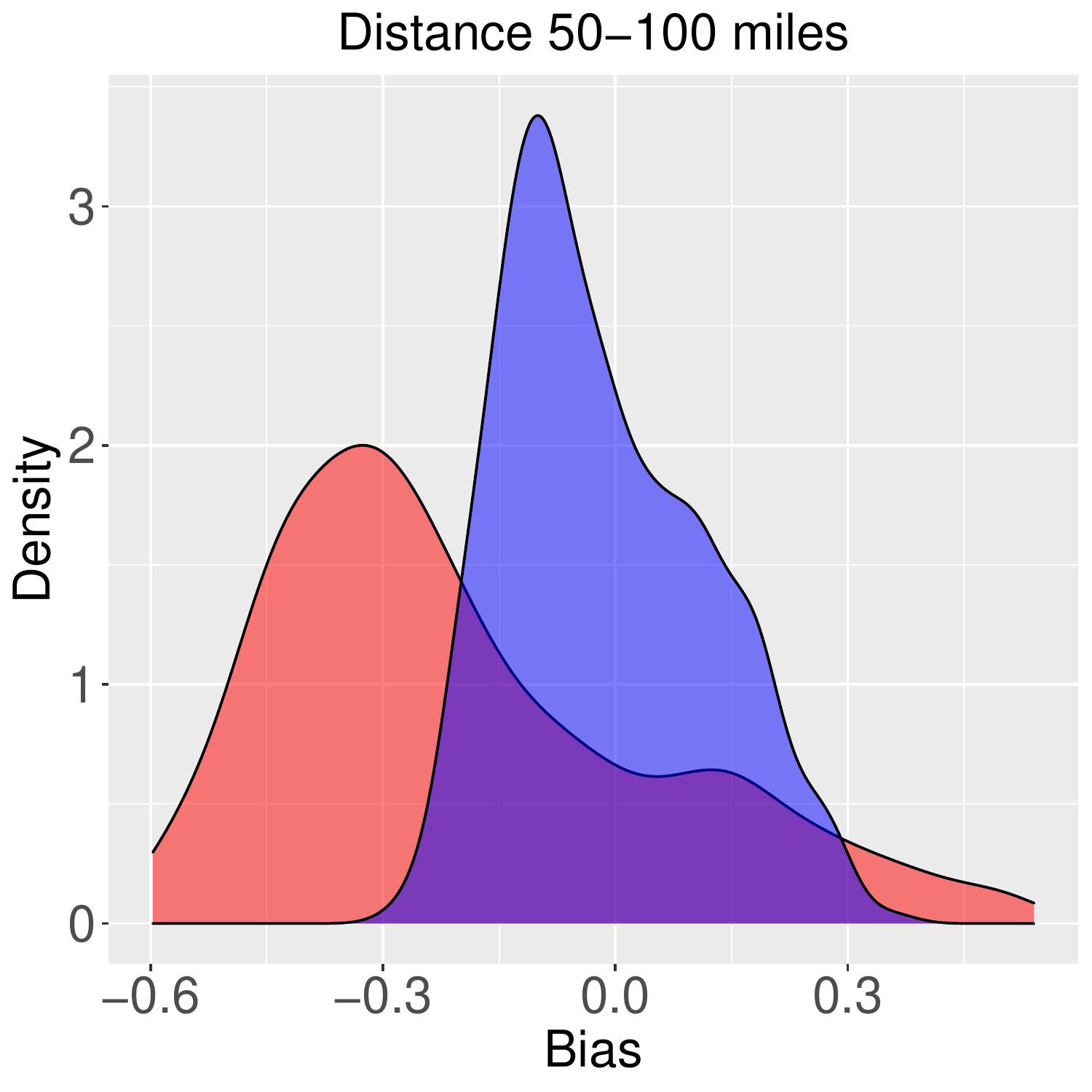}
\includegraphics[width=0.32\linewidth]{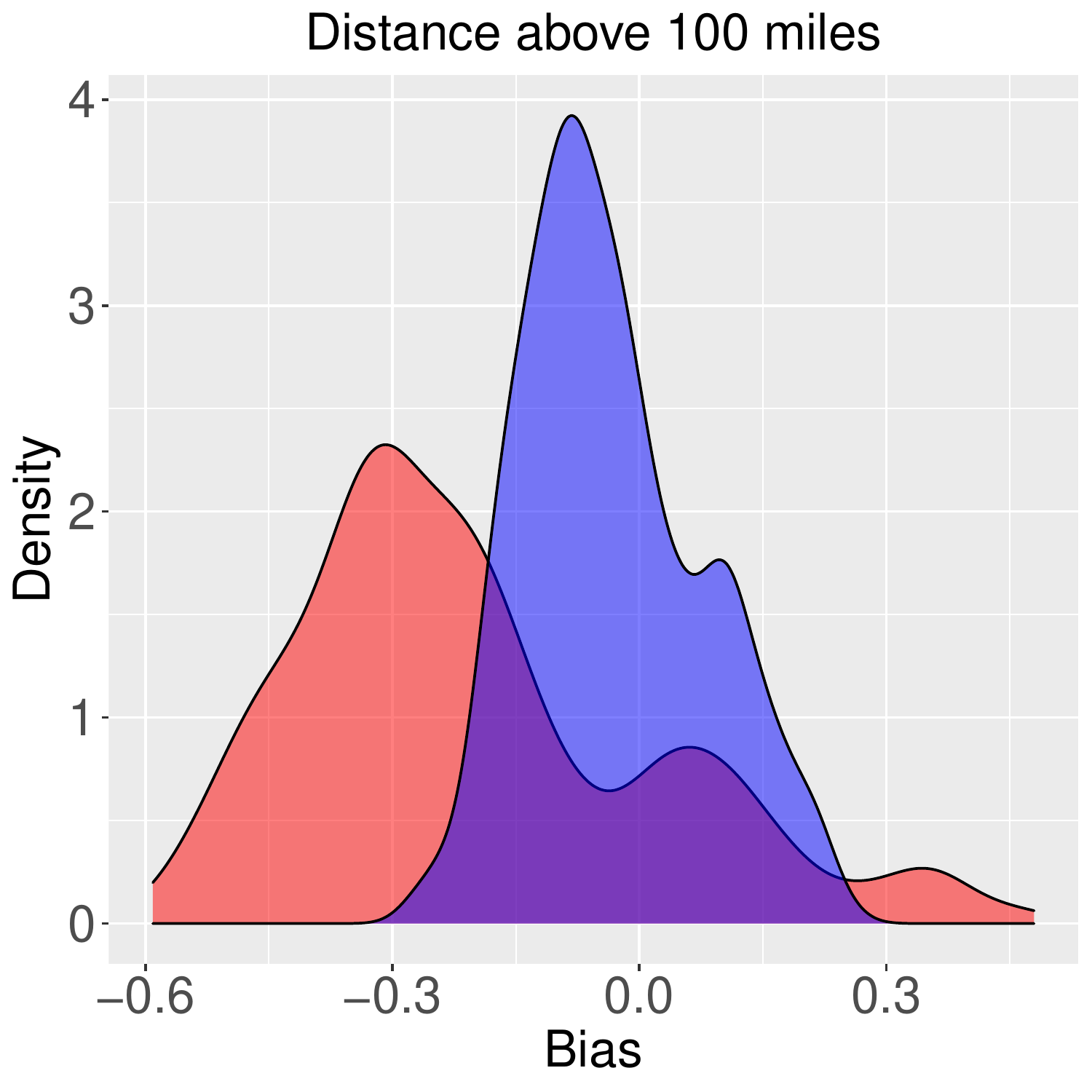}
\vspace{-2mm}
\caption{Kernel densities of the biases in estimating the Spearman's rank correlations between each pair of stations for the three cases-- Case 1: two (different) stations are closer than 50 miles (left), Case 2: distance between the two stations is between 50--100 miles (middle), and Case 3: distance between the two stations is above 100 miles (right), based on the models STP-DPM (blue) and GP (red).}
\label{fig_diagnostics_correl}
\end{figure}

Further, we present the kernel densities of the biases in fitting the pairwise Spearman's rank correlations in Figure \ref{fig_diagnostics_correl}. To assess the performances in both the cases-- when two stations are closer to each other or far from each other, we divide the pairs of stations into three categories-- Case 1: two (different) stations are closer than 50 miles (presented in the left panel), Case 2: distance between the two stations is between 50--100 miles (presented in the middle panel), and Case 3: distance between the two stations is above 100 miles (presented in the right panel). For the 61 stations we consider, the largest distance between two stations is 204.14 miles. While the biases are generally closer to zero for the STP-DPM model, GP underestimates the correlation heavily. The average biases for the three cases are -0.010, -0.016, and -0.033, respectively, for STP-DPM, and -0.166, -0.194, and -0.200, respectively, for GP.

\begin{figure}[h]
\includegraphics[width=0.32\linewidth]{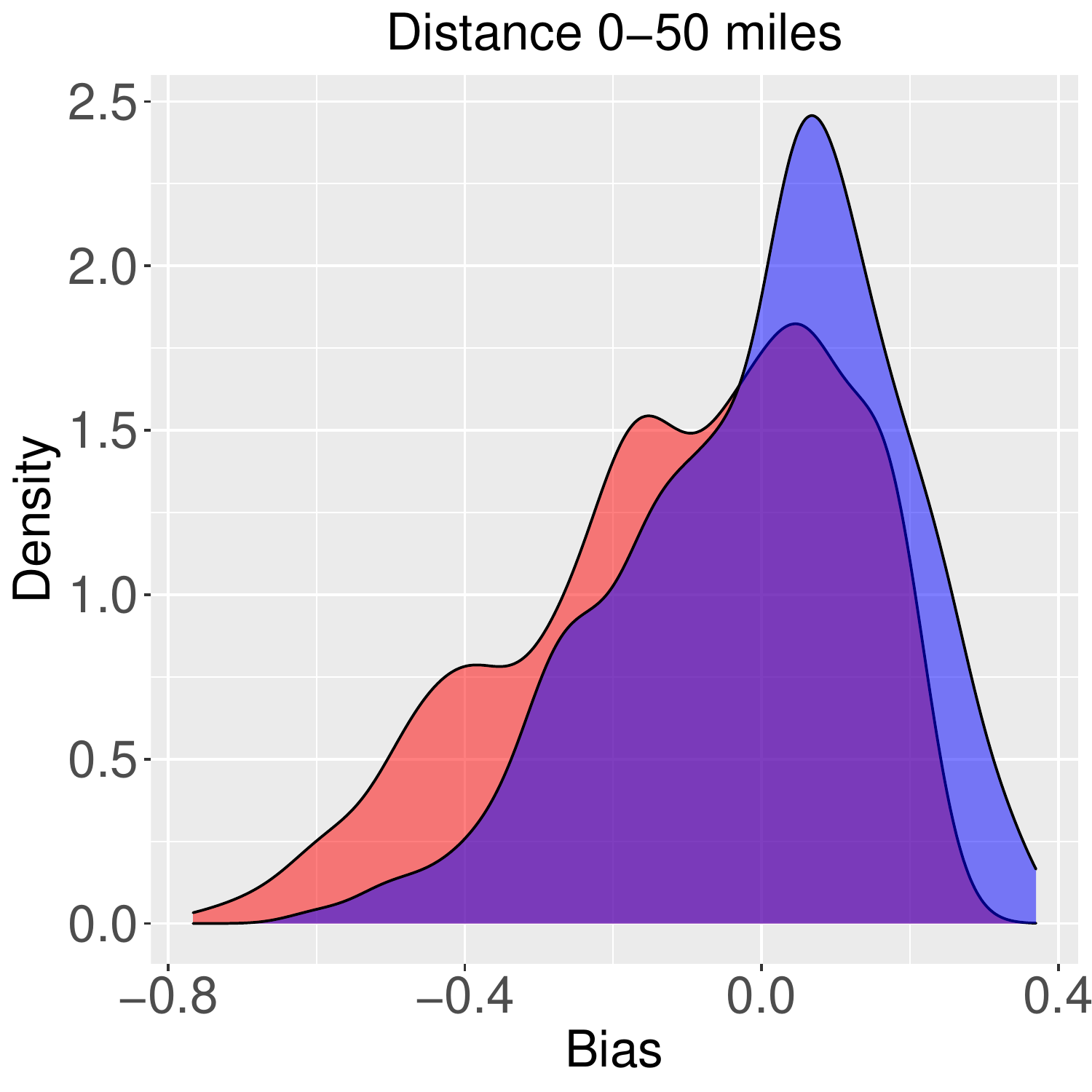} 
\includegraphics[width=0.32\linewidth]{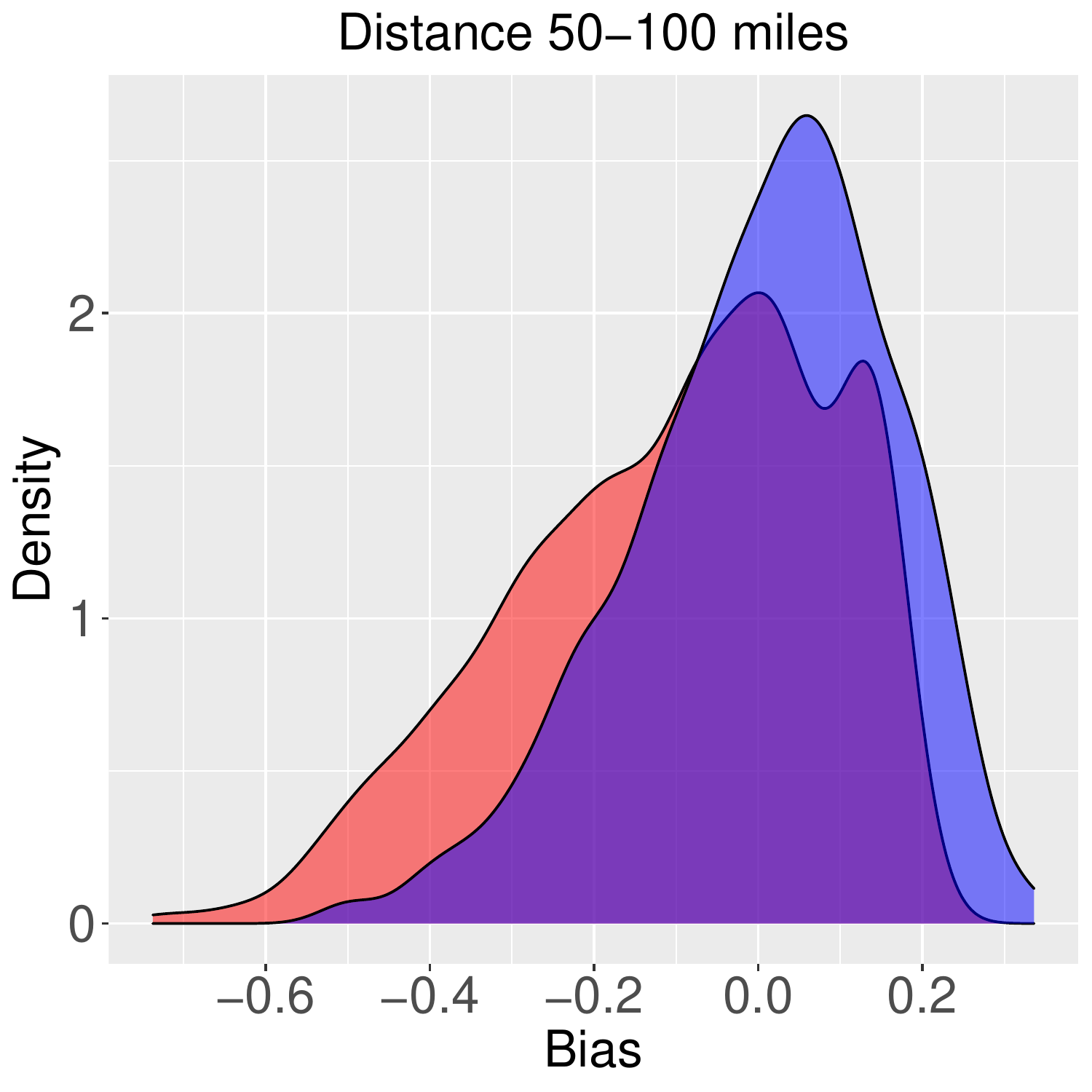}
\includegraphics[width=0.32\linewidth]{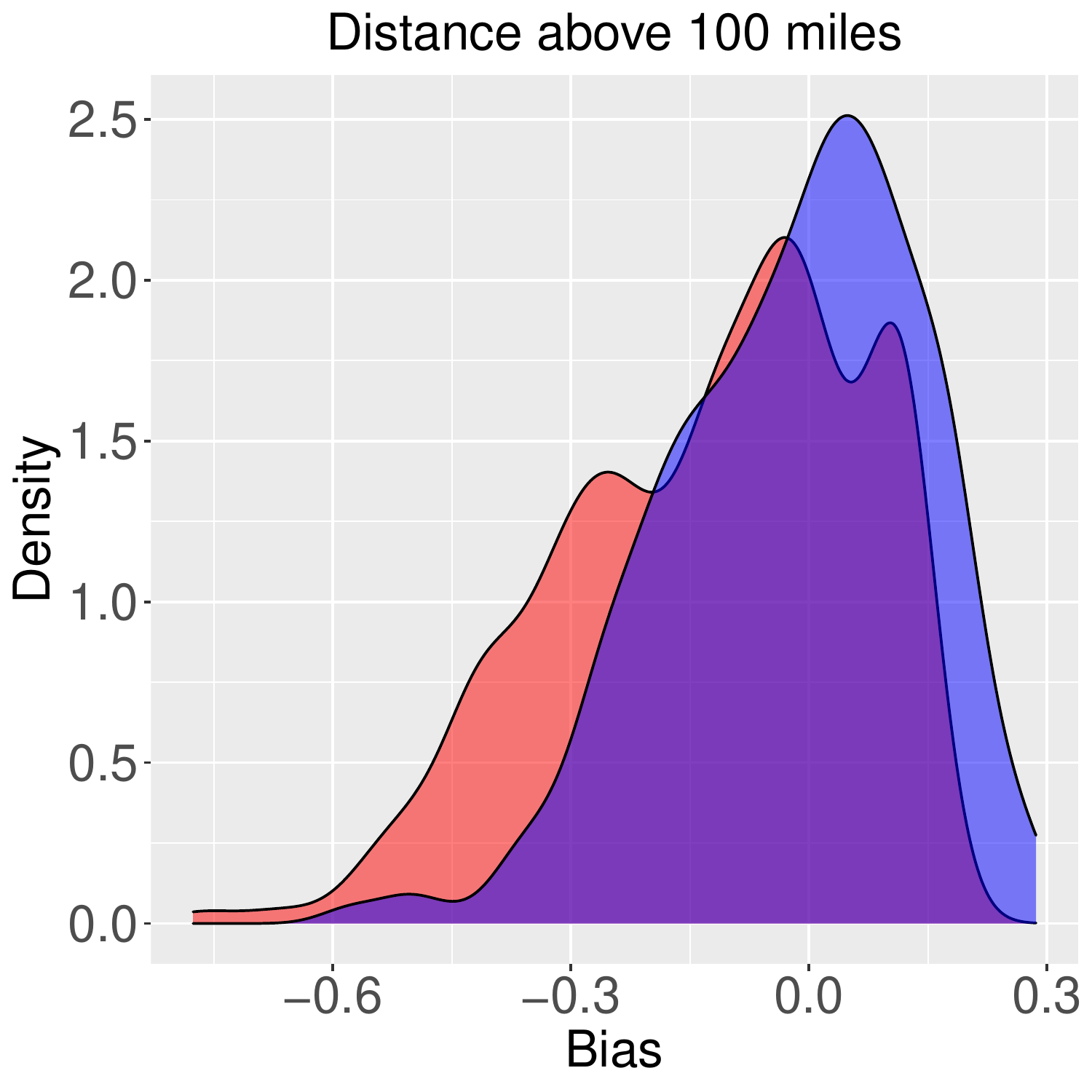}
\vspace{-2mm}
\caption{Kernel densities of the biases in estimating the tail-dependence measure $\chi_u$ for $u=0.95$, between each pair of stations for the three cases-- Case 1: two (different) stations are closer than 50 miles (left), Case 2: distance between the two stations is between 50--100 miles (middle), and Case 3: distance between the two stations is above 100 miles (right), based on the models STP-DPM (blue) and GP (red).}
\label{fig_diagnostics_chi}
\end{figure}

Further, we present the kernel densities of the biases in estimating the tail-dependence measure $\chi_u$ for $u=0.95$, between each pair of stations, for the three cases similar to that of Spearman's rank correlations, in Figure \ref{fig_diagnostics_chi}. Unlike the inference for correlation, here the biases are generally higher and also involves high variance (not shown) of the estimators, which is natural for tail-inference. The kernel densities of the biases are approximately centered around zero for the STP-DPM model, while GP underestimates $\chi_u$ for most of the pairs, possibly due to the thin joint tails of the Gaussian processes. The average biases for the three cases are -0.002, 0.002, and -0.020, respectively, for STP-DPM, and -0.107, -0.103, and -0.124, respectively, for GP.  


\bibliographystyle{rss}
\bibliography{hazra_et_al_2020}
\end{document}